\documentclass[11pt]{article}
\usepackage{natbib}
\usepackage{amsfonts}
\usepackage{amssymb}
\usepackage{amsmath}
\usepackage{lscape}
\usepackage{color}
\usepackage[left=1.7cm, right=2.5cm, top=1.7cm, bottom=1.7cm]{geometry}

\definecolor{DarkO}{cmyk}{0.0, 0.43, 0.90, 0.13}

\DeclareMathOperator*{\argmax}{arg\,max}
\DeclareMathOperator*{\argmin}{arg\,min}

\begin{document}

\title{Computing Bayes: Bayesian Computation\medskip\ \\
from 1763 to the 21st Century\thanks{{\footnotesize The first author
presented a historical overview of Bayesian computation, entitled `Computing
Bayes: Bayesian Computation from 1763 to 2017!', at `Bayes on the Beach'
(Queensland, November, 2017) and Monash University (March, 2018). She would
like to thank participants at both presentations for inspiring the eventual
writing of this paper. Martin and Frazier have been supported by Australian
Research Council Discovery Grant No. DP170100729 and the Australian Centre
of Excellence in Mathematics and Statistics. Frazier has also been supported
by Australian Research Council Discovery Early Career Researcher Award
DE200101070. Robert has been partly supported by a senior chair (2016-2021)
from l'Institut Universitaire de France and by a Prairie chair from the
Agence Nationale de la Recherche (ANR-19-P3IA-0001). The authors would also
like to thank the Editor, an associate editor and two anonymous reviewers
for very constructive and insightful comments on an earlier draft of the
paper.}}}
\author{Gael M. Martin\thanks{{\footnotesize Corresponding author:
gael.martin@monash.edu.}}, David T. Frazier and Christian P. Robert}
\maketitle

\begin{abstract}
\baselineskip14ptThe Bayesian statistical paradigm uses the language of
probability to express uncertainty about the phenomena that generate
observed data. Probability distributions thus characterize Bayesian
analysis, with the rules of probability used to transform prior probability
distributions for all unknowns --- parameters, latent variables, models ---
into posterior distributions, subsequent to the observation of data.
Conducting Bayesian analysis requires the evaluation of integrals in which
these probability distributions appear. Bayesian computation is all about
evaluating such integrals in the typical case where no analytical solution
exists. This paper takes the reader on a chronological tour of Bayesian
computation over the past two and a half centuries. Beginning with the
one-dimensional integral first confronted by Bayes in 1763, through to
recent problems in which the unknowns number in the millions, we place all
computational problems into a common framework, and describe all
computational methods using a common notation. The aim is to help new
researchers in particular --- and more generally those interested in
adopting a Bayesian approach to empirical work --- {make sense of the
plethora of computational techniques that are now on offer; understand when
and why different methods are useful; and see the links that do exist,
between them all.}

\bigskip


\bigskip

\emph{MSC2010 Subject Classification}: 62-03, 62F15, 65C60
\end{abstract}

\newpage

\section{The Beginning}

\baselineskip17.5pt

\textit{December 23 1763: London.} Richard Price reads to the Royal Society
a paper penned by a past Fellow, the late Thomas Bayes:\smallskip

\begin{center}
`\textit{An Essay Towards Solving a Problem in the Doctrine of Chances.}%
'\smallskip
\end{center}

\noindent With that {reading}, the concept of `inverse probability' --- 
\textit{Bayesian inference} as we know it now --- has its first public
airing.

To our modern eyes, the problem tackled by Bayes in his essay is a simple
one: If one performs $n$ independent Bernoulli trials, with a probability, $%
\theta $, of success on each trial, what is the probability --- given $n$
outcomes --- of $\theta $ lying between two values, $a$ and $b$? The answer
Bayes offered is equally simple to re-cast in modern terminology. Define $%
Y_{i}|\theta \sim i.i.d.$ $Bernoulli(\theta )$, $i=1,2,...,n$; record the
observed sequence of successes ($Y_{i}=1$) and failures ($Y_{i}=0$) as $%
\mathbf{y}=(y_{1},y_{2},...,y_{n})^{\prime }$; denote by $p(\mathbf{y}%
|\theta )$ the likelihood function for $\theta $; and invoke a Uniform
prior, $p(\theta )$, on the interval $(0,1)$. Bayes sought:%
\begin{equation}
\mathbb{P}(a<\theta <b|\mathbf{y})=\int\limits_{a}^{b}p(\theta \mathbf{|y}%
)d\theta ,  \label{Bayes_prob}
\end{equation}%
where $p(\theta \mathbf{|y})$ denotes the posterior probability density
function (pdf) for $\theta $,%
\begin{equation}
p(\theta \mathbf{|y})=\frac{p(\mathbf{y}|\theta )p(\theta )}{p(\mathbf{y})},
\label{posterior_pdf}
\end{equation}%
$p(\mathbf{y})=\int_{0}^{1}p(\mathbf{y}|\theta )p(\theta )d\theta $ defines
the marginal likelihood, and the scale factor $\left[ p(\mathbf{y})\right]
^{-1}$ in (\ref{posterior_pdf}) ensures that $p(\theta |\mathbf{y})$
integrates to one. Given the Bernoulli assumption for $Y$, the Uniform prior
on $\theta $, and defining $x=\Sigma _{i=1}^{n}y_{i}$, $p(\theta \mathbf{|y})
$ has a closed-form representation as the Beta pdf, 
\begin{equation}
p(\theta \mathbf{|y})=\left[ B(x+1,n-x+1)\right] ^{-1}\theta ^{x}(1-\theta
)^{n-x},  \label{beta_pdf}
\end{equation}%
where $B(x+1,n-x+1)=\Gamma (x+1)\Gamma (n-x+1)/\Gamma
(n+2)=\int_{0}^{1}\theta ^{x}(1-\theta )^{n-x}d\theta $ is the Beta
function, and $\Gamma (x)$ is the Gamma function. Bayesian inference ---
namely, quantification of uncertainty about an unknown $\theta $,
conditioned on known data, $\mathbf{y}$ --- thus first emerges as the
analytical solution to a particular inverse probability problem.\footnote{%
Bayes expressed this problem in terms of its equivalent representation as
one of deriving the probability of $a<\theta <b$ conditional on the value of 
$x$, where $X|\theta \sim $ $\mathcal{B}(n,\theta ).$ We have chosen to
denote the conditioning values explicitly as a \textit{sample} of n
(Bernoulli) observations, $\mathbf{y}$, in order to establish the notation $%
p(\theta \mathbf{|y})$ from the outset. Due to the sufficiency of $X$, $%
p(\theta \mathbf{|}x)$ is of course equivalent to $p(\theta \mathbf{|y})$.
Bayes also cast this problem in physical terms: as one in which balls were
rolled across a square table, or plane. Over time his pictorial
representation of the problem has come to be viewed as a `billiard table',
despite Bayes making no overt reference to such an item in his essay. For
this, and other historical anecdotes, see \cite{stigler:1986} and \cite%
{fienberg:2006}.}

Bayes, however, did not seek the pdf in (\ref{beta_pdf}) \textit{per se}.
Rather, he sought to evaluate the probability in (\ref{Bayes_prob}) which,
for either $a\neq 0$ or $b\neq 1$, {involved evaluation of the }incomplete{\
Beta function. Except for the case when either }$x$ {or }$(n-x)$ {were
small, a closed-form solution to }(\ref{Bayes_prob}) {eluded Bayes. }Hence,
along with the elegance of (\ref{posterior_pdf}) --- `Bayes' theorem' {as it
is now commonly known} --- and the availability of the analytical {expression%
} in (\ref{beta_pdf}), {came} the need to {approximate}, {or }\textit{compute%
}, {the} quantity {of interest in} (\ref{Bayes_prob}). The quest for a
computational {solution to a} Bayesian problem was\ thus born.

\section{Bayesian Computation in a Nutshell}

\subsection{The Computational Problem\label{general}}

Bayes' probability of interest in (\ref{Bayes_prob}) can, of course, be
expressed as a posterior expectation, $\mathbb{E}(I_{[a,b]}\mathbf{|y)}=$%
\newline
$\int_{\Theta }\mathbb{I}_{[a,b]}p(\theta \mathbf{|y})d\theta ,$ where the
indicator function $\mathbb{I}_{[a,b]}$ equals $1$ if $a<\theta <b$, and
equals $0$ otherwise. Generalizing at this point to any problem with unknown 
$\boldsymbol{\theta }=(\theta _{1},\theta _{2},...,\theta _{p})^{\prime }\in 
$ ${\Theta }$ and joint posterior pdf $p(\boldsymbol{\theta }\mathbf{|y})$,
most Bayesian quantities of interest are posterior\textit{\ }expectations of
some function $g(\boldsymbol{\theta })$ and, hence, can be expressed as,

\begin{equation}
\mathbb{E}(g(\boldsymbol{\theta }\mathbf{)|y)}=\int_{{\Theta }}g(\boldsymbol{%
\theta }\mathbf{)}p(\boldsymbol{\theta }\mathbf{|y})d\boldsymbol{\theta }%
\mathbf{.}  \label{gen_expect}
\end{equation}%
In addition to posterior probabilities like that of Bayes, familiar examples
include posterior moments, such as $\mathbb{E}(\boldsymbol{\theta }\mathbf{%
|y)}=\int_{{\Theta }}\boldsymbol{\theta }p(\boldsymbol{\theta }\mathbf{|y})d%
\boldsymbol{\theta }\ $and $\text{Var}(\boldsymbol{\theta }\mathbf{|y)}%
=\int_{{\Theta }}\left[ \boldsymbol{\theta }-\mathbb{E}(\boldsymbol{\theta }%
\mathbf{|y)}\right] \left[ \boldsymbol{\theta }-\mathbb{E}(\boldsymbol{%
\theta }\mathbf{|y)}\right] ^{\prime }p(\boldsymbol{\theta }\mathbf{|y})d%
\boldsymbol{\theta }\mathbf{,}$ plus marginal quantities like $p(\theta
_{1}^{\ast }|\mathbf{y})=\int_{{\Theta }}p(\theta _{1}^{\ast }|\theta
_{2},...,\theta _{p},\mathbf{y})p(\boldsymbol{\theta }\mathbf{|y})d%
\boldsymbol{\theta }$ (for $\theta _{1}^{\ast }$ a point in the support of $%
p(\theta _{1}|\mathbf{y})$). However, (\ref{gen_expect}) also subsumes the
case where $g(\boldsymbol{\theta }\mathbf{)}=$ $p(y_{n+1}^{\ast }|%
\boldsymbol{\theta }\mathbf{,y})$ (with $y_{n+1}^{\ast }$\textbf{\ }in the
support of the `out-of-sample' random variable, $y_{n+1}$), in which case (%
\ref{gen_expect}) defines the \textit{predictive} distribution for $y_{n+1}$%
, $p(y_{n+1}^{\ast }|\mathbf{y}).$ Further, it encompasses $g(\boldsymbol{%
\theta }\mathbf{)}=$ $L(\boldsymbol{\theta },d)$,\textbf{\ }for $L(%
\boldsymbol{\theta },d)$ the loss function associated with a decision $d$,
in which case (\ref{gen_expect}) is the quantity minimized in Bayesian
decision theory (\citealp{berger:1985}; \citealp{robert:2001}). Finally,
defining $g(\boldsymbol{\theta }\mathbf{)}=$ $p(\mathbf{y}|\boldsymbol{%
\theta },\mathcal{M})$ as the likelihood function that explicitly conditions
on the model, $\mathcal{M}$ say, the marginal likelihood of $\mathcal{M},$ $%
p(\mathbf{y}|\mathcal{M})$, is the expectation,%
\begin{equation}
\mathbb{E}(g(\boldsymbol{\theta }\mathbf{)|}\mathcal{M}\mathbf{)}=\int_{{%
\Theta }}g(\boldsymbol{\theta }\mathbf{)}p(\boldsymbol{\theta }|\mathcal{M})d%
\boldsymbol{\theta },  \label{gen_expect_prior}
\end{equation}%
with respect to the\ prior, $p(\boldsymbol{\theta }|\mathcal{M})$. The ratio
of (\ref{gen_expect_prior}) to the comparable quantity for an alternative
model defines the \textit{Bayes factor }for use in choosing between the two
models. In summary then, the key quantities that underpin all Bayesian
analysis --- inference, prediction, decision theory and model choice --- can
be expressed as expectations.

The need for numerical computation arises simply because analytical
solutions to (\ref{gen_expect}) and (\ref{gen_expect_prior}) are rare.
Indeed, Bayes' original problem highlights that a solution to (\ref%
{gen_expect}) can elude us \textit{even} when the posterior pdf itself has a
closed form. Typically, the posterior does not possess a closed form, as the
move from the generative problem (the specification of $p(\mathbf{y}|%
\boldsymbol{\theta })$) to the inverse problem (the production of $p(%
\boldsymbol{\theta }\mathbf{|y})$), yields a posterior that is known only up
to a constant of proportionality, as 
\begin{equation}
p(\boldsymbol{\theta }\mathbf{|y})\propto p(\mathbf{y}|\boldsymbol{\theta }%
)p(\boldsymbol{\theta });  \label{Bayes_proport}
\end{equation}%
exceptions to this including when $p(\mathbf{y}|\boldsymbol{\theta })$ is
from the exponential family, and either a natural conjugate, or convenient
noninformative prior is adopted {(}as in Bayes'{\ problem)}. The
availability of $p(\boldsymbol{\theta }\mathbf{|y})$ only up to the
integrating constant\textit{\ immediately}{\ precludes} the analytical
solution of (\ref{gen_expect}), for any $g(\boldsymbol{\theta })$. By
definition, a lack of knowledge of the integrating constant implies that the
marginal likelihood for the model in (\ref{gen_expect_prior}) is
unavailable. Situations where the likelihood function itself does not have a
closed form render the analytical solution of (\ref{gen_expect}) and (\ref%
{gen_expect_prior}) an even more distant dream. Hence the need for
computational\textit{\ }solutions.

\subsection{The Computational Solutions\label{sol}}

Despite their large number, it is useful to think about all Bayesian
computational techniques falling into one or more of three broad categories:

\begin{enumerate}
\item[\textit{1)}] \textit{Deterministic integration methods}

\item[\textit{2)}] \textit{Simulation methods}

\item[\textit{3)}] \textit{Approximation (including asymptotic) methods}%
\medskip
\end{enumerate}

\noindent Whilst all techniques are applicable to both the posterior
expectation in (\ref{gen_expect}) and the prior expectation in (\ref%
{gen_expect_prior}), we give emphasis throughout the paper to the
computation of (\ref{gen_expect}); reserving discussion of the computation
of (\ref{gen_expect_prior}) until Section \ref{mp}.

In brief, the methods in \textit{1) }define $L$ grid-points, $\boldsymbol{%
\theta }_{1},$ $\boldsymbol{\theta }_{2},...,\boldsymbol{\theta }_{L}$, to
span the support of $\boldsymbol{\theta }$, compute $g(\boldsymbol{\theta }%
_{l}\mathbf{)}p(\boldsymbol{\theta }_{l}\mathbf{|y})$, for $l=1,2,...,L$,
and estimate (\ref{gen_expect}) as a weighted sum of these $L$ values of the
integrand. Different deterministic numerical integration (or quadrature)
rules are based on different choices for $\boldsymbol{\theta }_{l}$, $%
l=1,2,...,L$, and different formulae for the weights. The methods in \textit{%
2) }use simulation\textit{\ }to produce $M$ posterior draws of $g(%
\boldsymbol{\theta }\mathbf{)}$, $g(\boldsymbol{\theta }^{(i)}\mathbf{)}$,
\thinspace $i=1,2,...,M$, a (weighted) mean of which is used to estimate (%
\ref{gen_expect}). Different simulation methods are distinguished by the way
in which the draws are produced and weighted. Finally, the methods in 
\textit{3) }involve replacing the integrand in (\ref{gen_expect}) with an
approximation of some sort, and evaluating the resultant integral. Different
approximation methods are defined by the choice of replacement for the
integrand, with the nature of this replacement determining the way in which
the final integral is computed. In particular, certain methods approximate%
\textbf{\ }$p(\boldsymbol{\theta }\mathbf{|y})$ itself and\textbf{\ }use
simulation from this approximate posterior\textbf{\ }to evaluate the
resultant integral. That is, the categories \textit{1)} and \textit{2)}\ are
certainly not mutually exclusive. \textit{Asymptotic }approximation methods
replace the integrand with an expansion that is accurate for large $n$, and
yield an estimate of (\ref{gen_expect}) (via analytical means) that is
accurate asymptotically.

\subsection{The Aim and Scope of this Review\label{scope}}

The aim of this review is to provide readers --- in particular those new to
the Bayesian paradigm --- with some insights into questions like: `Why are
there so many different ways of performing Bayesian computation?', `What are
the connections between them'?, `When does one use one approach, and when
another?' and `Are computational problems now different from computational
problems of the past?'

To achieve this aim we have made two key decisions: \textit{i)} to describe
all methods using a common notation; and \textit{ii)} to place the evolution
of computational methods in a historical context. In so doing, we are able
to present a coherent chronological narrative about Bayesian computation.
Specifically, all methods can be seen to be, in essence, attempting to
compute integrals like (\ref{gen_expect}) and (\ref{gen_expect_prior}); the
use of a common notation makes that clear. However, important details of
those integrals have changed over time: the dimension of $\boldsymbol{\theta 
}$ (i.e. the number of$\ $`unknowns'), the dimension of $\mathbf{y}$ (i.e.
the `size' of the data), and the nature of the integrand itself. Computation
has evolved accordingly, and the chronological ordering helps make sense of
that evolution. Hence, while we do make reference, where helpful, to the
above categories of computational methods, the over-arching structure that
we adopt is one of chronology, as understanding \textit{when} a
computational method has appeared aids in the appreciation of \textit{why}.

Excessive formalism, and extensive theoretical detail is avoided in order to
make the paper as accessible as possible, in particular to researchers whose
knowledge of Bayesian computation is not comprehensive. Whilst our
referencing is reasonably thorough, we have still been selective; directing
readers to key review papers, handbook chapters and other texts, for a more
complete coverage of published work.\ We also defer to those other resources
for descriptions of the dedicated software that is available for
implementing specific computational techniques.

Importantly, to render the scope of the paper manageable, we have also had
to be selective about the coverage of methods. As an overall principle, we
give focus to computational developments that have ultimately been linked to
the need to solve Bayesian problems. Hence, whilst deterministic numerical
integration remains an important tool in the Bayesian arsenal, and despite
the recent explosion of probabilistic numerics creating new connections
between Bayesian concepts and numerical integration (Briol \textit{et al}.,
2019), we make scant reference to developments in \textit{1) }(See also %
\citealp{davis1975numerical}, \citealp{naylor:smith:1982}, and %
\citealp{vanslette2019simple}, for relevant coverage). Instead, our focus is
primarily on the techniques in categories \textit{2) and 3) }that have
either had their genesis within, or been transformational for, Bayesian
analysis. We also consider computation only in the case of a parametric
model, $p(\boldsymbol{y}|\boldsymbol{\theta })$, with a finite\textit{\ }set
of unknowns, and do not attempt to review Bayesian computation in
nonparametric settings. (See \citealp{ghosal2017fundamentals}, for a
thorough coverage of computation in that sphere.)

In addition, we do not cover Bayesian optimization (%
\citealp{gutmann2016bayesian}; \citealp{frazier2018tutorial}), the Bayesian
bootstrap (\citealp{rubin1981}), or the Bayesian empirical likelihood (%
\citealp{lazar:2003}; \citealp{chib2018bayesian}). Other than brief mentions
made in Sections \ref{basic_pm} and \ref{IS_ML}, we do not discuss
sequential Monte Carlo (SMC) in any detail, referring the reader to \cite%
{naesseth2019elements} for a recent review. The coverage of more recent
developments in \textit{Markov chain Monte Carlo }(MCMC) (in Section \ref%
{advances}) is also very summary in nature. We categorize the primary goals
of such developments, and defer to existing reviews for all details. Given
our decision not to cover SMC, our citation of newer developments in \textit{%
importance sampling }(IS) is less thorough, although IS\ methods that focus
on estimation of the marginal likelihood are treated in Section \ref{IS_ML}.
We refer to the reader to \cite{hoogerheide2009simulation} and \cite%
{Tokdar2010} for discussion of more recent advances in IS.

\subsection{The Structure of this Review}

We begin, in Section \ref{timeline}, by returning to Bayes' integral in (\ref%
{Bayes_prob}), briefly discussing the nature of the computational problem.
We then use this as a springboard for pinpointing four particular points in
time during the two centuries (or so) subsequent to 1763: 1774, 1953, 1964
and 1970. These time points correspond, in turn, to four publications --- by
Laplace, Metropolis \textit{et al., }Hammersley and Handscomb, and Hastings,
respectively --- in which computational methods that produce estimates of
integrals like that of Bayes, were proposed. Whilst only the method of
Laplace was explicitly set within the context of inverse probability (or
Bayesian inference), all five methods of computing integrals can be viewed
as harbingers of what was to come in Bayesian computation \textit{per se}.

In Section \ref{70s}, we look at Bayesian computation in the late 20th
century, during which time the inexorable rise in the speed, and
accessibility, of computers led to the pre-eminence of \textit{%
simulation-based }computation. Whilst significant advances were made in
econometrics (\citealp{kloek1978bayesian}; \citealp{bauwens:1985}; %
\citealp{geweke:1989}) and signal processing \citep{gordon:salmon:smith:1993}
using the principles of IS, the `revolution' was driven primarily by MCMC
algorithms. As many treatments of Bayesian computation have covered this
period, we keep our coverage of this period very brief, deferring most\
details to more specialized reviews and seminal articles.

In contrast, the coverage in Section \ref{second} --- of what we term a
`second computational revolution' --- is much more extensive, given that we
bring together in one place, and using a common notational framework, the
large variety of computational methods that have evolved during the 21st
century. We begin with \textit{pseudo-marginal} methods, including particle
MCMC, before covering the main \textit{approximate }methods: approximate
Bayesian computation (ABC), Bayesian synthetic likelihood (BSL), variational
Bayes (VB) and integrated nested Laplace approximation\textit{\ }(INLA). One
goal is to link the development of these new techniques to the increased
complexity --- and size --- of the empirical problems being analyzed. A
second goal is to draw out insightful links \textit{and} differences between
all and, in so doing, pinpoint when and why each technique has value. This
provides some context for the \textit{hybrid} computational methods that we
then review. Section \ref{second} is completed by a brief summary of
important modifications and refinements of MCMC that have occurred since its
initial appearance, including Hamiltonian up-dates, adaptive sampling, and
coupling; developments that are, again, motivated by the challenges
presented by modern problems, most notably, the need to process huge data
sets and/or to infer high-dimensional unknowns.

We round off the review in Section \ref{mp} by switching focus from
parameter inference to model choice and prediction, and to the role of
computation therein.\textit{\ }We then end the paper with Section \ref%
{future}, bravely entitled: `The Future', in which we identify certain key
computational challenges that remain, and the directions in which solutions
to those challenges are being sought.

\section{Some Early Chronological Signposts\label{timeline}}

\subsection{1763: Bayes' Integral}

Bayes' desire was to {evaluate} the probability in (\ref{Bayes_prob}). As
noted {above}, {for either} $a\neq 0$ {or} $b\neq 1$, {this required
evaluation of the incomplete Beta function. For either }$x$ {or }$(n-x)$ {%
small}, Bayes {proposed} a Binomial expansion and term-by-term integration {%
to give an exact solution (his `Rule 1')}. {However, for }$x$ {and }$(n-x)$ {%
both large, this approach was infeasible: prompting Bayes (and,
subsequently, Price himself; \citealp{price:1764})} to {resort to producing
upper and lower bounds for (\ref{Bayes_prob}) using quadrature}. Indeed, {%
Stigler (1986a) speculates} that the inability to produce {an approximation
to (\ref{Bayes_prob})} that was sufficiently accurate may explain Bayes'
reluctance to publish his work and, {perhaps}, the lack of attention it
received subsequent to its (posthumous) presentation {by Price }in 1763%
\textbf{\ }and publication the following year {in \cite{bayes:1764}}.%
\footnote{%
On November 10, 1763, Price sent an edited and annotated version of Bayes'
essay to the {Secretary of the} Royal Society, with his own Appendix added.
Price read the essay to the Society on December 23, as noted earlier. The
essay and appendix were subsequently published in 1764, in the \textit{%
Philosophical Transactions of the Royal Society of London}. The front matter
of the issue appears here:
https://royalsocietypublishing.org/cms/asset/f005dd95-c0f8-45b2-8347-0296a93c4272/front.pdf. The publication has been reprinted since, including in 
\cite{Bayes:Biometrika1958}, with a biographical note by G.A. Barnard.
Further historical detail on the important role played by Price in the
dissemination of Bayes' ideas can be found in \cite{Hooper:2013} and \cite%
{stigler2018}. As the submission of Bayes' essay by Price, and his
presentation to the Royal Society occurred in 1763, and Volume 53 of the 
\textit{Philosophical Transactions} in which the essay appears\ is `For the
Year 1763', Bayes' essay is often dated 1763. We follow \cite{stigler:1986}
in using the actual publication date of 1764.}

Whilst the integral that Bayes wished to compute was a very particular one,
it was representative of the general hurdle that needed to be overcome if
the principle of inverse probability were to be a useful practical tool. In
brief, inference\textit{\ }about $\theta $ was expressed in probabilistic
terms and, hence, required either the direct computation of probability
intervals, or the computation of distributional moments of some sort.
Ironically, the choice of the Bernoulli model, possibly the simplest process
for generating data `forward' (conditional on $\theta $) that Bayes could
have assumed, exacerbated this problem, given that the `inversion' problem
does not possess the simplicity of the generative problem. What was required
was a solution that was, in large measure, workable no matter what the
nature of the generative model, and the first solution came via the 1774 
\textit{`M\'{e}moire sur la probabilit\'{e} des causes par les \'{e}v\'{e}%
nemens'} by Pierre Laplace.

\subsection{1774: Laplace and His Method of Asymptotic Approximation \label%
{Laplace}}

Laplace envisaged an experiment in which $n$ tickets were drawn with
replacement from an urn containing a given proportion of white and black
tickets. Recasting his analysis in our notation, $\theta $ is the
probability of drawing a white ticket, $\mathbf{y}=(y_{1},y_{2},...,y_{n})^{%
\prime }$ denotes the sequence of white tickets ($Y_{i}=1$) and black
tickets ($Y_{i}=0$) in the $n$ independent draws of $Y|\theta $, and $x=$ $%
\Sigma _{i=1}^{n}y_{i}$ is the number of white tickets drawn. Laplace's aim
was to show that, for arbitrary $w$: $\mathbb{P}(\left\vert \frac{x}{n}%
-\theta \right\vert <w|\mathbf{y})=\mathbb{P}(\frac{x}{n}-w<\theta <\frac{x}{%
n}+w|\mathbf{y})\rightarrow 1$ as $n\rightarrow \infty .$ That is, Laplace
wished to demonstrate \textit{posterior consistency}: concentration of the
posterior onto the true proportion of white tickets in the urn, $\theta _{0}=%
\underset{n\rightarrow \infty }{\lim }\frac{x}{n}$. Along the way, however,
he stumbled upon the same problem as had Bayes: computing the following
probability of a Beta random variable, 
\begin{equation}
\mathbb{P}(a<\theta <b|\mathbf{y})=\left[ B(x+1,n-x+1)\right]
^{-1}\int\limits_{a}^{b}\theta ^{x}(1-\theta )^{n-x}d\theta ,  \label{lp}
\end{equation}%
with $a=\frac{x}{n}-w\neq 0$ and $b=\frac{x}{n}+w\neq 1.$ Laplace's genius
(allied with the power of asymptotics!) was to recognize that the
exponential of the integrand in (\ref{lp}) has the bulk of its mass in the
region of its mode, as $n$ gets large, and that the integral can be computed
in closed form in this case. This enabled him to prove (in modern notation)
that $\mathbb{P}(\left\vert \theta _{0}-\theta \right\vert >w|\mathbf{y}%
)=o_{p}(1),$ where $p$ denotes the probability law for $\mathbf{y}$.

The route he took to this proof, however, involved approximating the Beta
posterior with a Normal distribution, which (under regularity) is an
approach that can be used to provide a large sample approximation of
virtually \textit{any} posterior probability.\ Specifically, express an
arbitrary posterior probability as%
\begin{equation}
\mathbb{P}(a<\theta <b|\mathbf{y})=\int\limits_{a}^{b}p(\theta \mathbf{|y}%
)d\theta =\int\limits_{a}^{b}\exp \left\{ nf(\theta )\right\} d\theta ,
\label{la_prob}
\end{equation}%
where $f(\theta )=\log \left[ p(\theta \mathbf{|y})\right] /n$, and assume
appropriate regularity for $p(\mathbf{y}|\theta )$ and $p(\theta ).$ What is
now referred to as the \textit{Laplace asymptotic approximation} involves
first taking a second-order Taylor series approximation of $f(\theta )$
around its mode, $\widehat{\theta }$: $f(\theta )\approx f(\widehat{\theta }%
)+\frac{1}{2}f^{^{\prime \prime }}(\widehat{\theta })(\theta -\widehat{%
\theta })^{2}$, where $f^{\prime }(\widehat{\theta })=0$ by construction.
Defining $\sigma ^{2}=-[nf^{^{\prime \prime }}(\widehat{\theta })]^{-1}$, {%
and }substituting {the expansion }into (\ref{la_prob}) then yields%
\begin{equation}
\begin{array}{cl}
\mathbb{P}(a<\theta <b|\mathbf{y}) & \approx \exp \left\{ nf(\widehat{\theta 
})\right\} \int\limits_{a}^{b}\exp \left\{ -\frac{1}{2\sigma ^{2}}(\theta -%
\widehat{\theta })^{2}\right\} d\theta \\ 
& =\exp \left\{ nf(\widehat{\theta })\right\} \sqrt{2\pi \sigma ^{2}}\times
\{\Phi \lbrack \frac{b-\widehat{\theta }}{\sigma }]-\Phi \lbrack \frac{a-%
\widehat{\theta }}{\sigma }]\},%
\end{array}
\label{la}
\end{equation}%
where $\Phi (.)$ denotes the standard Normal cumulative distribution
function (cdf).\footnote{%
Of course, buried within the symbol `$\approx $' in (\ref{la}) is a rate of
convergence that is a particular order of $n$, and is probabilistic if
randomness in $\mathbf{y}$ is acknowledged. See \cite{tierney:kadane:1986}
and Robert and Casella (2004) for futher elaboration; and \cite%
{ghosal1995convergence} and \cite{vandervaart:1998} for more formal
demonstrations of the conditions under which a posterior distribution
converges in probability to a Normal distribution, and the so-called
Bernstein-von Mises theorem --- the modern day version of Laplace's
approximation --- holds.}

With (\ref{la}), Laplace had thus devised a general way of implementing
inverse probability: probabilistic statements about an unknown parameter, $%
\theta $, conditional on data generated from any (regular) model, could now
be made, at least up to an error of approximation. Whilst his focus was
solely on the computation of {a specific posterior} {probability}, and in a
single parameter setting, {his method} was eventually used to approximate
general posterior expectations of the form in (\ref{gen_expect}) (%
\citealp{lindley1980bayesian}; \citealp{tierney:kadane:1986}; %
\citealp{tierney:kass:kadane:1989}) and, indeed, applied as an integral
approximation method in its own right \citep{bruijn1961asymptotic}. The
approach also underpins the modern INLA technique to be discussed in Section %
\ref{inla} \citep{rue:martino:chopin:2009}.\footnote{%
Stigler (1975, Section 2) states that he has found no documentary evidence
that Laplace's ideas on inverse probability, as presented in the 1774
publication, including his own statement of `Bayes' theorem' in (\ref%
{posterior_pdf}), were informed by Bayes' earlier ideas. See Stigler (1986a,
Chapter 3) for discussion of Laplace's later extensions of Bayes' theorem to
the case of a non-Uniform prior, and See \cite{stigler:1975}, \cite%
{stigler:1986}, {\cite{stigler:Laplace1774}} and \cite{fienberg:2006} {on
matters of attribution}. The first recorded reference to Bayes' prior claim
to inverse probability is in the preface, written by Condorcet, to Laplace's
later 1781 publication: `\textit{M\'{e}moire sur les probabilit\'{e}s}'.}

Meanwhile, it would take 170-odd years for the \textit{next} major advance
in the computation of probability integrals to occur; an advance that would
eventually transform the way in which problems in inverse probability could
be tackled. This development was based on a new form of thinking and,
critically, required a platform on which such thinking could operate:
namely, machines that could \textit{simulate }repeated\textit{\ }random
draws of $\boldsymbol{\theta }$ from $p(\boldsymbol{\theta }\mathbf{|y})$,
or from some representation thereof. Given a sufficient number of such
draws, and the correct use of them, an estimate of (\ref{gen_expect}) could
be produced that --- unlike the Laplace approximation --- would be accurate
for any sample size, $n$, and would require less analytical input. This
potential to accurately estimate (\ref{gen_expect}) for essentially any
problem, and any given sample size, was the catalyst for a flourishing of
Bayesian inference in the late 20th century and beyond. The 1953 publication
in the \textit{Journal of Chemical Physics }by Metropolis, Rosenbluth,
Rosenbluth, Teller and Teller: \textit{`Equation of State Calculations by
Fast Computing Machines'}, was a first major step in this journey.\footnote{%
With reference to the mechanical simulation of a random variable, we
acknowledge the earlier 1870s' invention of the \textit{quincunx} by Francis
Galton. This machine used the random dispersion of metal shot to illustrate
(amongst other things) draws from a hierarchical Normal model and regression
to the mean. Its use can thus be viewed as the first illustration of the
conjugation of a Normal likelihood and a Normal prior. See \cite%
{stigler:1986} for more details, including Galton's graphical illustration
of his machine in a letter to his cousin (and Charles Darwin's son), George
Darwin.}

\subsection{1953: Monte Carlo Simulation and the Metropolis Algorithm\label%
{MC}}

The convergence of the idea of simulating random draws from a probability
distribution, and the production of such draws by computing machines,
occurred in the scientific hothouse of the Los Alamos Laboratory, New
Mexico, in the 1940s and 1950s; the primary impetus being the need to
simulate physical processes, including neutrons in the fissile material in
atomic bombs. We refer the reader to \cite{liu01}, \cite{hitchcock:2003}, {%
\cite{Gub2005}} {and }\cite{robert:casella:2011} for reviews of this period,
including details of the various personalities who played a role therein.%
\footnote{%
We make particular mention here of John and Klara von Neumann, and Stanislav
Ulam, with the latter co-authoring the 1949 publication in the \textit{%
Journal of the Americal Statistical Association: `The Monte Carlo Method' }%
with\ Nicholas Metropolis. We also note the controversy concerning the
respective contributions of the five authors of the 1953 paper (who included
two married couples). {On this particular point, we refer the reader to the
informative 2005 article by Gubernatis, in which Marshall Rosenbluth gives a
bird's eye account of who did what, and when. The article brings to light
the important roles played by both Adriana Rosenbluth and Mici Teller.}} Our
focus here is simply on the nature of the problem that was at the heart of 
\cite{metropolis:1953}, the solution proposed, and the ultimate importance
of that solution to Bayesian computation.

In short, the authors wished to compute an expectation of the form, 
\begin{equation}
\mathbb{\mathbb{E}}(g(\mathbf{x))=}\int_{\mathcal{X}}g(\mathbf{x)}p(\mathbf{x%
})d\mathbf{x,}  \label{metrop}
\end{equation}%
where $p(\mathbf{x})$ denotes the so-called Boltzmann distribution of a set, 
$\mathbf{x,}$ of $N$ particles on $\mathbb{R}^{2}$. (See %
\citealp{robert:casella:2011}, Section 2.1, for all details.) Two particular
characteristics of (\ref{metrop}) are relevant to us here: \textit{i)} the
integral is of very high dimension, $2N$, with $N$ large; and \textit{ii)} $%
p(\mathbf{x})$ is generally known only up to its integrating constant. The
implication of \textit{i)} is that a basic rectangular integration method,
based on $L$ grid-points in each of the $2N$ directions, is infeasible,
having a computational burden of $L^{2N}$ or, equivalently, an approximation
error of $O(L^{-1/2N})$ (\citealp{kloek1978bayesian}). The implication of 
\textit{ii)} is that a Monte Carlo\textit{\ }(MC) estimate of (\ref{metrop}%
), based on $M$ $i.i.d.$ direct draws from $p(\mathbf{x})$, $\mathbf{x}%
^{(i)} $, $i=1,2,...,M$: $\widehat{E}_{MC}(g(\mathbf{x))}=\frac{1}{M}%
\sum\nolimits_{i=1}^{M}g(\mathbf{x}^{(i)}\mathbf{),}$ with approximation
error of $O(M^{-1/2})$ independent of dimension, is not available.\footnote{%
The authors actually make mention of a \textit{naive }Monte Carlo method,
based on \textit{Uniform} sampling over the $2N$ dimensional space, followed
by a reweighting of the Uniform draws by a kernel of $p(\mathbf{x}).$ The
idea is dismissed, however, as `not practical'. In modern parlance, whilst
this method would yield an $O(M^{-1/2})$ approximation error, the constant
term within the order would be large, since the Uniform distribution used to
produce draws of $\mathbf{x}$ differs substantially from the \textit{actual}
distribution of $\mathbf{x}$, $p(\mathbf{x}).$}

Features \textit{i)} and \textit{ii)} --- either individually or in tandem
--- broadly characterize the posterior expectations in (\ref{gen_expect})
that are the focus of this review. Hence the relevance to Bayesian
computation of the solution offered by \cite{metropolis:1953} to the
non-Bayesian problem in (\ref{metrop}); and we describe their solution with
direct reference to (\ref{gen_expect}) {and the notation used therein}.

Specifically, the authors advocate computing an integral such as (\ref%
{gen_expect}) via the simulation of a \textit{Markov chain: }$\boldsymbol{%
\theta }^{(i)}$, $i=1,2,...,M$, with \textit{invariant distribution} $p(%
\boldsymbol{\theta }\mathbf{|y})$. The draw at iteration $i+1$ in the chain
is created by taking the value at the $ith$ iteration, $\boldsymbol{\theta }%
^{(i)}$, and perturbing it according to a random walk:\textit{\ }$%
\boldsymbol{\theta }^{c}=\boldsymbol{\theta }^{(i)}+\delta \boldsymbol{%
\varepsilon }$, where each element of $\boldsymbol{\varepsilon }$ is drawn
independently from $U(-1,1)$, and $\delta $ `tunes' the algorithm.\footnote{%
\cite{metropolis:1953} actually implemented their algorithm one element of $%
\boldsymbol{\theta }$ at a time, as a harbinger of the Gibbs sampler to
come. See \cite{robert:casella:2011} for more details.} The `candidate' draw 
$\boldsymbol{\theta }^{c}$ is accepted as draw $\boldsymbol{\theta }^{(i+1)}$
with probability: 
\begin{equation}
\alpha =\min \{p^{\ast }(\boldsymbol{\theta }^{c}\mathbf{|y})/p^{\ast }(%
\boldsymbol{\theta }^{(i)}\mathbf{|y}),1\},  \label{MH_RW_ratio}
\end{equation}%
where $p^{\ast }$ is a kernel of $p.$ Using the theory of reversible Markov
chains, it can be shown (see, for example, \citealp{tierney94}) that use of (%
\ref{MH_RW_ratio}) to determine the $(i+1)th$ value in the chain does indeed
produce a dependent sequence of draws with invariant distribution $p(%
\boldsymbol{\theta }\mathbf{|y})$. Hence, subject to convergence to $p(%
\boldsymbol{\theta }\mathbf{|y})$ (conditions for which were verified by the
authors\ for their particular problem) these draws can be be used to
estimate (\ref{gen_expect}) as the sample mean, 
\begin{equation}
\overline{g(\boldsymbol{\theta }\mathbf{)}}=\frac{1}{M}\sum%
\limits_{i=1}^{M}g(\boldsymbol{\theta }^{(i)}\mathbf{),}  \label{metrop_est}
\end{equation}%
and an appropriate weak law of large numbers (WLLN) and central limit
theorem (CLT) invoked to prove the $\sqrt{M}$-consistency and limiting
normality of the {estimator}. (See \citealp{geyer2011introduction}, for
details.)

Due to the (positive) autocorrelation in the Markov chain, the variance of
the \textit{Metropolis estimator} (as it would become known) is larger than
that of the (infeasible) MC estimate in (\ref{MH_RW_ratio}), computed as in (%
\ref{metrop_est}), but using $i.i.\dot{d}$ draws from $p(\boldsymbol{\theta }%
\mathbf{|y})$, namely:%
\begin{equation}
\sigma _{MC}^{2}=\text{Var}(g(\boldsymbol{\theta }))/M,  \label{MC_variance}
\end{equation}%
{expressed here for the case of scalar }$g(\boldsymbol{\theta })$. However,
as is clear from (\ref{MH_RW_ratio}), the Metropolis MCMC algorithm requires
knowledge of $p(\boldsymbol{\theta }\mathbf{|y})$ only up to the normalizing
constant, and does \textit{not }require direct simulation from $p(%
\boldsymbol{\theta }\mathbf{|y})$ itself. It is this\textit{\ }particular
feature that would lend the technique its great power in the decades to come.%
\footnote{\cite{dongarra2000guest} rank the Metropolis algorithm as one of
the 10 algorithms \textquotedblleft with the greatest influence on the
development and practice of science and engineering in the 20th
century\textquotedblright .}

\subsection{1964: Hammersley and Handscomb: Importance Sampling\label{hh}}

The obviation of the need to directly\textit{\ }sample from $p(\boldsymbol{%
\theta }\mathbf{|y})$ also characterizes importance sampling, and underlies
its eventual importance in solving difficult Bayesian computational
problems. Nevertheless, \cite{hammersley:handscomb:1964} did not emphasize
this feature but, rather, introduced the concept of IS for the express
purpose of variance reduction in simulation-based estimation of integrals.%
\footnote{%
One could in fact argue that a similar aim motivated Metropolis and
co-authors, given that they drew a sharp contrast (in effect) between the
efficiency of their method and that of the naive Monte Carlo technique based
on Uniform sampling.} Again, the focus was not on Bayesian integrals, but we
describe the method in that setting.

In brief, given an `importance' ({or }`proposal') density, $q(\boldsymbol{%
\theta }\mathbf{|y})$, that preferably mimics $p(\boldsymbol{\theta }\mathbf{%
|y})$ well, and $M$ $i.i.d.$ draws, $\boldsymbol{\theta }^{(i)}$, from $q(%
\boldsymbol{\theta }\mathbf{|y})$, an IS estimate of (\ref{gen_expect}) is $%
\overline{g(\boldsymbol{\theta }\mathbf{)}}_{IS}=\frac{1}{M}%
\sum\limits_{i=1}^{M}g(\boldsymbol{\theta }^{(i)}\mathbf{)}w(\boldsymbol{%
\theta }^{(i)})\mathbf{,}$ where $w(\boldsymbol{\theta }^{(i)})=p(%
\boldsymbol{\theta }^{(i)}\mathbf{|y})/q(\boldsymbol{\theta }^{(i)}\mathbf{|y%
}).$ In the typical case where $p(\boldsymbol{\theta }^{(i)}\mathbf{|y})$ is
available only up to the integrating constant, and $w(\boldsymbol{\theta }%
^{(i)})$ cannot be evaluated as a consequence, the estimate is modified as 
\begin{equation}
\overline{g(\boldsymbol{\theta }\mathbf{)}}_{IS}=\sum\limits_{i=1}^{M}g(%
\boldsymbol{\theta }^{(i)}\mathbf{)}w(\boldsymbol{\theta }^{(i)})\big/%
\sum_{j=1}^{M}w(\boldsymbol{\theta }^{(i)}),  \label{is_est_2}
\end{equation}%
with the weights re-defined as $w(\boldsymbol{\theta }^{(j)})=p^{\ast }(%
\boldsymbol{\theta }^{(i)}\mathbf{|y})/q^{\ast }(\boldsymbol{\theta }^{(i)}%
\mathbf{|y})$, for kernels, $p^{\ast }(\boldsymbol{\theta }^{(i)}\mathbf{|y}%
) $ and $q^{\ast }(\boldsymbol{\theta }^{(i)}\mathbf{|y})$, of $p(%
\boldsymbol{\theta }\mathbf{|y})$ and $q(\boldsymbol{\theta }\mathbf{|y})$
respectively. Once again, and under regularity conditions pertaining to the 
\textit{importance density }$q(\boldsymbol{\theta }\mathbf{|y)}$, asymptotic
theory can be invoked to prove that (\ref{is_est_2}) is a $\sqrt{M}$%
-consistent estimator of $\mathbb{\mathbb{E}}(g(\boldsymbol{\theta }\mathbf{%
)|y)}$ \citep{geweke:1989}. A judicious choice of $q(\boldsymbol{\theta }%
\mathbf{|y)}$ is able to yield a sampling variance that is less than (\ref%
{MC_variance}) in some cases, as befits the original motivation of IS as a
variance reduction method. {(See \citealp{geweke:1989}, and %
\citealp{robert:casella:2004}, for discussion.}) Critically however, like
the Metropolis method, (\ref{is_est_2}) serves as a feasible estimate of $%
\mathbb{\mathbb{E}}(g(\boldsymbol{\theta }\mathbf{)|y)}$ when $p(\boldsymbol{%
\theta }\mathbf{|y})$ cannot be easily simulated; hence the significance of
IS in Bayesian computation. Moreover, its maintenance of independent\textit{%
\ }draws, allied with its re-weighting of draws from an approximating
density, has led to the emergence of IS as a vehicle for implementing\ SMC
algorithms, like particle filtering, to be referenced in Section \ref{pseudo}
in the context of particle MCMC.

\subsection{1970: Hastings and his Generalization of the Metropolis
Algorithm \label{hast copy(1)}}

The final publication that we pinpoint during the 200-odd year period
subsequent to 1763, is the 1970 \textit{Biometrika }paper, `\textit{Monte
Carlo Sampling Methods Using Markov Chains and Their Applications', }by
Wilfred Keith Hastings. Whilst \cite{metropolis:1953} proposed the use of
MCMC sampling to compute particular integrals in statistical mechanics, it
was the Hastings paper that elevated the concept to a general one, and
introduced it to the broader statistics community. Included in the paper is
also the first mention of what would become known as the \textit{%
Metropolis-within-Gibbs sampler} \citep{robert:casella:2011}. Once again,
the author's focus was not a Bayesian integral \textit{per se}; however we
describe the method in that context.

In contrast to Metropolis and co-authors, \cite{hastings:1970} acknowledges
up-front that the need to know $p(\boldsymbol{\theta }\mathbf{|y})$ only up
to the integrating constant is a compelling feature of an MCMC-based
estimate of (\ref{gen_expect}). Hastings also generalizes the acceptance
probability in (\ref{MH_RW_ratio}) to one that accommodates a general
`candidate' distribution $q(\boldsymbol{\theta }\mathbf{|y)}$ from which $%
\boldsymbol{\theta }^{c}$ is drawn, as:%
\begin{equation}
\alpha =\min \left\{ \left[ p^{\ast }(\boldsymbol{\theta }^{c}\mathbf{|y})/q(%
\boldsymbol{\theta }^{(i)}|\boldsymbol{\theta }^{c},\mathbf{y)}\right] \Big/%
\left[ p^{\ast }(\boldsymbol{\theta }^{(i)}\mathbf{|y})/q(\boldsymbol{\theta 
}^{c}|\boldsymbol{\theta }^{(i)},\mathbf{y)}\right] ,1\right\} ,
\label{hastings}
\end{equation}%
which clearly collapses to (\ref{MH_RW_ratio}) when $q(\boldsymbol{\theta }%
\mathbf{|y)}$ is symmetric (in $\boldsymbol{\theta }^{c}$ and $\boldsymbol{%
\theta }^{(i)}$), as in the original random walk proposal of \cite%
{metropolis:1953}. Importantly, the more general algorithm allows for a
targeted choice of $q(\boldsymbol{\theta }\mathbf{|y)}$ that reduces the
need for tuning and which can, potentially, reduce the degree of dependence
in the chain and, hence, the variance of the estimate of $\mathbb{E}(g(%
\boldsymbol{\theta }\mathbf{)|y).}$ Hastings formalizes the standard error
of this estimate using time series theory, explicitly linking, for the first
time, the autocorrelation in the Markov draws to the efficiency of the
MCMC-based estimate of (\ref{gen_expect}). Crucially, the author tackles the
issue of dimension by advocating the treatment of one element (or several
elements) of $\boldsymbol{\theta }$ at a time, conditional on all remaining
elements.

In summary, all of the important ingredients from which the huge smorgasbord
of future MCMC algorithms would eventually be constructed --- for the 
\textit{express} purpose of solving Bayesian problems --- were now on the
table, via this particular paper.

\section{The Late 20th Century: Gibbs Sampling \& the MCMC Revolution\label%
{70s}}

Whilst the role that \textit{could }be played by simulation in computation
was thus known by the 1970s, the computing technology needed to exploit that
knowledge lagged behind.\footnote{%
Many readers may be too young to remember the punchcards! But there was a
time when RAND's 1955 \textit{A Million Random Digits with 100,000 Normal
Deviates} was more than an entry for sarcastic Amazon comments, as producing
this million digits took more than two months at the time.} Over the next
two decades, however, things changed. Indeed, \textit{two} developments now
went hand in hand to spawn a remarkable expansion in simulation-based
Bayesian computation:\textit{\ i)} the increased speed and availability of
computers, including personal desktop computers (\citealp{ceruzzi:2003}),
and \textit{ii)} the collective recognition that MCMC draws from a joint
posterior, $p(\boldsymbol{\theta }\mathbf{|y})$, could be produced via
iterative sampling from lower dimensional, and often standard, \textit{%
conditional }posteriors. When allied with both the concept of \textit{%
augmentation, }and an understanding of the theoretical properties of
combinations of MCMC algorithms, \textit{ii) }would lead to \textit{Gibbs
sampling} (with or without \textit{Metropolis-Hastings }(MH) subchains)
becoming the work-horse of Bayesian computation in the 1990s.

An MH algorithm `works', in the sense of producing a Markov chain that
converges to the required distribution $p(\boldsymbol{\theta }\mathbf{|y})$,
due to the form of the acceptance probability in (\ref{hastings}) (or the
nested version in (\ref{MH_RW_ratio})). More formally, the algorithm, as
based on candidate density $q(\boldsymbol{\theta }\mathbf{|y)}$, and
acceptance probability as defined in (\ref{hastings}), defines a \textit{%
transition kernel }with invariant distribution, $p(\boldsymbol{\theta }%
\mathbf{|y}).$ The `Gibbs sampler' similarly yields a Markov chain with
invariant distribution, $p(\boldsymbol{\theta }\mathbf{|y})$, but via a
transition kernel that is defined as the product of full conditional
posteriors\textit{\ }associated with the joint. For the simplest case of a
two-dimensional vector $\boldsymbol{\theta }=(\theta _{1},\theta
_{2})^{\prime }$, the steps of the Gibbs algorithm are as follows: first,
specify an initial value for $\theta _{2}$, $\theta _{2}^{(0)}$; second, for 
$i=1,2,...,M$, cycle iteratively through the two conditional distributions,
drawing respectively: $\theta _{1}^{(i)}$ from $p_{1}(\theta
_{1}^{(i)}|\theta _{2}^{(i-1)},\mathbf{y})$, and $\theta _{2}^{(i)}$ from $%
p_{2}(\theta _{2}^{(i)}|\theta _{1}^{(i)},\mathbf{y)}$. Given the
satisfaction of the required convergence conditions (which essentially place
sufficient regularity on the conditionals), the draws $\boldsymbol{\theta }%
^{(i)}=(\theta _{1}^{(i)},\theta _{2}^{(i)})^{\prime }$, $i=1,2,...,M$,
converge in distribution to the joint posterior distribution as $%
M\rightarrow \infty $, and can be used to produce a $\sqrt{M}$-consistent
estimator of (\ref{gen_expect}) in the form of (\ref{metrop_est}). Extension
to higher-dimensional problems is obvious, although decisions about how to
`block' the parameters, and thereby define the conditionals, now play a role %
\citep{roberts:sahu:1997}.\footnote{%
The Gibbs sampler can be viewed as a special case of a `multiple-block' MH
sampler, in which the candidate values for each block of parameters are
drawn directly from their full conditional distributions and the acceptance
probability in (each blocked version of) (\ref{hastings}) is equal to one.
(See, for example, Chib, 2011). See also \cite{tran2018common} for further
discussion of this point --- conducted in the context of a generalized MH
framework, in which many of the algorithms to be discussed in Sections \ref%
{pseudo} and \ref{advances} are also nested.}

Gibbs thus exploits the simplicity yielded by conditioning: whilst joint and
marginal posterior distributions are usually complex in form, (full)
conditional posteriors are often standard and, hence, able to be simulated
from directly. While one may find hints in both \cite{hastings:1970} and 
\cite{besag:1974}, this point was first made clearly by \cite{geman:1984},
who also coined the phrase `Gibbs sampling' because their problem used Gibbs
random fields in image restoration (named, in turn, after the physicist,
Josiah Willard Gibbs). However, the later paper by \cite{gelfand:smith90} is
generally credited with bringing this transformational idea to the attention
of the broader statistical community, and illustrating its broad
applicability.

The idea of Gibbs sampling overlapped with a related proposal by \cite%
{tanner87}: that of `augmenting' the set of unknowns ($\boldsymbol{\theta }$
in our notation) with latent data, $\mathbf{z}=(z_{1},z_{2},...,z_{n})^{%
\prime }$, to yield conditionals --- $p(\boldsymbol{\theta }|\mathbf{z},%
\mathbf{y})$ and $p(\mathbf{z}|\boldsymbol{\theta },\mathbf{y})$ --- that
facilitate the production of a simulation-based estimate of $p(\boldsymbol{%
\theta }\mathbf{|y})$; with $p(\boldsymbol{\theta }|\mathbf{z},\mathbf{y})$,
in particular, often being standard. The melding of these two ideas, i.e.
sampling via conditionals \textit{per se}, and yielding more tractable
conditionals through the process of augmentation, enabled the analysis of
complex models that had thus far eluded Bayesian treatment, due to their
dependence on high-dimensional vectors of latent variables; selected
examples being: \cite{carlin:polson:stoffer:1992}, \cite{carter:kohn:1994}, 
\cite{fruhwirth-schnatter:1994} and \cite{jacquier94}.\ However, it also led
to the realization that \textit{artificial }latent variables could be
judiciously introduced into a model for the sole purpose of producing
tractable conditional posteriors over the augmented space, thereby opening
up a whole range of additional models to a Gibbs-based solution (e.g. %
\citealp{albert:chib:1993b}; \citealp{dieb:robe:1994}; {\citealp{higdon1998};%
} \citealp{kim1998svl}; \citealp{damien:wakefield:walker:1999}). {The 
\textit{slice sampler} (\citealp{roberts:rosenthal:1999}; \citealp{neal:2003}%
) is one particularly notable, and generic, way of generating an MCMC
algorithm via this principle of auxiliary variable augmentation.}

Of course, in most high-dimensional problems --- and in particular those in
which latent variables feature --- certain conditionals remain nonstandard,
such that direct simulation from them is not possible. Critically though,
the reduced dimension renders this a simpler problem than sampling from the
joint itself: via either the inverse cumulative distribution function
technique \citep{devroye:1986} --- approximated in the `Griddy Gibbs'
algorithm\textbf{\ }of \cite{ritter:tanner:1992} --- or by embedding an MH
algorithm within the outer Gibbs loop (a so-called `Metropolis-within-Gibbs'
algorithm).\footnote{%
We refer the reader to: \cite{besag:green:1993}, \cite{smith:roberts:1993}
and \cite{chib_greenberg_1996} for early reviews of MCMC sampling; \cite%
{casella:george:1992} and \cite{chibandgreenberg:1995} for descriptions of
the Gibbs and MH algorithms (respectively) that are useful for
practitioners; \cite{robert2015metropolishastings}, \cite{betancourt:2018}
and \cite{dunson2019hastings} for more recent reviews; and \cite%
{andrieu2004computational} and \cite{robert:casella:2011} for historical
accounts of MCMC sampling.}

\section{The 21st Century: A Second Computational Revolution\label{second}}

\subsection{Why Did We Need a Second One?\label{why}}

The advent of accessible, fast computers in the last two decades of the 20th
century, allied with the methodological and theoretical developments
referenced above, led to an explosion in the use of simulation-based
Bayesian computation, with variants of MCMC leading the charge. The impact
of these developments was felt across a huge array of fields --- genetics,
biology, neuroscience, astrophysics, image analysis, ecology, epidemiology,
education, economics, marketing and finance, to name but some --- and
brought Bayesian analysis into the statistical mainstream. The two 2011
Handbooks: \textit{Handbook of Markov Chain Monte Carlo }(%
\citealp{brooks:etal:2011})\ and \textit{The Oxford Handbook of Bayesian
Econometrics} (\citealp{geweke2011handbook}), further highlight the wide
spectrum of fields, and broad scope of empirical problems to which MCMC
algorithms were (and continue to be)\textbf{\ }applied; as do certain
contributions to the series of vignettes edited by Mengersen and Robert for 
\textit{Statistical Science} (2014, Vol 29, No. 1), under the theme of `Big
Bayes Stories'.

Despite their unquestioned power and versatility however, these original
simulation techniques did have certain limitations; with these limitations
to become more marked as the empirical problems being tackled became more
ambitious; and this despite a concurrent rise in computing power (parallel
computing, access to GPUs etc.) over recent decades. With reference to the
posterior pdf in (\ref{Bayes_proport}), two characteristics are worthy of
note. First, as already highlighted, in all but the most stylized problems $%
p(\boldsymbol{\theta }\mathbf{|y})$ is available only\textit{\ }up to its
integrating constant, and cannot be directly simulated. Second,
representation of $p(\boldsymbol{\theta }\mathbf{|y})$ only as a kernel, $%
p^{\ast }(\boldsymbol{\theta }\mathbf{|y})\propto p(\mathbf{y}|\boldsymbol{%
\theta })p(\boldsymbol{\theta })$, still requires closed forms for $p(%
\mathbf{y}|\boldsymbol{\theta })$ and $p(\boldsymbol{\theta }).$ With
reference to $p(\mathbf{y}|\boldsymbol{\theta })$, this means that, for any $%
\boldsymbol{\theta }$, $p(\mathbf{y}|\boldsymbol{\theta })$ needs to be able
to be evaluated at the observed $\mathbf{y}$. The MCMC and IS simulation
methods obviate the first problem by drawing \textit{indirectly }from $p(%
\boldsymbol{\theta }\mathbf{|y})$ via another distribution (or set of
distributions) from which simulation is feasible. However, these methods
still require evaluation of $p(\mathbf{y}|\boldsymbol{\theta })$: in the
computation of the importance weights in (\ref{is_est_2}), in the
computation of the acceptance probability in (\ref{MH_RW_ratio}) or (\ref%
{hastings}), and in the implementation of any Gibbs-based algorithm, in
which the conditional posteriors are required either in full form or at
least up to a scale factor.

The assumption that $p(\mathbf{y}|\boldsymbol{\theta })$ can be evaluated is
a limitation for two reasons. First, some DGPs do not admit pdfs in closed
form; examples being: probability distributions defined by quantile or
generating functions (\citealp{devroye:1986}; \citealp{peters2012likelihood}%
), continuous time models in finance with unknown transition densities %
\citep{gallant1996moments}, dynamic equilibrium models in economics %
\citep{calvet2015accurate}, certain deep learning models in machine learning %
\citep{goodfellow2014generative}; complex astrophysical models %
\citep{jennings2017astroabc}; and\ DGPs for which the normalizing constant
is unavailable, such as Markov random fields in spatial modelling (%
\citealp{rue:held:2005}; \citealp{stoehr2017review}). Second, pointwise
evaluation of $p(\mathbf{y}|\boldsymbol{\theta })$ (at any $\boldsymbol{%
\theta }$) (in the case where $p(\boldsymbol{\cdot }|\boldsymbol{\theta })$
has a closed form) entails an $O(n)$ computational burden; meaning that the
MCMC and IS methods described above are \textit{not scalable} to so-called
`big (or tall) data' problems \citep{bardenet2017markov}.

Just as important are the challenges that arise when the dimension of the
unknowns themselves is very large (the so-called `high-dimensional'
problem); for instance, when a model contains a very large number of latent
variables over which integration is required (e.g. %
\citealp{tavare:balding:griffith:donnelly:1997}; %
\citealp{braun2010variational}; \citealp{beaumont:2010}; %
\citealp{lintusaari2017fundamentals}; \citealp{johndrow2019mcmc}). In such
cases, standard MCMC methods --- even if feasible in principle --- may not
(as highlighted further in Section \ref{advances}) enable an accurate
estimate of (\ref{gen_expect}) to be produced in finite computing time;\
i.e. such methods are \textit{not necessarily scalable} in the dimension of
the unknowns.

Each of the techniques discussed in Sections \ref{pseudo}, \ref{approx} and %
\ref{hybrid} relieves the investigator of one or more of these `burdens';
although, as we shall see, the relief is not costless. In particular, the
approximation\textit{\ }methods covered in Section \ref{approx} and \ref%
{hybrid}, whilst enabling some problems to be tackled that would be
intractable for MCMC and IS, do not produce an estimate of (\ref{gen_expect}%
) that is \textit{exact }up to simulation error but, instead, produce an
estimate that is only ever \textit{approximate}.

Finally, we provide a very brief overview in Section \ref{advances} of
further advances made in MCMC methods \textit{per se}, primarily over the
past decade or so; advances designed, in large measure, to improve the
accuracy\ with which these dependent chains estimate posterior quantities of
interest such as (\ref{gen_expect}), most notably in large-scale problems.

\subsection{Pseudo-Marginal Methods\label{pseudo}}

\subsubsection{The basic idea\label{basic_pm}}

Referencing the concept of data augmentation introduced in Section \ref{70s}%
: given draws from the joint posterior of $\boldsymbol{\theta }$ and $%
\mathbf{z}$, $p(\boldsymbol{\theta },\mathbf{z}|\mathbf{y})$, the draws of $%
\boldsymbol{\theta }$ can be used to produce an exact simulation-based
estimate of $p(\boldsymbol{\theta }\mathbf{|y})$ or any associated quantity,
for large enough $M.$ Again, the latent states, $\mathbf{z}$, may be either
intrinsic to the model, or introduced `artificially' as a computational
device, as highlighted therein. Initially, draws of $(\boldsymbol{\theta },%
\mathbf{z})$ were produced via Gibbs-based MCMC schemes, with a variety of
MH algorithms (based on alternative candidates) used to sample from $p(%
\mathbf{z}|\boldsymbol{\theta }\mathbf{,y})$, in the typical case where this
conditional could not be simulated directly (see \citealp{fearnhead2011mcmc}%
, and \citealp{giordani2011bayesian}, for reviews). As highlighted in
Section \ref{why} however, depending on the problem, such MH-within-Gibbs
schemes can be slow to explore the joint space of $(\boldsymbol{\theta },%
\mathbf{z)}$ and, hence, to produce an accurate estimate of $p(\boldsymbol{%
\theta }\mathbf{|y})$.

The (combined) insight of \cite{beaumont:2003} and \cite%
{andrieu:roberts:2009} was to recognize that draws of $\mathbf{z}$ could be
used in a potentially more effective way to yield an estimate of the
`marginal' of interest: $p(\boldsymbol{\theta }\mathbf{|y})$ (and any
integral of the form of (\ref{gen_expect})). Key to this insight is the
following observation. Use $\mathbf{u\in }$ $\mathcal{U}$ to denote all of
the canonical (problem-specific) random variables that are used to generate $%
\mathbf{z}$. If draws of\textbf{\ $\mathbf{u}$ }can be used to produce an 
\textit{unbiased }estimate of the likelihood function, $p(\mathbf{y}|%
\boldsymbol{\theta })$, then an MCMC scheme applied to the joint space $(%
\boldsymbol{\theta }\mathbf{,u})$, can target the required invariant
distribution, $p(\boldsymbol{\theta }\mathbf{|y}).$ An informal
demonstration of this result is straightforward. Define $h(\mathbf{u})$ as
the distribution of the random variables underpinning the generation of $%
\mathbf{z}$ (independently of the prior $p(\boldsymbol{\theta })$), and let $%
h(\mathbf{y}|\boldsymbol{\theta }\mathbf{,u})$ denote an estimate of the
likelihood $p(\mathbf{y}|\boldsymbol{\theta })$, that is unbiased in the
sense that $E_{\mathbf{u}}[h(\mathbf{y}|\boldsymbol{\theta },\mathbf{u})]=p(%
\mathbf{y}|\boldsymbol{\theta })$. Then we have that $h(\boldsymbol{\theta }%
\mathbf{|y})\propto \int\nolimits_{\mathcal{U}}h(\mathbf{y}|\boldsymbol{%
\theta }\mathbf{,u})p(\boldsymbol{\theta })h(\mathbf{u})d\mathbf{u}=p(%
\boldsymbol{\theta })E_{\mathbf{u}}[h(\mathbf{y}|\boldsymbol{\theta },%
\mathbf{u})]=p(\boldsymbol{\theta })p(\mathbf{y}|\boldsymbol{\theta }%
)\propto p(\boldsymbol{\theta }\mathbf{|y})$.

Use of $h(\mathbf{y}|\boldsymbol{\theta }\mathbf{,u})$ within an MH
algorithm amounts to replacing the acceptance probability in (\ref{hastings}%
) with 
\begin{equation}
\alpha =\min \left\{ \left[ h(\mathbf{y}|\boldsymbol{\theta }^{c},\mathbf{u}%
^{c})p(\boldsymbol{\theta }^{c})/q(\boldsymbol{\theta }^{(i)}|\boldsymbol{%
\theta }^{c},\mathbf{y)}\right] Big/\left[ h(\mathbf{y}|\boldsymbol{\theta }%
^{(i)},\mathbf{u}^{(i)})p(\boldsymbol{\theta }^{(i)})/q(\boldsymbol{\theta }%
^{c}|\boldsymbol{\theta }^{(i)},\mathbf{y)}\right] ,1\right\} ,  \label{pmmh}
\end{equation}%
where $\boldsymbol{\theta }^{c}$ is proposed from $q(\boldsymbol{\theta }%
^{c}|\boldsymbol{\theta }^{(i)},\mathbf{y)}$ and $\mathbf{u}^{(i)}$ and $%
\mathbf{u}^{c}$ are independent draws from $h(\mathbf{u}).$ The use of an
estimate of the likelihood in (\ref{pmmh}) prompted use of the term
`pseudo'-marginal MH (PMMH) by \cite{andrieu:roberts:2009} although, as
noted, this replacement still yields a chain with an invariant distribution
equal to the correct marginal, $p(\boldsymbol{\theta }\mathbf{|y})$, when
the estimate is unbiased.

When a likelihood estimate is produced specifically via the use of \textit{%
particle filtering} in a state space model (SSM), the term \textit{particle}
MCMC (PMCMC) has also been coined \citep{andrieu:doucet:holenstein:2010}.%
\footnote{%
Whilst not a pseudo-marginal method, particle filtering has also been used
to provide an estimate of $p(\mathbf{z}|\boldsymbol{\theta ,y})$ in a Gibbs
scheme for an SSM --- so-called `particle Gibbs' %
\citep{andrieu:doucet:holenstein:2010}.} Whilst we omit details of the use
of filtering to estimate a likelihood function (see reviews in %
\citealp{doucet:defreitas:gordon:2001}, and \citealp{giordani2011bayesian}),
we do remark that particle filtering does involve the sequential application
of IS, with independent, but differentially weighted draws of the latent
states (from both `filtered' and `prediction' state distributions) being the
outcome. As such, much of the early work on IS, including the impact of
proposal choice on the efficiency of the sampler, has assumed a renewed
importance in filtering-based settings, including PMCMC.

Whilst unbiasedness of the likelihood estimate is required for a general
PMMH algorithm to `work', the variance of the estimate also affects the
performance of the sampler and, hence, the simulation efficiency of any
estimate of (\ref{gen_expect}) that is produced. However, improving the
precision of the likelihood estimator by increasing the number of draws of $%
\mathbf{u}$ used in its production comes at a computational cost, and an
`optimal' number of draws that balances computational cost with an
acceptable mixing of the chain needs to be sought. See \cite{pitt2012some}, 
\cite{doucet2015efficient} and \cite{deligiannidis2018correlated} for
discussion of the optimal structuring and tuning of pseudo-marginal
algorithms.

\subsubsection{The benefits\label{ben}}

The benefits of pseudo-marginal schemes are three-fold. First, in cases
where the parameters and latent states are strongly correlated, use of a
PMMH scheme rather than a Gibbs-based scheme (based on $p(\mathbf{z}|%
\boldsymbol{\theta }\mathbf{,y})$ and $p(\boldsymbol{\theta }|\mathbf{z,y})$%
), {may reap efficiency gains} (conditional on appropriate `tuning' choices,
as flagged above). Linked to this, avoidance of the need to sample from $p(%
\mathbf{z}|\boldsymbol{\theta }\mathbf{,y})$ obviates the need to make
choices regarding the blocking of $\mathbf{z}$ and the proposal densities
for those blocks. Second, in cases where \textit{only} forward simulation of
the latent process is possible, and point-wise evaluation of $p(\mathbf{y},%
\mathbf{z}|\boldsymbol{\theta })$ is infeasible as a result, PMMH remains
possible. For example, in an SSM an estimate of $p(\mathbf{y}|\boldsymbol{%
\theta })$ can be based on the bootstrap particle filter, for which only
simulation from the transition density $p(z_{t}|z_{t-1},\boldsymbol{\theta })
$ (not evaluation thereof) is required. Third, in cases where the dimension
of $\mathbf{y}$ is very large, an unbiased estimate of $p(\mathbf{y}|%
\boldsymbol{\theta })$ based on appropriately selected \textit{subsamples }%
of data can be used to produce a valid PMMH scheme, at a much smaller
computational cost than any scheme that requires full evaluation of $p(%
\mathbf{y}|\boldsymbol{\theta })$ (\citealp{bardenet2017markov}; %
\citealp{quiroz2018speeding}; \citealp{quiroz2019speeding}).

\subsection{Approximate Bayesian Inference\label{approx}}

The goal of all simulation-based computational methods discussed thus far,
including the pseudo-marginal techniques, has been to estimate the posterior
expectation in (\ref{gen_expect}) `exactly', at least up to an order $%
O(M^{-1/2})$, where $M$ is the number of draws that defines the simulation
scheme. The alternative methods do, of course, differ one from the other in
terms of the constant term that quantifies the precise error of
approximation. Hence, it may be the case that even for a very large $M$, a
nominally `exact' method (despite being `tuned' optimally) has an
approximation error that is non-negligible;\ a point that we revisit in
Section \ref{advances}\textbf{.} Nevertheless, the convention in the
literature is to refer to all simulation methods outlined to this point as 
\textit{exact}, typically without qualification.\footnote{%
We note that we have omitted any mention herein of so-called `quasi-Monte
Carlo' integration schemes, which aim for exactness at a faster rate than $%
O(M^{-1/2})$. See \cite{lemieux2009monte} for a review of such methods, \cite%
{chen2011} for the extension to quasi-MCMC algorithms, and \cite%
{gerber:chopin:2015} for an entry on sequential quasi-Monte Carlo.}

In contrast, when applying an \textit{approximation} method (using the
taxonomy from Section \ref{scope}), investigators make no claim to
exactness, other than citing the asymptotic (in $n$) accuracy of the Laplace
methods (in Sections \ref{Laplace} and \ref{inla}), or the asymptotic
validity of certain other approximations \citep{fearnhead2018asymptotics}.
That is, for finite $n$ at least, such methods are only ever acknowledged as
providing an approximation to (\ref{gen_expect}), with that approximation
perhaps claimed to be as `accurate' as possible, given the relevant choice
variables that characterize the method; but not more.

So what benefits do such techniques offer, in return for sacrificing the
goal of exact inference? With reference to the methods discussed below: ABC
and BSL both completely obviate the need to evaluate $p(\mathbf{y}|%
\boldsymbol{\theta })$ and, in so doing, open up to Bayesian treatment a
swathe of empirical problems --- so-called \textit{doubly-intractable }%
problems\ --- that would otherwise not be amenable to Bayesian analysis. In
computing (\ref{gen_expect}), both methods replace the posterior in the
integrand, $p(\boldsymbol{\theta }\mathbf{|y})$, with an approximation
produced via simulation. A simulation-based estimate of the integral, $%
\overline{g(\boldsymbol{\theta }\mathbf{)}}=(1/M)\sum\nolimits_{i=1}^{M}g(%
\boldsymbol{\theta }^{(i)}\mathbf{)}$, is then produced using draws, $%
\boldsymbol{\theta }^{(i)}$, from this approximate posterior.\footnote{%
It can also be argued that these are simulation methods for an exact, albeit
different posterior, using either a degraded version of the observations or
a projection of them via a nonsufficient statistic\ \citep{wilkinson:2013}.}
In contrast, VB and INLA both require evaluation of $p(\mathbf{y}|%
\boldsymbol{\theta })$, but reap computational benefits in certain types of
problems (in particular those of high-dimension) by replacing --- at least
in part --- \textit{simulation} with (in some cases closed-form) \textit{%
optimization}. In the case of VB, the posterior $p(\boldsymbol{\theta }%
\mathbf{|y})$ used to define (\ref{gen_expect}) is replaced by an
approximation produced via the calculus of variations. Depending on the
nature of the problem, including the `variational family' from which the
`optimal' approximation is produced, the integral is computed in either
closed form or via a simulation step. With INLA, the approximation of $p(%
\boldsymbol{\theta }\mathbf{|y})$ is chosen in such a way that (\ref%
{gen_expect}) can be computed with the aid of low-dimensional deterministic
integration. Taking a wider perspective, these approximate methods may also
be considered as novel approaches to inference and, hence, evaluated as
such, rather than being viewed \textit{solely} from the vantage point of%
\textbf{\ }computation.

\subsubsection{Approximate Bayesian computation (ABC)\label{abc}}

From its initial beginnings as a practical approach for inference in
population genetics models with intractable likelihoods (%
\citealp{tavare:balding:griffith:donnelly:1997}; %
\citealp{pritchard:seielstad:perez:feldman:1999}), ABC has grown in
popularity and is now commonly applied in numerous fields;{\ its broad
applicability highlighted by the more than 11,000 citations garnered on
Google Scholar since 2000. }As such, not only do several reviews of the area
exist (\citealp{marin:pudlo:robert:ryder:2011}; %
\citealp{sisson2011likelihood}), but the technique has recently reached
`handbook status', with the publication of \cite{sisson2018handbook}; and it
is to those resources that we refer the reader for extensive details on the
method, application and theory of ABC. We provide only the essence of the
approach here, including its connection to other computational methods.

The aim of ABC is to approximate $p(\boldsymbol{\theta }\mathbf{|y})$ in
cases where --- despite the complexity of the problem preventing the \textit{%
evaluation} of $p(\mathbf{y|}\boldsymbol{\theta })$ --- $p(\mathbf{y|}%
\boldsymbol{\theta })$ (and $p(\boldsymbol{\theta })$) can still be \textit{%
simulated}. The simplest (accept/reject) form of the algorithm proceeds as
follows: first, simulate $\boldsymbol{\theta }^{i}$, $i=1,2,...,M$, from $p(%
\boldsymbol{\theta })$, and artificial data $\mathbf{x}^{i}$ from $p(%
\boldsymbol{\cdot }|\boldsymbol{\theta }^{i})$; second, use $\mathbf{x}^{i}$
to construct ({a vector of}) simulated summary statistics $\eta (\mathbf{x}%
^{i})$, which is then compared against the {(vector of}) \textit{observed}
statistics $\eta (\mathbf{y})$ using a distance $d\{\cdot ,\cdot \}$; third,
retain all values of $\boldsymbol{\theta }^{i}$ that yield simulated
statistics, $\eta (\mathbf{x}^{i})$, for which $d\{\eta (\mathbf{x}%
^{i}),\eta (\mathbf{y})\}\leq \varepsilon $, for some small tolerance $%
\varepsilon $.

ABC thus produces draws of $\boldsymbol{\theta }$ from a posterior that
conditions not on the full data set $\mathbf{y}$, but on statistics $\eta (%
\mathbf{y})$ (with dimension less than $n$) that summarize the key
characteristics of $\mathbf{y.}$ Only if $\eta (\mathbf{y})$ are sufficient
for conducting inference on $\boldsymbol{\theta }$, and for $\varepsilon
\rightarrow 0$, does ABC provide draws from the exact posterior $p(%
\boldsymbol{\theta }\mathbf{|y})$. In practice, the complexity of the models
to which ABC is applied implies --- almost by definition --- that {a {%
low-dimensional} set of sufficient statistics is unavailable}, and the
implementation of the method (in finite computing time) requires a non-zero
value for $\varepsilon $, and a given number of draws, $M.$ As such, draws
from the ABC algorithm provide (via kernel density methods) a
simulation-based approximation of $p(\boldsymbol{\theta }\mathbf{|}\eta (%
\mathbf{y}))$, which we denote by $\widehat{p}_{\varepsilon }(\boldsymbol{%
\theta }\mathbf{|}\eta (\mathbf{y})).$

The difference between $\widehat{p}_{\varepsilon }(\boldsymbol{\theta }%
\mathbf{|}\eta (\mathbf{y}))$ and the unattainable $p(\boldsymbol{\theta }%
\mathbf{|y})$ has two components: the difference between the `partial'
posterior, $p(\boldsymbol{\theta }\mathbf{|}\eta (\mathbf{y}))$, and $p(%
\boldsymbol{\theta }\mathbf{|y})$, and the difference between $\widehat{p}%
_{\varepsilon }(\boldsymbol{\theta }\mathbf{|}\eta (\mathbf{y}))$ and $p(%
\boldsymbol{\theta }\mathbf{|}\eta (\mathbf{y})).$ The first difference is
the critical one, and depends on the informativeness, or otherwise, of the
chosen summaries; loosely speaking, the `closer' is $\eta (\mathbf{y})$ to
being sufficient for $\boldsymbol{\theta }$, the `closer' is $p(\boldsymbol{%
\theta }\mathbf{|}\eta (\mathbf{y}))$ to $p(\boldsymbol{\theta }\mathbf{|y})$%
. Attention has been given to maximizing the information content of the
summaries in some sense (e.g. \citealp{joyce:marjoram:2008}; %
\citealp{blum:2010}; \citealp{fearnhead:prangle:2012}). This includes the
idea of defining $\eta (\mathbf{y})$ as (some function of) the maximum
likelihood estimator (MLE) of the parameter vector of an approximating
`auxiliary' model; thereby producing summaries that are --- via the
properties of the MLE --- close to being \textit{asymptotically} sufficient,
depending on the accuracy of the approximating model (%
\citealp{drovandi:pettitt:faddy:2011}; \citealp{drovandi2015bayesian}; %
\citealp{martin2019auxiliary}). This approach mimics, in the Bayesian
setting, the frequentist methods of indirect inference %
\citep{gourieroux:monfort:renault:1993} and efficient method of moments %
\citep{gallant1996moments} using, as it does, an approximating model to
produce feasible inference about an intractable true model. Whilst the price
paid for the approximation in the frequentist case is reduced sampling
efficiency, in the Bayesian case the cost is posterior inference that is
conditioned on insufficient summaries, and is `partial' inference as a
consequence.

Regarding the second difference, at its simplest level: the smaller is $%
\varepsilon $ and the larger is $M$, the more accurate will $\widehat{p}%
_{\varepsilon }(\boldsymbol{\theta }\mathbf{|}\eta (\mathbf{y}))$ be as a
kernel density estimator of $p(\boldsymbol{\theta }\mathbf{|}\eta (\mathbf{y}%
))$, for any given choice of $\eta (\mathbf{y})$, with the dimension of both 
$\eta (\mathbf{y})$ and $\boldsymbol{\theta }$ affecting accuracy (%
\citealp{blum:etal:2013}; \citealp{FMRR2016}; \citealp{nott2018high}). For
given $M$ (and, hence, a given computational burden), modifications of the
basic accept/reject algorithm that improve the accuracy with which $p(%
\boldsymbol{\theta }\mathbf{|}\eta (\mathbf{y}))$ is estimated by $\widehat{p%
}_{\varepsilon }(\boldsymbol{\theta }\mathbf{|}\eta (\mathbf{y}))$ have been
proposed which, variously, involve post-sampling corrections of the draws (%
\citealp{be02}; \citealp{blum:2010}), the insertion of\ MCMC and/or SMC
steps (\citealp{marjoram:etal:2003}; \citealp{sisson:fan:tanaka:2007}; %
\citealp{beaumont:cornuet:marin:robert:2009}), or the use of randomized
quasi-Monte Carlo, rather than (standard) Monte Carlo in the simulation of $%
\boldsymbol{\theta }^{i}$, $i=1,2,...,M$ \citep{buchholz2019improving}.

Before concluding this section, we note that recent work has begun to
explore the use of ABC methods that do not rely on summary statistics but,
instead, match empirical measures calculated from the observed and simulated
data using appropriate metrics. In particular, we highlight the work of \cite%
{Bernton2019} based on the Wasserstein distance, but note that similar
approaches using alternative distances have also been proposed (e.g., %
\citealp{frazier2020robust}, and \citealp{nguyen2020approximate}). In the
case of a scalar $\theta $, and if the data is also scalar valued, \textit{%
Wasserstein ABC} amounts to selecting draws of $\theta $ based on matching
all $n$ order statistics in the observed sample $\mathbf{y}$. While the use
of the Wasserstein distance within ABC alleviates the need to choose
summaries, it is unclear exactly when this approach can be expected to
deliver superior inference to that obtained using summary statistics.
Indeed, the general performance of distance-based ABC methods relative to
methods based on very informative summaries is still a matter requiring
further investigation.

\subsubsection{Bayesian synthetic likelihood (BSL)\label{bsl}}

(Summary statistic-based) ABC thus targets $p(\boldsymbol{\theta }\mathbf{|}%
\eta (\mathbf{y}))\propto p(\eta (\mathbf{y})\mathbf{|}\boldsymbol{\theta }%
)p(\boldsymbol{\theta }),$ with $p(\boldsymbol{\theta }\mathbf{|}\eta (%
\mathbf{y}))$ itself, for nonsufficient $\eta (\mathbf{y})$, being an
approximate representation of $p(\boldsymbol{\theta }\mathbf{|y}).$ It is
clear then that, embedded within the {simplest accept/reject ABC algorithm},
based on a tolerance $\varepsilon $, is a likelihood function of the form, 
\begin{equation}
p_{\varepsilon }(\eta (\mathbf{y})|\boldsymbol{\theta })=\int_{\mathcal{X}}p(%
\mathbf{x}|\boldsymbol{\theta })\mathbb{I}\left( d\{\eta (\mathbf{y}),\eta (%
\mathbf{x})\}\leq \varepsilon \right) d\mathbf{x}.  \label{abc_like}
\end{equation}%
For a given draw $\boldsymbol{\theta }^{i}$, and associated $\eta (\mathbf{x}%
^{i})$, (\ref{abc_like}) is approximated by its simulation counterpart, $%
\widehat{p}_{\varepsilon }(\eta (\mathbf{y})|\boldsymbol{\theta }^{i})$ $=%
\mathbb{I}\left( d\{\eta (\mathbf{y}),\eta (\mathbf{x}^{i})\}\leq
\varepsilon \right) ,$ which can implicitly be viewed as a nonparametric
estimator, based on a Uniform kernel, for the quantity of interest $%
p_{\varepsilon }(\eta (\boldsymbol{\mathbf{y}})|\boldsymbol{\theta })$.
Following \cite{andrieu:roberts:2009}, and as illustrated in detail by \cite%
{bornn2017use}, $\widehat{p}_{\varepsilon }(\eta (\mathbf{y})|\boldsymbol{%
\theta }^{i})$ can serve as a likelihood estimate within a form of
pseudo-marginal MCMC scheme (referred to as ABC-MCMC by the authors) for
sampling from $p_{\varepsilon }(\boldsymbol{\theta }\mathbf{|}\eta (\mathbf{y%
}))\propto p_{\varepsilon }(\eta (\mathbf{y})|\boldsymbol{\theta })p(%
\boldsymbol{\theta }),$ where in this context we take `pseudo-marginal MCMC'
to mean an MCMC scheme that replaces the intractable likelihood, $%
p_{\varepsilon }(\eta (\mathbf{y})|\boldsymbol{\theta })$, within the MH
ratio by an unbiased estimator, $\widehat{p}_{\varepsilon }(\eta (\mathbf{y}%
)|\boldsymbol{\theta }^{i})$. (See also \citealp{marjoram:etal:2003}.)
However, in contrast with other results in the pseudo-marginal literature, 
\cite{bornn2017use} demonstrate that the efficiency of the MCMC chain so
produced is not necessarily improved by using more than one draw of $\eta (%
\mathbf{x}^{i})$ for a given draw $\boldsymbol{\theta }^{i}.$

Bayesian synthetic likelihood (BSL)\textit{\ }\citep{price2018bayesian} also
targets a posterior for $\boldsymbol{\theta }$ that conditions on $\eta (%
\mathbf{y})$, and requires only\ simulation\textit{\ }from $p(\mathbf{y|}%
\boldsymbol{\theta })$\textit{\ }(not its evaluation) in so doing.\textit{\ }%
However, in contrast to the nonparametric likelihood estimate that is
implicit in ABC, BSL (building on \citealp{wood:2010}) adopts a Gaussian
parametric approximation to $p(\eta (\mathbf{y})|\boldsymbol{\theta })$,%
\begin{equation}
p_{a}(\eta (\mathbf{y})|\boldsymbol{\theta })=\mathcal{N}\left[ \eta (%
\mathbf{y});\mu (\boldsymbol{\theta }),\Sigma (\boldsymbol{\theta })\right]
,\;\mu (\boldsymbol{\theta })=\mathbb{E}[\eta (\mathbf{y})],\;\Sigma (%
\boldsymbol{\theta })=\text{Var}\left[ \eta (\mathbf{y})\right] .
\label{bsl_likelihood}
\end{equation}%
Use of this parametric kernel leads to the \textit{ideal} BSL posterior, 
\begin{equation}
p_{a}(\boldsymbol{\theta }\mathbf{|}\eta (\mathbf{y}))\propto p_{a}(\eta (%
\mathbf{y})|\boldsymbol{\theta })p(\boldsymbol{\theta }),  \label{bsl_1}
\end{equation}%
where the subscript `$a$' highlights that (\ref{bsl_1}) is still an
approximation to $p(\boldsymbol{\theta }\mathbf{|}\eta (\mathbf{y}))$, due
to the Gaussian approximation, $p_{a}(\eta (\mathbf{y})|\boldsymbol{\theta }%
) $, of $p(\eta (\mathbf{y})\mathbf{|}\boldsymbol{\theta })$.

In general, however, the mean and variance-covariance matrix of $\eta (%
\mathbf{y})$ are unknown and must be estimated via simulation. Given $%
\mathbf{x}_{j}\sim i.i.d.$ $p(\boldsymbol{\cdot }|\boldsymbol{\theta })$, $%
j=1,\dots ,m$, we can estimate $\mu (\boldsymbol{\theta })$ and $\Sigma (%
\boldsymbol{\theta })$ in (\ref{bsl_likelihood}) via their empirical Monte
Carlo averages, $\mu _{m}(\boldsymbol{\theta })=\frac{1}{m}%
\sum_{j=1}^{m}\eta (\mathbf{x}_{j})$ and $\Sigma _{m}(\boldsymbol{\theta })=%
\frac{1}{m-1}\sum_{j=1}^{m}\mathcal{(}\eta (\mathbf{x}_{j})-\mu _{m}(%
\boldsymbol{\theta }))(\eta (\mathbf{x}_{j})-\mu _{m}(\boldsymbol{\theta }%
))^{\prime },$ and thereby define%
\begin{equation}
p_{a,m}(\eta (\mathbf{y})\mathbf{|}\boldsymbol{\theta })=\int_{\mathcal{X}}%
\mathcal{N}\left[ \eta (\mathbf{y});\mu _{m}(\boldsymbol{\theta }),\Sigma
_{m}(\boldsymbol{\theta })\right] \prod_{j=1}^{m}p(\eta (\mathbf{x}_{j})|%
\boldsymbol{\theta })d\mathbf{x}_{1}\dots d\mathbf{x}_{m},
\label{bsl_likelihood_2}
\end{equation}%
and the associated \textit{target} BSL posterior, 
\begin{equation}
p_{a,m}(\boldsymbol{\theta }\mathbf{|}\eta (\mathbf{y}))\propto p_{a,m}(\eta
(\mathbf{y})\mathbf{|}\boldsymbol{\theta })p(\boldsymbol{\theta }).
\label{bsl_2}
\end{equation}%
Note that, even for a single draw $\eta (\mathbf{x}_{j})$, $\mathbf{x}%
_{j}\sim p(\boldsymbol{\cdot} |\boldsymbol{\theta })$, we have that $%
\mathcal{N}\left[ \eta (\mathbf{y});\mu _{m}(\boldsymbol{\theta }),\Sigma
_{m}(\boldsymbol{\theta })\right] $ is an unbiased estimate of (\ref%
{bsl_likelihood_2}). Hence, with $p_{a,m}(\boldsymbol{\theta }\mathbf{|}\eta
(\mathbf{y}))$ then accessed via an MCMC algorithm, and with arguments in 
\cite{drovandi2015bayesian} used to show that $p_{a,m}\rightarrow p_{a}$ as $%
m\rightarrow \infty $, BSL can yield a form of pseudo-marginal\ MCMC method.

\subsubsection{ABC versus BSL}

ABC and BSL both invoke two forms of approximation: one, replacement\textbf{%
\ }of the full dataset $\mathbf{y}$ by the summaries $\eta (\mathbf{y})$;
two, approximation of the likelihood\textbf{\ }$p(\eta ($\textbf{$\mathbf{y}%
)|\boldsymbol{\theta }$}$)$. It is the difference in the second form of
approximation that distinguishes the two approaches and leads to potential
differences in performance. It is helpful to characterize these differences
in terms of \textit{i)} asymptotic (in $n$) behaviour and \textit{ii)}
computational efficiency. This then enables us to provide some guidelines as
to when, and why, one might use one method over the other. We consider 
\textit{i)} and\textit{\ ii)} in turn.

\textit{i) }As ABC has evolved into a common approach to inference,
attention has turned to its asymptotic validation. This work demonstrates
that, under certain conditions on $\eta (\mathbf{y})$, $\varepsilon $ and $M$%
, the ABC posterior $\widehat{p}_{\varepsilon }(\boldsymbol{\boldsymbol{%
\theta }}\mathbf{|}\eta (\mathbf{y}))$ concentrates onto the true vector $%
\boldsymbol{\boldsymbol{\theta }}_{0}$ (i.e. is Bayesian consistent),
satisfies a Bernstein von Mises (BvM) theorem (i.e. is asymptotically
Gaussian with credible sets that have the correct level of frequentist
asymptotic coverage) and yields an ABC posterior mean with an asymptotically
Gaussian sampling distribution. (See \citealp{FMRR2016}, for this full suite
of results, and \citealp{LF2016b}, \citealp{LF2016a}, and %
\citealp{frazier2020model}, for related work.) Moreover, the conditions on $%
\eta (\mathbf{y})$ under which these results are valid are surprisingly
weak, requiring only the existence of at least a polynomial moment
(uniformly in the parameter space). In addition, we note that the ABC
posterior mean is asymptotically as efficient as the maximum likelihood
estimator based on $p(\eta (\mathbf{y})|\boldsymbol{\theta })$.

The required conditions on the tolerance, $\varepsilon $, for these results
to be in evidence can be ordered in terms of the speed with which $\epsilon
\rightarrow 0$ as $n\rightarrow \infty $: stronger results, such as a valid
BvM, require faster rates of decay for $\varepsilon $ than weaker results,
such as posterior concentration. Such a taxonomy is important since the
chosen tolerance $\varepsilon $ largely determines the computational effort
required for $\widehat{p}_{\varepsilon }(\boldsymbol{\theta }|\eta (\mathbf{y%
}))$ to be an accurate estimate of $p(\boldsymbol{\theta }|\eta (\mathbf{y}))
$. Broadly speaking, the smaller is\textbf{\ }$\varepsilon $, the smaller is%
\textbf{\ }$|\widehat{p}_{\varepsilon }(\boldsymbol{\theta }|\eta (\mathbf{y}%
))-p(\boldsymbol{\theta }|\eta (\mathbf{y}))|$. However, a smaller choice of 
$\varepsilon $ requires a larger number of simulations (i.e., a larger value
of $M$) and, hence, a greater computational burden. For instance, if we wish
for credible sets obtained by $\widehat{p}_{\varepsilon }(\boldsymbol{\theta 
}|\eta (\mathbf{y}))$ to be valid in the frequentest sense, $M$ is required
to diverge faster than $n^{\text{dim}(\eta )/2}$ (Corollary 1 in %
\citealp{FMRR2016}).

In contrast to ABC, as BSL is based on the Gaussian parametric approximation
to the likelihood $p(\eta (\mathbf{y})|\boldsymbol{\theta })$, it does not
require any choice of tolerance. However, in order for the BSL posterior $%
p_{a,m}(\boldsymbol{\theta }|\eta (\mathbf{y}))$ to be a reasonable
approximation to $p(\boldsymbol{\theta }|\eta (\mathbf{y}))$, the Gaussian
approximation must be reasonable. More specifically, the summaries $\eta (%
\mathbf{y})$ and $\eta (\mathbf{z})$ themselves must satisfy a CLT
(uniformly in the case of the latter) (see \citealp{frazier2019bayesian} for
details), and the variance of the summaries must be consistently estimated
by $\Sigma _{m}(\boldsymbol{\theta })$ for some value of $\boldsymbol{\theta 
}$, as $m$ (the number of data sets drawn for a given draw of $\boldsymbol{%
\theta }$) increases. If, moreover, we wish $p_{a,m}(\boldsymbol{\theta }%
|\eta (\mathbf{y}))$ to deliver asymptotically correct frequentest coverage,
additional conditions on the summaries and $m$ are required. In particular, 
\cite{frazier2019approximate} demonstrate that if the summaries exhibit an
exponential moment, then correct uncertainty quantification is achieved so
long as $m/\log (n)$\textbf{$\rightarrow \infty $}.

Under the restrictions delineated above for $\eta (\mathbf{y})$, $%
\varepsilon $, $M$ and $m$, the results of \cite{FMRR2016} and \cite%
{frazier2019bayesian} can then be used to deduce that the ABC and BSL
posteriors are asymptotically equivalent, in the sense that $\int |p_{a,m}(%
\boldsymbol{\theta }|\eta (\mathbf{y}))-\widehat{p}_{\varepsilon }(%
\boldsymbol{\theta }|\eta (\mathbf{y}))|d\boldsymbol{\theta }\overset{p}{%
\rightarrow }0$ as $n\rightarrow \infty .$ That is, in large samples, and
under regularity, we could expect the results obtained by both methods to be
comparable. However, the above discussion makes plain that BSL requires much
stronger conditions on the summaries than does ABC to produce equivalent
asymptotic behaviour. Hence, in the case of summaries that have thick tails,
non-Gaussian features, or non-standard rates of convergence, ABC would seem
to be the better choice.

\textit{ii) }Once computational efficiency is taken into account however,
the comparison between the two methods becomes more nuanced. \cite%
{frazier2019bayesian} use theoretical arguments to compare the computational
efficiency of BSL and accept/reject ABC, and demonstrate that BSL does not
pay the same penalty for summary statistic dimension as does ABC. In
particular, the BSL acceptance probability is asymptotically non-vanishing,
and does not depend on the dimension of the summaries, neither of which is
true for accept/reject ABC, even under an optimal choice for $M$. Given
this, when the summaries are approximately Gaussian, BSL is likely to be
more computationally efficient than standard ABC.\footnote{%
BSL can often be implemented using the random walk MH algorithm, and often
with minimal tuning required in practice (\citealp{price2018bayesian}). See
also \cite{frazier2019robust} for a slice sampling approach to sampling the
BSL posterior.} That being said, if one considers post-sampling corrections
to ABC (such as those cited in Section \ref{abc}), or the newer ABC-Gibbs
schemes (Section \ref{hybrid}), whose specific goal is to mitigate the curse
of dimensionality associated with the summaries, then these
\textquotedblleft corrected\textquotedblright\ ABC approaches are likely to
be more efficient or comparable to standard BSL on computational grounds.
What is still undocumented is the relative performance of corrected\ ABC
methods and the more computationally efficient varieties of BSL, such as the
shrinkage approaches of \cite{ong2018likelihood} or the whitening approach
of \cite{priddle2019efficient}.

\subsubsection{Variational Bayes (VB)\label{vb}}

The two approximate methods discussed thus far, ABC and BSL, target an
approximation of the posterior that is (in a standard application of the
methods) conditioned on a vector of low-dimensional summary statistics. As
such, and most particularly when $\eta (\mathbf{y})$ is not sufficient for $%
\boldsymbol{\theta }$, these methods do not directly target the exact
posterior $p(\boldsymbol{\theta }|\mathbf{y})$, nor any expectation, (\ref%
{gen_expect}), defined with respect to it. In contrast, VB methods are a
general class of algorithms that produce an approximation to the posterior $%
p(\boldsymbol{\theta }|\mathbf{y})$ --- and hence (\ref{gen_expect}) --- 
\textit{directly}, by replacing simulation with optimization.

The idea of VB is to search for the best approximation to the posterior $p(%
\boldsymbol{\theta }|\mathbf{y})$ over a class of densities $\mathcal{Q}$,
referred to as the variational family, and where $q(\boldsymbol{\theta })$
indexes elements in $\mathcal{Q}$. The most common approach to VB is to find
the best approximation to the exact posterior, in the class $\mathcal{Q}$,
by minimizing the KL divergence between $q(\boldsymbol{\theta })$ and the
posterior $p(\boldsymbol{\theta }|\mathbf{y})$, which defines such a density
as the solution to the following optimal optimization problem, 
\begin{equation}
q^{\ast }(\boldsymbol{\theta }):=\argmin_{q(\boldsymbol{\theta })\in 
\mathcal{Q}}\text{KL}\left[ q(\boldsymbol{\theta })|p(\boldsymbol{\theta }|%
\mathbf{y})\right] ,  \label{opt_1}
\end{equation}%
where 
\begin{equation}
\text{KL}\left[ q(\boldsymbol{\theta })|p(\boldsymbol{\theta }|\mathbf{y})%
\right] =\int \log (q(\boldsymbol{\theta }))q(\boldsymbol{\theta })d%
\boldsymbol{\theta }-\int \log (p(\boldsymbol{\theta }|\mathbf{y}))q(%
\boldsymbol{\theta })d\boldsymbol{\theta }\equiv \mathbb{E}_{q}[\log (q(%
\boldsymbol{\theta }))]-\mathbb{E}_{q}[\log (p(\boldsymbol{\theta },\mathbf{y%
}))]+\log (p(\mathbf{y})).  \label{KL}
\end{equation}%
Of course, the normalizing constant $p(\mathbf{y})$ is, in all but most
simple problems (for which VB would not be required!), unknown; and the
quantity in (\ref{KL}) inaccessible as a result. Rather, the approach
adopted is to define the so-called evidence lower bound (ELBO),%
\begin{equation}
\text{ELBO}[q(\boldsymbol{\theta })]:=\mathbb{E}_{q}[\log (p(\boldsymbol{%
\theta },\mathbf{y}))]-\mathbb{E}_{q}[\log (q(\boldsymbol{\theta }))],
\label{elbo}
\end{equation}%
where KL$\left[ q(\boldsymbol{\theta })|p(\boldsymbol{\theta }|\mathbf{y})%
\right] $ is equivalent to $-$ELBO$\left[ q(\boldsymbol{\theta })\right] $
up to the unknown constant, $\log (p(\mathbf{y}))$,\ with the latter not
dependent on $q(\boldsymbol{\theta })$. Hence, we can obtain the variational
density by solving an optimization problem that is equivalent to that in (%
\ref{opt_1}): 
\begin{equation}
q^{\ast }(\boldsymbol{\theta }):=\argmax_{q(\boldsymbol{\theta })\in 
\mathcal{Q}}\text{ELBO}[q(\boldsymbol{\theta })].  \label{vb_1}
\end{equation}

The beauty of VB is that, for \textit{certain} problems, including certain
choices of the class $\mathcal{Q}$, the optimization problem in (\ref{vb_1})
can either yield a closed-form solution, or be solved relatively quickly
with various numerical algorithms; (see \citealp{ormerod2010explaining}, %
\citealp{blei2017variational}, and \citealp{zhang2018advances}, for
reviews). Most importantly, given that --- by design --- the variational
family is defined in terms of standard forms of distributions, replacement
of $p(\boldsymbol{\theta }|\boldsymbol{y})$\ by $q^{\ast }(\boldsymbol{%
\theta })$ in (\ref{gen_expect}) yields an expectation that is either
available in closed form, or amenable to a relatively simple
simulation-based solution.

VB \textit{truly} shines in cases where $\boldsymbol{\theta }$ is
high-dimensional, and an efficient MCMC algorithm may well be simply out of
reach. Indeed, in such cases, the family $\mathcal{Q}$ can be chosen in such
a way that the resulting posterior approximations remain tractable even when
the dimension of $\boldsymbol{\theta }$ is in the thousands, or the tens of
thousands (\citealp{braun2010variational}; \citealp{kabisa2016online}; %
\citealp{wand2017fast}; \citealp{koop2018variational}).

Finally, the link between (\ref{KL}) and (\ref{elbo}) makes it clear that
maximizing (\ref{elbo}) to yield $q^{\ast }(\boldsymbol{\theta })$ produces,
as a by-product, a lower bound on the logarithm of the `evidence', or
marginal likelihood, $p(\mathbf{y}).$ Hence, ELBO$\left[ q^{\ast }(%
\boldsymbol{\theta })\right] $ serves as an estimate of the quantity which,
as noted in Section \ref{general} and discussed further in Section \ref%
{ml_sect}, underpins model choice.

Recently, several authors have analyzed the asymptotic properties of VB
methods; see, for example, \cite{wangblei2019b,wangblei2019a} and \cite%
{zhang2017convergence}. The most complete treatment can be found in %
\citeauthor{zhang2017convergence}, wherein the authors demonstrate that the
rate at which the VB posterior concentrates is bounded above by the
following two components: \textit{i)} the concentration rate of the exact
posterior, and \textit{ii)} the approximation error incurred by the chosen
variational family. This novel decomposition highlights the fundamental
importance of the variational family that is used to approximate the
posterior, something that is not present in other results on the asymptotic
behavior of VB. Interestingly, while \citeauthor{zhang2017convergence}
deliver a convenient upper bound in a general context, they also demonstrate
that in specific examples, such as Gaussian sequence models and sparse
linear regression models, the VB posterior can display concentration rates
that are actually faster than those obtained by the exact posterior, owing
to the fact that VB performs a type of `internal regularization' as a
consequence of the algorithm's optimization step.\footnote{\cite%
{huggins2019validated} propose a method for validating the accuracy of VB
posterior approximations using alternative (nonasymptotic) principles. See
also \cite{yu2019assessment} (and earlier references therein) for practical
validation approaches that are relevant to approximate posteriors in general.%
}

\subsubsection{Integrated nested Laplace approximation (INLA)\label{inla}}

We complete our review of 21st century approximate computational methods
with a reminder of a computational innovation from the 18th! In 1774, in a
quest to illustrate posterior consistency, Laplace produced an asymptotic
(in $n$) approximation to a particular posterior probability. Not only did
this result represent the first step in the development of Bayesian
asymptotic \textit{theory}, it also provided a simple practical solution to
the \textit{computation }of general posterior expectations like that in (\ref%
{gen_expect}). In 1986, Tierney and Kadane revived and formalized the
Laplace approximation: using it to yield an asymptotic approximation (of a
given order) of any posterior expectation of the form of (\ref{gen_expect}),
including (in the multiple parameter case) marginal posterior densities.
(See also \citealp{tierney:kass:kadane:1989}.) Two decades later, \cite%
{rue:martino:chopin:2009} took the method further: adapting it to
approximate marginal posteriors (and {general expectations like those in (%
\ref{gen_expect})}) in latent Gaussian models (LGMs). With the authors using
a series of \textit{nested }Laplace approximations, allied with
low-dimensional numerical \textit{integration}, they termed their method 
\textit{integrated nested Laplace approximation, }or INLA for short. Since
the LGM class encompasses a large range of empirically relevant models ---
including, generalized linear models, non-Gaussian state space (or hidden
Markov) models, and spatial, or spatio-temporal models --- a computational
method tailored-made for such a setting is sufficiently broad in its
applicability to warrant detailed consideration herein. In common with VB,
and as follows from the use of Laplace approximations evaluated at modal
values, INLA eschews simulation for optimization (in addition to using
low-dimensional deterministic integration methods).

Deferring to \cite{rue:martino:chopin:2009}, \cite{martino2019integrated}
and \cite{wood2019simplified} for all details (including of the LGM
structure), we provide here the \textit{key} steps of INLA. To adhere to our
goal of notational consistency, we reference the $p$-dimensional vector of
static parameters in the LGM as $\boldsymbol{\theta }$, the $n$-dimensional
vector of latent, random parameters as $\mathbf{z}$, and the $n$-dimensional
vector of observed {data} {as} $\mathbf{y}$, and express the model as: 
\begin{equation*}
\mathbf{y|z,}\boldsymbol{\theta }\sim \prod\nolimits_{i=1}^{n}p(y_{i}|z_{i},%
\boldsymbol{\theta })\qquad{\ }\mathbf{z|}\boldsymbol{\theta} \sim \mathcal{N%
}(0,Q^{-1}(\boldsymbol{\theta }))\qquad{\ }\boldsymbol{\theta }\sim p(%
\boldsymbol{\theta }),
\end{equation*}%
where $Q(\boldsymbol{\theta })$ is the precision matrix of the latent
Gaussian field, assumed --- for computational feasibility --- to be sparse.
Whilst the LGM allows for each element of $\mathbf{y}$ to be of dimension $%
d\geq 1$, we keep our description simple by assuming $d=1.$ The goal of the
authors is to approximate the marginal posteriors; $p(\theta _{j}|\mathbf{y}%
) $, $j=1,2,..,p$, and $p(z_{i}|\mathbf{y})$, $i=1,2,..,n.$ The problems
envisaged are those in which $p$ is small and $n$ is large (potentially in
the order of millions), with MCMC algorithms deemed to be possibly
unsuitable as a consequence, due to the scale of both $\mathbf{z}$ and $%
\boldsymbol{y}$.

Beginning with the expression of $p(\boldsymbol{\theta }|\mathbf{y})$ as%
\begin{equation}
p(\boldsymbol{\theta }|\mathbf{y})=\frac{p(\mathbf{z},\boldsymbol{\theta }|%
\mathbf{y})}{p(\mathbf{z}|\boldsymbol{\theta },\mathbf{y})}\propto \frac{p(%
\mathbf{z},\boldsymbol{\theta },\mathbf{y})}{p(\mathbf{z}|\boldsymbol{\theta 
},\mathbf{y})}=\frac{p(\mathbf{y|z,}\boldsymbol{\theta })p(\mathbf{z}|%
\boldsymbol{\theta })p(\boldsymbol{\theta })}{p(\mathbf{z}|\boldsymbol{%
\theta },\mathbf{y)}},  \label{theta}
\end{equation}%
and recognizing that the proportionality sign arises due to the usual lack
of integrating constant (over $\mathbf{z}$ and $\boldsymbol{\theta }$), the
steps of the algorithm (in its simplest form) are as follows. First, on the
assumption that all components of the model can be evaluated and, hence,
that the numerator is available, $p(\boldsymbol{\theta }|\mathbf{y})$ in (%
\ref{theta}) is approximated as%
\begin{equation}
\widetilde{p}(\boldsymbol{\theta }|\mathbf{y})\propto \frac{p(\mathbf{y|%
\widehat{\mathbf{z}}}(\mathbf{\boldsymbol{\theta }})\mathbf{,}\boldsymbol{%
\theta })p(\mathbf{\widehat{\mathbf{z}}}(\mathbf{\boldsymbol{\theta }})|%
\boldsymbol{\theta })p(\boldsymbol{\theta })}{p_{G}(\mathbf{\widehat{\mathbf{%
z}}}(\mathbf{\boldsymbol{\theta }})|\boldsymbol{\theta },\mathbf{y})}.
\label{approx_1}
\end{equation}%
The denominator in (\ref{approx_1}) represents a Gaussian approximation of $%
p(\mathbf{z}|\boldsymbol{\theta },\mathbf{y})$, $p_{G}(\mathbf{z}|%
\boldsymbol{\theta },\mathbf{y})=\mathcal{N}(\widehat{\mathbf{z}}(%
\boldsymbol{\theta }),\widehat{\Sigma }(\boldsymbol{\theta }))$, evaluated
at the mode, $\widehat{\mathbf{z}}(\boldsymbol{\theta })$, of $p(\mathbf{z},%
\boldsymbol{\theta },\mathbf{y})$ (at a given value of $\boldsymbol{\theta }$%
), where $\widehat{\Sigma }(\boldsymbol{\theta })$ is the inverse of the
Hessian of $-\log p(\mathbf{z},\boldsymbol{\theta },\mathbf{y)}$ with
respect to $\mathbf{z}$, also evaluated at $\widehat{\mathbf{z}}(\boldsymbol{%
\theta }).$ The expression in (\ref{approx_1}) can obviously be further
simplified to%
\begin{equation}
\widetilde{p}(\boldsymbol{\theta }|\mathbf{y})\propto p(\mathbf{y|\mathbf{%
\widehat{\mathbf{z}}(\boldsymbol{\theta }}),}\boldsymbol{\theta })p(\mathbf{%
\widehat{\mathbf{z}}(\boldsymbol{\theta }})|\boldsymbol{\theta }\mathbf{)}p(%
\boldsymbol{\theta })\left\vert \widehat{\Sigma }(\boldsymbol{\theta }%
)\right\vert ^{1/2}.  \label{approx_2}
\end{equation}%
With appropriate adjustments made for notation, and noting that the
expression is given up to the integrating constant only, (\ref{approx_2})
can be seen to be identical to the Laplace approximation of a marginal
density in Tierney and Kadane (1986, equation (4.1)). (%
\citealp{rue:martino:chopin:2009} discuss the circumstances in which the
order of approximation proven in \citealp{tierney:kadane:1986} applies to
the LGM setting.)

With marginal posterior for the $ith$ element of $\mathbf{z}$ defined as%
\begin{equation}
\widetilde{p}(z_{i}|\mathbf{y})=\int_{{\Theta }}\widetilde{p}(z_{i}|%
\boldsymbol{\theta },\mathbf{y)}\widetilde{p}(\boldsymbol{\theta }|\mathbf{y}%
)d\boldsymbol{\theta },  \label{marg_zi}
\end{equation}%
a second application of a Laplace approximation would yield%
\begin{equation}
\widetilde{p}(z_{i}|\boldsymbol{\theta },\mathbf{y)}\propto p(\mathbf{y|%
\widehat{\mathbf{z}}}_{-i}\mathbf{(}\boldsymbol{\theta },z_{i})\mathbf{,}%
\boldsymbol{\theta })p(\mathbf{\widehat{\mathbf{z}}}_{-i}\mathbf{(}%
\boldsymbol{\theta },z_{i})|\boldsymbol{\theta }\mathbf{)}p(\boldsymbol{%
\theta })\left\vert \widehat{\Sigma }_{-i}(\boldsymbol{\theta }%
,z_{i})\right\vert ^{1/2},  \label{approx_3}
\end{equation}%
where $\widehat{\mathbf{z}}_{-i}(\boldsymbol{\theta },z_{i})$ is the mode of 
$p(\mathbf{z}_{-i},z_{i},\boldsymbol{\theta },\mathbf{y})$ (at given values
of $\boldsymbol{\theta }$ and $z_{i}$, with $\mathbf{z}_{-i}$ denoting all
elements of $\mathbf{z}$ other than the $ith$); and where $\widehat{\Sigma }%
_{-i}(\boldsymbol{\theta },z_{i})$ is the inverse of the Hessian of $-\log p(%
\mathbf{z}_{-i},z_{i},\boldsymbol{\theta },\mathbf{y)}$ with respect to $%
\mathbf{z}_{-i}$, also evaluated at $\widehat{\mathbf{z}}_{-i}(\boldsymbol{%
\theta },z_{i}).$ Computation of (\ref{approx_3}) for each $z_{i}$ would,
however, involve $n$ optimizations (over $\mathbf{z}_{-i}$) plus $n$
specifications of the high-dimensional matrix $\widehat{\Sigma }_{-i}(%
\boldsymbol{\theta },z_{i}).$ \cite{rue:martino:chopin:2009} avoid this
computational burden by modifying the approximation in (\ref{approx_3}) in a
number of alternative ways, all details of which are provided in the
references cited above. Once a representation of $\widetilde{p}(z_{i}|%
\boldsymbol{\theta },\mathbf{y)}$ is produced, (\ref{marg_zi}) is computed
using a deterministic numerical integration scheme defined over a grid of
values for the low-dimensional $\boldsymbol{\theta }.$

Defining the marginal posterior for the $jth$ element of $\boldsymbol{\theta 
}$ as $\widetilde{p}(\theta _{j}|\mathbf{y})=\int_{{\Theta }_{-j}}\widetilde{%
p}(\boldsymbol{\theta }|\mathbf{y})d\boldsymbol{\theta }_{-j},$ where $%
\boldsymbol{\theta }_{-j}$ denotes all elements of $\boldsymbol{\theta }$
excluding $\theta _{j}$, this integral is computed using deterministic
integration over $\boldsymbol{\theta }_{-j}.$ Finally, if required, {the
marginal likelihood}, $p(\mathbf{y})$ can be approximated by computing the
normalizing constant in (\ref{approx_2}), $\int_{{\Theta }}p(\mathbf{y|%
\mathbf{\widehat{\mathbf{z}}}}(\mathbf{\mathbf{\boldsymbol{\theta }}})%
\mathbf{,}\boldsymbol{\theta })p(\mathbf{\widehat{\mathbf{z}}}(\mathbf{%
\boldsymbol{\theta }})|\boldsymbol{\theta }\mathbf{)}p(\boldsymbol{\theta }%
)\left\vert \widehat{\Sigma }(\boldsymbol{\theta })\right\vert ^{1/2}d%
\boldsymbol{\theta },$ using deterministic integration over $\boldsymbol{%
\theta }$.\footnote{{All steps of the algorithm are provided in a dedicated
package, R-INLA, for the general LGM framework, with particular packages
also available for implementing INLA in more specific models nested within
the LGM class; see \cite{martino2019integrated} for a listing of all such
packages.}}

\subsection{Hybrid Methods\label{hybrid}}

We remind the reader at this point of the following:\textit{\ i)} whilst ABC
and BSL are advantageous when $p(\mathbf{y}|\boldsymbol{\theta })$ cannot be
evaluated, a large dimension for $\boldsymbol{\theta }$ (and, hence, for $%
\eta (\mathbf{y})$) causes challenges (albeit to differing degrees) for
both;\ \textit{ii)} VB is much better equipped to deal with high-dimensional 
$\boldsymbol{\theta },$ but requires the evaluation of $p(\boldsymbol{\theta 
},\mathbf{y})$ and, thus, $p(\mathbf{y}|\boldsymbol{\theta })$; \textit{iii) 
}pseudo-marginal MCMC circumvents a challenging evaluation of $p(\mathbf{y}|%
\boldsymbol{\theta })$ by using an unbiased estimate, $h(\mathbf{y}|%
\boldsymbol{\theta },\mathbf{u})$, as defined in Section \ref{basic_pm}.
Recently, hybrid algorithms that meld aspects of all four methods --- ABC,
BSL, VB and pseudo-marginal MCMC --- have been used to deal with settings in
which the likelihood is intractable \textit{and} $\boldsymbol{\theta }$ is
high-dimensional.

\cite{tran2017variational} devise a VB method for use when the likelihood is
intractable, coining the technique `VBIL'. To appreciate the principles of
the method, consider that the variational approximation is indexed by a
finite dimensional parameter $\lambda $, so that $\mathcal{Q}:=\{\lambda \in
\Lambda :q_{\lambda }\}$. The variational approximation is then obtained by
maximizing the ELBO, $\mathcal{L}(\lambda ):=\text{ELBO}[q_{\lambda }(%
\boldsymbol{\theta })]$, over $\Lambda $. VBIL replaces the intractable
likelihood $p(\mathbf{y}|\boldsymbol{\theta })$ with an estimator $h(\mathbf{%
y}|\boldsymbol{\theta },\mathbf{u})$, such that $E_{\mathbf{u}}[h(\mathbf{y}|%
\boldsymbol{\theta },\mathbf{u})]=p(\mathbf{y}|\boldsymbol{\theta })$, and
considers as target distribution the joint posterior 
\begin{equation*}
h(\boldsymbol{\theta },z|\mathbf{y})\propto \pi (\boldsymbol{\theta })h(%
\mathbf{y}|\boldsymbol{\theta },\mathbf{u})\exp (z)g(z|\boldsymbol{\theta }),%
\text{ where }z:=\log h(\mathbf{y}|\boldsymbol{\theta },\mathbf{u})-\log p(%
\mathbf{y}|\boldsymbol{\theta }),
\end{equation*}%
and where $g(z|\boldsymbol{\theta })$ denotes the distribution of $z|%
\boldsymbol{\theta }$. Given that $h(\mathbf{y}|\boldsymbol{\theta },\mathbf{%
u})$ is, by construction, an unbiased estimator of $p(\mathbf{y}|\boldsymbol{%
\theta })$, it follows that marginalizing over $z$ in $h(\boldsymbol{\theta }%
,z|\mathbf{y})$, yields the posterior distribution of interest, namely $p(%
\boldsymbol{\theta }|\mathbf{y})$. \citeauthor{tran2017variational} then
minimize $\text{KL}[q_{\lambda }(\boldsymbol{\theta },z)|p(\boldsymbol{%
\theta },z|\mathbf{y})]$ over\textbf{\ }the augmented space of $(\boldsymbol{%
\theta },z)$, using as the variational family $\mathcal{Q}$ distributions of
the form $q_{\lambda }(\boldsymbol{\theta },z)=q_{\lambda }(\boldsymbol{%
\theta })g(z|\boldsymbol{\theta })$. Whilst, in general, minimization of $%
\text{KL}[q_{\lambda }(\boldsymbol{\theta },z)|p(\boldsymbol{\theta },z|%
\mathbf{y})]$ is not the same as minimization of $\mathcal{L}(\lambda )$,
the authors demonstrate the two solutions \textit{do} correspond under
particular tuning regimes for $h(\mathbf{y}|\boldsymbol{\theta },\mathbf{u})$%
.

Following \cite{tran2017variational}, \cite{ong2018variational} propose an
alternative VB method for intractable likelihood problems. The authors begin
with the recognition that establishing the conditions under which the
minimizers of $\text{KL}[q_{\lambda }(\boldsymbol{\theta },z)|p(\boldsymbol{%
\theta },z|\mathbf{y})]$ and $\mathcal{L}(\lambda )$ coincide is
non-trivial, and that in certain types of problems it may be difficult to
appropriately tune $h(\mathbf{y}|\boldsymbol{\theta },\mathbf{u})$ so that
they coincide. This acknowledgement then prompts them to construct a
variational approximation of a simpler target, namely the BSL posterior in %
\eqref{bsl_likelihood}. By focusing on the (simpler) approximate posterior,
rather than the exact posterior $p(\boldsymbol{\theta }|\mathbf{y})$, the
approach of \citeauthor{tran2017variational} can be recycled using any
unbiased estimator of the synthetic likelihood, $p_{a}(\eta (\mathbf{y})|%
\boldsymbol{\theta })$ --- which we recall is nothing but a Normal
likelihood with unknown mean and variance-covariance matrix --- of which
several closed-form examples exist. Moreover, since the approach of %
\citeauthor{ong2018variational} does not rely on the random variables $%
\mathbf{u}$ in order for its likelihood estimate to be unbiased, no tuning
of $h(\mathbf{y}|\boldsymbol{\theta },\mathbf{u})$ is required, and the
minimizers of $\text{KL}[q_{\lambda }(\boldsymbol{\theta },z)|p(\boldsymbol{%
\theta },z|\mathbf{y})]$ and $\mathcal{L}(\lambda )$ will always coincide.

While useful, it must be remembered that the approach of \cite%
{ong2018variational} targets only the \textit{partial} posterior $p(%
\boldsymbol{\theta }|\eta (\mathbf{y}))$. Furthermore, given the discussion
in Section \ref{bsl}, the approach is likely to perform poorly when the
summaries used to construct the unbiased estimator of the synthetic
likelihood $p_{a}(\eta (\mathbf{y})|\boldsymbol{\theta })$ are non-Gaussian.
Given that, by definition, the problem is a high-dimensional one, thereby
requiring a large collection of summaries, the Gaussian approximation for $%
\eta (\mathbf{y})$ may not be accurate.

Similar to the above, \cite{barthelme:chopin:2014} and \cite%
{barthelme2018divide} propose the use of variational methods to approximate
the ABC posterior. The approach of \citeauthor{barthelme:chopin:2014} is
based on `local' collections of summary statistics that are computed by
first partitioning the data into $b\leq n$ distinct `chunks', $\mathbf{y}%
_{1},\dots ,\mathbf{y}_{b}$, with possibly differing lengths and support,
and then computing the summaries $\eta (\mathbf{y}_{i})$ for each of the $b$
chunks. Using this collection of local summaries, the authors then seek to
compute an approximation to the following ABC posterior: 
\begin{equation}
p_{\varepsilon }(\boldsymbol{\theta }|\eta (\mathbf{y}))\propto p(%
\boldsymbol{\boldsymbol{\theta }})\prod_{i=1}^{b}\left\{ \int p(\mathbf{z}%
_{i}|\mathbf{y}_{1:i-1},\boldsymbol{\boldsymbol{\theta }})\mathbb{I}\left\{
\left\Vert \eta \left( \mathbf{z}_{i}\right) -\eta \left( \mathbf{y}%
_{i}\right) \right\Vert \leq \varepsilon \right\} d\mathbf{z}_{i}\right\} =p(%
\boldsymbol{\boldsymbol{\theta }})\prod_{i=1}^{b}\ell _{i}(\boldsymbol{%
\boldsymbol{\theta }}),  \label{abc_st}
\end{equation}%
which implicitly maintains that the `likelihood chunks', $\ell _{i}(%
\boldsymbol{\boldsymbol{\theta }})$, $i=1,2,...,b,$ are conditionally
independent.

The posterior in (\ref{abc_st}) is then approximated using expectation
propagation (EP) (see \citealp{bishop2006pattern}, Ch 10 for details). The
EP approximation seeks to find a tractable density $q_{\lambda }(\boldsymbol{%
\theta })\in \mathcal{Q}$ that is close to $p_{\varepsilon }(\boldsymbol{%
\theta }|\eta (\mathbf{y}))$ by minimizing $\text{KL}\left[ p_{\varepsilon }(%
\boldsymbol{\theta }|\eta (\mathbf{y}))|q_{\lambda }(\boldsymbol{\theta })%
\right] $. The reader will note that this minimization problem is actually
the reverse of the standard variational problem in \eqref{vb_1}, and is a
feasible variational problem because $p_{\varepsilon }(\boldsymbol{\theta }%
|\eta (\mathbf{y}))$ is accessible. Using a factorizable Gaussian
variational family with chunk-specific location and covariance parameters, $%
\mu _{i}$ and variance $\Sigma _{i}$, respectively, $i=1,\dots ,b$, i.e., $%
\mathcal{Q}:=\{\lambda =(\lambda _{1},\dots ,\lambda _{b})\in \Lambda
:q_{\lambda }(\boldsymbol{\theta }):=\prod_{i=1}^{b}q_{i,\lambda _{i}}(%
\boldsymbol{\theta })\}$, this minimization problem is solved iteratively by
minimizing the KL divergence between $\ell _{i}(\boldsymbol{\theta })$ and $%
q_{i,\lambda _{i}}(\boldsymbol{\theta })$ for $i=1,2,...,b$. A coordinate
ascent optimization approach allows the $i$-th variational component to be
updated by calculating (using Monte Carlo integration) the mean and variance
of $q_{i,\lambda _{i}}(\boldsymbol{\theta })$, based on data simulated from $%
\ell _{i}(\boldsymbol{\theta })$, conditional on $\boldsymbol{\theta }$
drawn from the variational approximation based on the remaining $j\neq i$
chunks.

By chunking data to create conditionally independent likelihood increments,
and by employing (conditionally independent) Gaussian approximations over
these chunks, EP-ABC creates a (sequentially updated) Gaussian
pseudo-posterior that serves as an approximation to the original ABC
posterior. Given that EP-ABC requires the posterior approximation to be
Gaussian (or more generally within the linear exponential family), the
resulting EP-ABC posterior may not be a reliable approximation to the ABC
posterior if the data has strong, or nonlinear, dependence, or (similar to
the problem identified for BSL) if (\ref{abc_st}) has non-Gaussian features,
such as thick tails, multimodality or boundary issues. Moreover, the need to
generate synthetic data sequentially according to different chunks\ of the
likelihood is unlikely to be feasible in models where there is strong or
even moderate serial dependence, and generation of new data requires
simulating the entire path history up to that point.

We complete this section on `hybrid methods' by noting that in state space
settings ABC principles have also been used to implement particle filtering,
in particular in cases where the measurement density has no closed form and,
hence, cannot be used to define the particle weights in the usual way. ABC
filtering can be used to estimate the likelihood function, as either a basis
for producing frequentist point estimates of $\boldsymbol{\theta }$\ (%
\citealp{jasra:etal:2012}; \citealp{calvet2015accurate}) or as an input into
a PMCMC scheme (\citealp{dean2014parameter}; \citealp{jasra2015approximate}%
). More generally, recent algorithms have combined the principles of ABC and
Gibbs sampling (\citealp{clarte2019component}; %
\citealp{rodrigues2019likelihood}), with the primary aim being to reduce the
impact of dimension on the accuracy of ABC. We reference also \cite%
{dehideniya2019synthetic} for the production of an approximate posterior via
the melding of BSL steps with a Laplace approximation.

\subsection{MCMC Algorithms Revisited\label{advances}}

Despite the rich pickings now on offer with all of the new (including
approximate) computational methods, it is \textit{far} from the case that
the stalwart of the late 20th century --- MCMC --- has run its race! Hence,
we complete this section with a brief overview of the key advances in MCMC
that have been made subsequent to the appearance of the first algorithms,
many of which have been motivated by the by the need to deal effectively
with large data sets and/or high-dimensional unknowns. Brevity is adopted,
not because this segment of the literature is not ripe with developments; in
fact, attempts to improve the performance of the original (Gibbs and MH)
schemes began early, have been continual, and engage a substantial number of
researchers. Rather, we choose to be brief simply because the fundamental
principles of the newer advances remain essentially faithful to the original
principles of MCMC, and those principles have already been covered herein.%
\footnote{%
We acknowledge here a slight inconsistency in our approach, by\ having
allocated (above) a full section to pseudo-marginal MCMC methods, methods
that also remain faithful to the fundamental principles of MCMC. However,
the goals of the pseudo-marginal literature are arguably broader than just
improving algorithmic performance, as we have touched on in Section \ref{ben}%
.}

Indeed, we begin with three reminders about MCMC algorithms:

\begin{enumerate}
\item \textit{First}, an \textit{MC-}MC algorithm is just that --- a \textit{%
Markov chain }Monte Carlo algorithm. As such, an MCMC scheme --- by design
--- produces a \textit{local }exploration of the target posterior, with the
location in the parameter space of any simulated draw being dependent\textit{%
\ }on the location of the previous draw, in a manner that reflects the
specific structure of the algorithm. Most notably, an MCMC algorithm with a
high degree of dependence will potentially be slow in exploring the high
mass region of the target posterior (or the `target set', in the language of %
\citealp{betancourt:2018}), with this problem usually being more severe the
larger is the dimension of the parameter space. Looked at through another
lens: for $M$ MCMC draws, the greater the degree of (typically positive)
dependence in those draws, the less efficient is the MCMC-based estimate of (%
\ref{gen_expect}), relative to an estimate based on $M$ $i.i.d.$ draws from
the target. This loss of efficiency is measured by the so-called \textit{%
inefficiency factor }(IF), defined ({in the case of scalar }$g(\boldsymbol{%
\theta })$) as the ratio of the \textit{MCMC standard error}, $\sigma _{MCMC}
$ with $\sigma _{MCMC}^{2}=\sqrt{\text{Var}(g(\boldsymbol{\theta }%
))[1+2\sum\nolimits_{l=1}^{\infty }\rho _{l}]/M}$ to the standard error
associated with $M$ $i.i.d.$ draws, $\sigma _{MC}^{2}$, with $\sigma
_{MC}^{2}$ as given in (\ref{MC_variance}), where $\rho _{l}$ is the lag-$l$
autocorrelation of the draws of $g(\boldsymbol{\theta })$ over the history
of the chain. This ratio, in turn, defines the \textit{effective sample size 
}of the MCMC algorithm, $ESS=M/[1+2\sum\nolimits_{l=1}^{\infty }\rho _{l}].$
Improving the efficiency of an MCMC algorithm, for any given value of $M$,
thus equates to increasing $ESS$ to its maximum possible value of $M$ by
reducing the dependence in the draws.

\item \textit{Second}: an MC-\textit{MC} algorithm is also a Markov chain%
\textit{\ Monte Carlo} algorithm. That is, under appropriate regularity it
produces a $\sqrt{M}$-consistent estimate of (\ref{gen_expect}), whatever
the degree of dependence in the chain, with the dependence affecting the
constant term implicit in the $O(M^{-1/2})$ rate of convergence, but not the
rate itself. Hence, in principle, any MCMC algorithm, no matter how
inherently inefficient, can produce an estimate of (\ref{gen_expect}) that
is arbitrarily accurate, simply through an increase in $M.$ However, an
increase in $M$ entails an increase in \textit{computational cost},
measured, say, by\textit{\ computing clock-time}. The extent of this
increase depends, in turn, on the (per-iteration) cost of generating a
(proposal/candidate) draw and, with an MH step, the cost of calculating the
acceptance probability. Both component costs will (for any algorithm)
clearly increase with the number of unknowns that need to be simulated, and
assessed, at each iteration. Either cost, or both, will also increase with
the sample size, given the need for pointwise evaluation of the likelihood
function across the elements of $\mathbf{y}$.

\item \textit{Third}: the very concept of efficiency is relevant only if the
Markov chain is (asymptotically in $M$) \textit{unbiased}, which depends
critically on draws being produced from the correct invariant distribution.
That is, the production of an accurate\textit{\ }MCMC-based estimate of (\ref%
{gen_expect}) depends, not just on reducing the degree of dependence in the
chain, or on increasing the number of draws, but on ensuring that the chain%
\textit{\ actually} explores the target set, and thereby avoids bias in the
estimation of (\ref{gen_expect}).\footnote{%
It is acknowledged in the literature that MCMC algorithms produce
potentially strong biases in their initial phase of `convergence' to the
typical set from an initial point in the parameter space. However, under
appropriate regularity, such biases are transient, and their impact on the
estimation of (\ref{gen_expect}) able to be eliminated by discarding a
sufficiently large number of `burn-in' or `warm-up' draws from the
computation. (See \citealp{robert:casella:2004}, and %
\citealp{gelman2011inference}, for textbook discussions of convergence,
including diagnostic methods.) Some of the more recent literature is
concerned with removing this transitory bias after a finite number of
iterations; e.g. \cite{jacob2019unbiased}. Other literature is concerned
with ensuring that an MCMC algorithm does not yield a bias that is \textit{%
non-transitory} due to the inability of the algorithm to effectively explore
the target set at all (within a meaningful time frame); see e.g. \cite%
{betancourt:2018}.}
\end{enumerate}

Hence, all advances in MCMC -\textit{\ }at their core --- aim to increase
the effectiveness\ with which an algorithm explores the high mass region of
the target posterior and, hence, the accuracy with which (\ref{gen_expect})
is estimated, by doing one (or more) of three things: reducing dependence in
the chain, reducing the computational cost per iteration of the chain (thus
enabling more draws to be produced), or eliminating bias. Focus is
increasingly directed towards developing algorithms that \textit{scale well}%
, in terms of the dimension of the data and/or the number of unknowns.

With our goal of brevity in mind, we simply list the main contenders here,
including certain key references or reviews, deflecting both to those
papers, and to the broad overviews of modern developments in MCMC in \cite%
{greenetal2015}, \cite{robert2018accelerating} and \cite{dunson2019hastings}
for all details. Of particular note is the recent survey on Bayesian methods
for `Big Data' in \cite{Jahan2020} (Section 5.1 being most pertinent), which
describes the precise manner in which certain of the methods cited below
(and others) tackle the problem of scale. We categorize the methods
according to whether improved performance is achieved (primarily): \textit{i)%
} via the exploitation of more geometric information about the target
posterior; \textit{ii)} by better choice of {proposals}; \textit{iii)} by
the use of parallel, batched, subsample, coupled or ensemble sampling
methods; or \textit{iv)} by {the explicit use of variance reduction methods.}

\begin{enumerate}
\item[\textit{i)}] Hamiltonian Monte Carlo (HMC) (\citealp{neal:2011}; %
\citealp{carpenter:etal:2017}; \citealp{betancourt:2018}); no U-turn
sampling (NUTS) (\citealp{hoffman2014no});\footnote{%
As described in \cite{neal:2011}, simulation methods based on Hamiltonian
dynamics can actually be viewed as having as long a history as MCMC itself.
The more modern manifestations of HMC, however, including NUTS, can be
viewed as Markov chain algorithms that simply explore the parameter space
more effectively than (say) a default random walk scheme. The probabilistic
programming platform Stan (\citealp{carpenter:etal:2017}) enables
implementation of NUTS, in addition to certain variants of VB.}
Metropolis-Adjusted Langevin algorithm (MALA) (%
\citealp{roberts1996exponential}; \citealp{roberts1998optimal}); stochastic
gradient MCMC \citep{nemeth2019stochastic}; piecewise deterministic Markov
processes (PDMP) (\citealp{bierkens2018piecewise}; %
\citealp{fearnhead:etal:2018}; \citealp{bierkens2019}).

\item[\textit{ii)}] Optimal scaling of random-walk MH %
\citep{roberts:gelman:gilks:1997}; adaptive sampling (%
\citealp{nott2005adaptive}; \citealp{roberts2009examples}; %
\citealp{rosenthal2011optimal}); {MCMC with ordered overrelaxation (%
\citealp{Neal1998}); simulated tempering, parallel tempering and tempered
transition methods} (\citealp{geyer91pt}; \citealp{marinarietparisi92}; %
\citealp{Neal1996};\textbf{\ }\citealp{Gramacy2010}; \citealp{geyer2011st}; %
\citealp{tawn2019weight}); delayed rejection sampling (%
\citealp{tierney:mira:1998}); delayed acceptance sampling (%
\citealp{christen:fox:2005}; \citealp{Golightly2015}; %
\citealp{wiqvist2018accelerating}; \citealp{banterle2019accelerating});
multiple try MCMC (\citealp{liuliang2000}; \citealp{bedard2012scaling}; %
\citealp{martino2018review}; \citealp{luo2019multiple}); taylored randomized
block MH (TaRB-MH) (\citealp{CHIB201019}); tempered Gibbs sampling (TGS) (%
\citealp{zanellaroberts2019}); quasi-stationary Monte Carlo and subsampling %
\citep{pollock:etal:2020}.

\item[\textit{iii)}] Parallelized MCMC (\citealp{jacob:robert:smith:2010}; %
\citealp{wang:dunson:2013}); subposterior (batched) methods (%
\citealp{neiswanger:wang:xing:2013}; \citealp{scott2016bayes}); subsampling
methods based on pseudo-marginal MCMC (\citealp{bardenet2017markov}; %
\citealp{quiroz2018speeding}; \citealp{quiroz2019speeding}); {perfect
sampling (\citealp{proppetwilson96}; \citealp{casella:lavine:2001}, %
\citealp{craiu2011perfection}; \citealp{huber2016perfect}); }unbiased MCMC
via coupling (\citealp{glynn:rhee:2014}; \citealp{Glynn:exact2016}; %
\citealp{middleton2018unbiased}; \citealp{jacob2019unbiased}); unbiased MCMC
for doubly-intractable problems using pseudo-marginal principles (%
\citealp{lyne2015}); ensemble MCMC (\citealp{iba0}; %
\citealp{cappe:guillin:marin:robert:2004}; \citealp{neal2011mcmc}).

\item[\textit{iv)}] Rao-Blackwellization (\citealp{casella:robert:1996}; %
\citealp{robert:casella:2004}; \citealp{douc:robert:2011}); {antithetic
variables (\citealp{Frigessi2000}; \citealp{craiu:meng:2005}); control
variates (\citealp{Dellaportas2012}; \citealp{baker2019}); }thinning (%
\citealp{owen2017statistically}).
\end{enumerate}

\section{The Role of Computation in Model Choice and Prediction\label{mp}}

Thus far, our focus has been on computing the posterior expectation in (\ref%
{gen_expect}) defined in the context of an assumed model. Other than the
brief reference made to the expectation that defines the marginal likelihood
in (\ref{gen_expect_prior}), to the lower bound on the marginal likelihood
yielded by the VB procedure, and to the deterministic approximation of it
produced by INLA, the issue of model uncertainty itself has not been
addressed; nor have the specific issues that arise when the expectation in (%
\ref{gen_expect}) defines the predictive distribution. We touch on these
topics below. In Section \ref{ml_sect} each model in the assumed model space
is treated separately, and the set of \textit{posterior model probabilities }%
so produced used to make decisions. In Section \ref{rj} model uncertainty is
directly incorporated via the principle of augmentation, with inference
about the model being a direct outcome. In Section \ref{pred} we look at
Bayesian prediction.

\subsection{Model Uncertainty and Marginal Likelihood Computation\label%
{ml_sect}}

We begin by adopting the simplest possible characterization of model
uncertainty, where the model space\textit{\ }is spanned by two models, $%
\mathcal{M}_{1}$ and $\mathcal{M}_{2}$, with prior probabilities, $p(%
\mathcal{M}_{1})$ and $p(\mathcal{M}_{2})$ respectively. A simple
application of the Bayesian calculus leads to the following expression for
the ratio of posterior model probabilities (or \textit{posterior odds ratio):%
}%
\begin{equation}
\frac{p(\mathcal{M}_{1}|\mathbf{y})}{p(\mathcal{M}_{2}|\mathbf{y})}=\frac{p(%
\mathcal{M}_{1})}{p(\mathcal{M}_{2})}\times \frac{p(\mathbf{y}|\mathcal{M}%
_{1})}{p(\mathbf{y}|\mathcal{M}_{2})},  \label{podds}
\end{equation}%
where 
\begin{equation}
p(\mathbf{y}|\mathcal{M}_{k})=\int_{{\Theta }_{k}}p(\mathbf{y|}\boldsymbol{%
\theta }_{k},\mathcal{M}_{k})p(\boldsymbol{\theta }_{k}|\mathcal{M}_{k})d%
\boldsymbol{\theta }_{k},  \label{ml}
\end{equation}%
$\boldsymbol{\theta }_{k}$ is the unknown parameter (vector) for model $%
\mathcal{M}_{k}$, $k=1,2$, and $p(\mathbf{y|}\boldsymbol{\theta }_{k},%
\mathcal{M}_{k})$ and $p(\boldsymbol{\theta }_{k}|\mathcal{M}_{k})$ define
respectively the likelihood and prior conditioned explicitly on $\mathcal{M}%
_{k}.$ The density in (\ref{ml}) defines, equivalently, the \textit{marginal
likelihood}, the \textit{marginal data density}, or the \textit{evidence} of
model $\mathcal{M}_{k}$, and the ratio of the two such densities on the
right-hand-side of (\ref{podds}) defines the \textit{Bayes factor}.

Computation of $p(\mathcal{M}_{1}|\mathbf{y})$ and $p(\mathcal{M}_{2}|%
\mathbf{y})$ proceeds via (\ref{podds}) allied with the restriction that $p(%
\mathcal{M}_{1}|\mathbf{y})+$ $p(\mathcal{M}_{2}|\mathbf{y})=1$ (with the
extension to $K>2$ models being obvious\textbf{)}. \textit{Model choice} can
be performed by invoking decision-theoretic arguments, and minimizing
expected posterior loss. This leads to $\mathcal{M}_{1}$ being chosen if $p(%
\mathcal{M}_{1}|\mathbf{y})/p(\mathcal{M}_{2}|\mathbf{y})$ exceeds the ratio
of losses of `Type 2' and `Type 1' errors. \textit{Model averaging }can also
be used, whereby the posterior expectation of a quantity of interest is
computed for both models, then averaged, using $p(\mathcal{M}_{1}|\mathbf{y}%
) $ and $p(\mathcal{M}_{2}|\mathbf{y})$ as the weights (an example of which
is given in Section \ref{pred}).\ Textbook illustrations of all such steps
can be found in \cite{zellner:1971}, \cite{koop2003bayesian} and \cite%
{robert:2007}.

Key to all of this is the evaluation of the two integrals in (\ref{ml}). As
noted in Section \ref{general}, analytical solutions to (\ref{ml}) are
available only for certain special cases. Whilst the VB approach to
computing $\mathbb{E}(g(\boldsymbol{\theta })|\boldsymbol{y})$ within the
context of a given model yields, as a bi-product, a lower bound on the
evidence for that model, we focus in this section on methods that target the
marginal likelihood \textit{directly}. We do not reproduce here the
INLA-based method for computing $p(\mathbf{y})$ that has been described in
Section \ref{inla}.

The integral in (\ref{ml}) is the prior\textit{\ }expectation in (\ref%
{gen_expect_prior}) defined for $\mathcal{M}_{k}$,\textbf{\ }$k=1,2.$\textbf{%
\ }Being a prior, rather than a posterior expectation has two consequences.
First, it is well-defined only if $p(\boldsymbol{\theta }_{k}|\mathcal{M}%
_{k})$ is a \textit{proper }density function. Hence, Bayes factors, and the
posterior odds ratios that they imply, cannot be computed with impunity
under (typically improper) non-informative, or objective, priors. Attempts
to incorporate objective prior information into Bayes factors have been made
({see \citealp{strahan:2014}, for a recent treatment and relevant referencing%
}); however the convention in the literature remains one of adopting
informative, proper priors in the computation of posterior model
probabilities. The \textit{second} consequence relates to computation: the
most direct approach to computing (\ref{ml}) via simulation, namely drawing $%
\boldsymbol{\theta }_{k}$ from $p(\boldsymbol{\theta }_{k})$, and averaging $%
p(\mathbf{y}|\boldsymbol{\theta }_{k})$ over the draws, is typically
inaccurate, as $p(\boldsymbol{\theta }_{k})$ will not necessarily have high
mass in the high-mass region of the likelihood function. All
simulation-based estimates of (\ref{ml}) thus use draws of $\boldsymbol{%
\theta }_{k}$ that are informed by the data in some way, to improve accuracy.

Alternative methods for simulation-based estimation of (\ref{ml}) have been
proposed, in addition to --- and sometimes combined with --- either partial
analytical solutions or asymptotic (Laplace) approximations. We refer the
reader to \cite{kass:raftery:1995}, \cite{geweke99}, \cite{marin:robert:2010}%
, \cite{chib2011introduction} and \cite{fourment201819} for reviews. We
simply emphasize below three distinct uses of simulation, all of which nest,
or can be linked to, a range of specific methods, not all of which we cover
here. (We refer the reader to Aridia \textit{et al}., 2012, for a useful
comparative study of some of the simulation methods we review below, plus
additional referencing.) Section headings that indicate the over-arching%
\textit{\ principle} underlying each approach are adopted.

\subsubsection{Importance sampling\label{IS_ML}}

As already noted, the original motivation of \cite{hammersley:handscomb:1964}
in using IS to compute integrals was that of variance reduction. The authors
highlight that a judicious choice of importance density can yield a more
accurate estimator of any integral than a\ `crude' Monte Carlo estimator, as
they term it, by targeting the parts of the support where the integrand is
`important'.

The problem of computing (\ref{ml}) can be approached in the same way.
Defining $w^{(i)}=p(\mathbf{y}|\boldsymbol{\theta }_{k}^{(i)})p(\boldsymbol{%
\theta }_{k}^{(i)})/$\newline
$q(\boldsymbol{\theta }_{k}^{(i)}|\mathbf{y}),$ $i=1,2,...,M,$ for draws, $%
\boldsymbol{\theta }_{k}^{(i)}$, from some suitable importance density $q(%
\boldsymbol{\cdot |}\mathbf{y})$, an IS approach produces $\widehat{p}(%
\mathbf{y}|\mathcal{M}_{k})=\sum\limits_{i=1}^{M}w^{(i)}/M$. Subject to
regularity on $q(\boldsymbol{\cdot }|\mathbf{y})$, the usual asymptotic (in $%
M$) arguments can be invoked to prove the consistency and asymptotic
normality of $\widehat{p}(\mathbf{y}|\mathcal{M}_{k})$ as an estimator of (%
\ref{ml}) (see \citealp{geweke99}). Note that, on the assumption that $p(%
\mathbf{y|}\boldsymbol{\theta }_{k}^{(i)},\mathcal{M}_{k}) $ and $p(%
\boldsymbol{\theta }_{k}^{(i)})$ are available in closed form, and that $q(%
\boldsymbol{\cdot |}\mathbf{y})$ is known in in its entirety (i.e. including
its integrating constant), no additional normalization step (like that in (%
\ref{is_est_2})) is required. See \cite{geweke1989exact}, \cite%
{gelfand:dey:1994} and \cite{raftery:1996} for early examples of this
approach, and \cite{geweke99} for illustration of a non-$i.i.d.$ version,
based on draws from an MH candidate distribution, $q(\boldsymbol{\cdot }|%
\mathbf{y})$.

The general principle of IS has, of course, two aspects to it: \textit{i)}
the use of draws simulated from the importance density to compute a weighted
mean; and \textit{ii)} evaluation of the importance density in the weight.
The `reciprocal IS' (RIS) method (\citealp{gelfand:dey:1994}; %
\citealp{fruhwirth:2004}) uses the second aspect, whilst taking draws from
the posterior itself. Simple calculations can be used to show that, for some 
$q(\boldsymbol{\cdot }|\mathbf{y})$ that is contained in the support of $p(%
\boldsymbol{\theta }_{k}\mathbf{|y})$, and defining $g(\boldsymbol{\theta }%
_{k}\mathbf{)}=q(\boldsymbol{\theta }_{k}\boldsymbol{|}\mathbf{y})/\left[ p(%
\mathbf{y}|\boldsymbol{\theta }_{k})p(\boldsymbol{\theta }_{k})\right] $, 
\begin{equation}
\mathbb{E}(g(\boldsymbol{\theta }_{k}\mathbf{)|y)}=\left[ p(\mathbf{y}|%
\mathcal{M}_{k})\right] ^{-1}.  \label{IS-rec}
\end{equation}%
Again, under stringent regularity conditions, including some on the support
of $q(\boldsymbol{\theta }_{k}\boldsymbol{|}\mathbf{y})$, draws from $p(%
\boldsymbol{\theta }_{k}\mathbf{|y})$, can be used to estimate (\ref{IS-rec}%
), and its reciprocal used as an estimate of (\ref{ml}) itself. The
`harmonic mean estimator' of \cite{newton:raftery:1994} is a special, if
infamous, case of (\ref{IS-rec}) in which $q(\boldsymbol{\theta }_{k}%
\boldsymbol{|}\mathbf{y})=p(\boldsymbol{\theta }_{k})$. The complete
untrustworthiness of the output when the prior has fatter tails than the
posterior \citep{neal:1994,neal:1999} have led to this method being largely
eschewed in the literature. The `bridge sampler', on the other hand,
provides a more robust version of (\ref{IS-rec}) by exploiting draws from 
\textit{both }the posterior and the weight density, $q(\boldsymbol{\theta }%
_{k}\boldsymbol{|}\mathbf{y})$ (\citealp{meng:wong:1996}; %
\citealp{meng:schilling:2002}; \citealp{fruhwirth:2004}; %
\citealp{gronau2017tutorial}). More recent versions of the RIS method retain
draws from the posterior, but use VB approximations of $p(\boldsymbol{\theta 
}_{k}\mathbf{|y})$ to define $q(\boldsymbol{\theta }_{k}\boldsymbol{|}%
\mathbf{y})$ (\citealp{fourment201819}; \citealp{hajargasht2018accurate}).

{Finally, building on the thermodynamic integration method from theoretical
physics, \cite{gelman:meng:1998} develop the `path sampler'. They illustrate
the sense in which} {the earlier IS schemes for estimating }(\ref{ml}), {%
followed by bridge sampling, and then path sampling, represent a natural
methodological lineage. See also \cite{rischard2018unbiased} for a recent
amalgam of path sampling with unbiased MCMC, and \cite{neal2001ais} for
related work using `annealed importance sampling' to estimate (\ref{ml}).}%
\footnote{%
On a related thread, note that SMC has often been advocated for estimating
the evidence %
\citep{doucet:godsill:andrieu:2000,fiel:wyse:2012,everitt:etal:2020}.
Similarly, techniques incorporating the intractable marginal likelihood as a
supplementary parameter can be traced back to \cite{geyer:1993}, with more
recent occurrences like noise-contrastive estimation %
\citep{gutmann:hyvaarinen:2012}, being based on a so-called `logistic trick'
that turns the approximation of the evidence into the estimation of the
intercept of a logistic classification program. See also \cite%
{barthelme:chopin:2015} and \cite{lyne2015} for related work.}

\subsubsection{Multiple runs of MCMC}

Given the definition of $p(\boldsymbol{\theta }_{k}\mathbf{|y})$, we can
produce a representation of the marginal likelihood as: 
\begin{equation}
p(\mathbf{y}|\mathcal{M}_{k})=\frac{p(\mathbf{y}|\boldsymbol{\theta }_{k})p(%
\boldsymbol{\theta }_{k})}{p(\boldsymbol{\theta }_{k}\mathbf{|y})}.
\label{chib_1}
\end{equation}%
The insight of Chib (1995) was to recognize that (\ref{chib_1})\ holds for
all $\boldsymbol{\theta }$. Hence, an estimate of $p(\mathbf{y}|\mathcal{M}%
_{k})$ is simply produced as: $\widehat{p}(\mathbf{y}|\mathcal{M}_{k})=p(%
\mathbf{y}|\boldsymbol{\theta }_{k}^{\ast})p(\boldsymbol{\theta }_{k}^{\ast
})/p(\boldsymbol{\theta }_{k}^{\ast }\mathbf{|y})$, where the convention is
to take $\boldsymbol{\theta }_{k}^{\ast }$ as some high posterior value.
Whilst the (common) availability of the likelihood and prior in closed form
renders the ordinates of the factors in the numerator readily accessible in
most cases, the denominator and, indeed, the value of $\boldsymbol{\theta }%
_{k}^{\ast }$ itself are, by the very nature of the problem, inaccessible
without further work. However, defining $\boldsymbol{\theta }_{k}^{\ast
}=(\theta _{k,1}^{\ast },\theta _{k,2}^{\ast },...,\theta _{k,p_{k}}^{\ast
})^{\prime }$, the joint posterior (evaluated at $\boldsymbol{\theta }%
_{k}^{\ast }$) can be decomposed as:%
\begin{equation}
p(\boldsymbol{\theta }_{k}^{\ast }\mathbf{|y})=p(\theta _{k,1}^{\ast
}|\theta _{k,2}^{\ast },...,\theta _{k,p_{k}}^{\ast },\mathbf{y})p(\theta
_{k,2}^{\ast }|\theta _{k,3}^{\ast },...,\theta _{k,p_{k}}^{\ast },\mathbf{y}%
)...p(\theta _{k,p_{k}}^{\ast }|\mathbf{y}).  \label{decom}
\end{equation}%
The last term on the right-hand-side of (\ref{decom}) can simply be
estimated in the usual way using a full run of an MCMC sampler. The
remaining conditionals can be estimated from additional applications of the
same simulation scheme, but with the appropriate sets of parameters held
fixed. Modification of the original (pure Gibbs) approach proposed in \cite%
{chib:1995} to cater for full conditionals that are not available in closed
form, by using MH steps, is detailed in \cite{chib:jeliazkov:2001}. Recent
work in \cite{chib_shin_tan_2020} highlights the effective use of
parallelization to reduce the computational cost of the additional `reduced'
MCMC runs.

\subsubsection{Nested sampling}

Nested sampling \citep{skilling:2007} is yet another method for producing a
simulation-based estimate of $p(\mathbf{y}|\mathcal{M}_{k}).$ Whilst it
gained an immediate foothold in astronomy \citep{mukherjee2006nsa} ---
possibly due to the availability of dedicated software like MultiNest and
Dynesty --- and remains popular in that field (\citealp{Feroz_2019}),\textbf{%
\ }it has not gained wide acceptance beyond that field. A cartoon
description of the method is as the simulation version of Lebesgue
integration, in that $M$ points are simulated on slices of the likelihood
function (delineated by points $t,$ $t-1$ in the support of $\boldsymbol{%
\theta }$), $\left\{ \boldsymbol{\theta };\ p(\mathbf{y}|\boldsymbol{\theta }%
_{t-1})\leq p(\mathbf{y}|\boldsymbol{\theta })\leq p(\mathbf{y}|\boldsymbol{%
\theta }_{t})\right\} ,$ with each slice having approximately a prior
probability, $\exp \{-(t-1)/M\}$, of occurring. As first shown in \cite%
{chopin:robert:2010}, nested sampling is a Monte Carlo method with a $\sqrt{M%
}$\ rate of convergence, whose performance relies on the ability to
efficiently simulate parameters within the above slices, which is
challenging when these are not connected.

\subsection{Reversible Jump MCMC\label{rj}}

All methods described above for computing $p(\mathbf{y}|\mathcal{M}_{k})$
have one thing in common: the marginal likelihood for \textit{each }model is
tackled as a separate computational exercise. Once each $p(\mathbf{y}|%
\mathcal{M}_{k})$ is computed, the posterior model probabilities follow, and
model choice, or model averaging, can proceed.

Alternatively, uncertainty about the model can be used to \textit{augment}
the unknowns, and a posterior sampler designed to target this augmented
space. This is the principle adopted in a range of papers, including those
that focus on variable selection in regression models, and the number of
components in finite mixture models; and we refer the reader to \cite%
{george:2000}, \cite{marin:mengersen:robert:2005} and \cite%
{chib2011introduction} for reviews and references. We focus here on one
particular approach only: that of \cite{green:1995}.

\cite{green:1995} characterizes the problem of an unknown model as one in
which the dimension of the (model-specific) unknowns \textit{varies},
depending on which model is in play. He thus designs an MCMC sampler that is
allowed to \textit{jump }between parameter spaces of differing dimensions;
coining the term: `reversible jump MCMC' (RJMCMC). At its core though,
Green's approach is one of augmentation, and can be viewed as a particular
application of the original idea of \cite{tanner87}, with the extra
complexity of dimension variation as the sampler traverses the augmented
space.

In brief, \cite{green:1995} assumes a countable collection of candidate
models $\mathcal{M}=\{\mathcal{M}_{1},\mathcal{M}_{2},...\}$, indexed by $%
k=1,2,.....$ Each model has a $p_{k}$-dimensional set of unknown parameters $%
\boldsymbol{\theta }_{k}$. Using obvious notation for the likelihood and
prior for the $kth$ model, and the prior $p(k)$ for the model index itself,
the target of the RJMCMC algorithm is then: $p(k,\boldsymbol{\theta }_{k}|%
\mathbf{y})=p(\mathbf{y}|k,\boldsymbol{\theta }_{k})p(\boldsymbol{\theta }%
_{k}|k)p(k)/p(\mathbf{y})$. The RJMCMC sampler moves between any two
parameters spaces by creating temporary auxiliary variables that bring the
dimensions of the augmented spaces to be equal, with a reversibility
constraint on the proposed moves between these models. Such draws from the
joint space of $\{k,\boldsymbol{\theta }_{k}\}$ can be used to compute any
particular $\mathbb{E}(g(\boldsymbol{\theta }_{k}\mathbf{)|y)}$ of interest.
Indeed, the draws can also be used to compute an expectation of the form: $%
\mathbb{E}(g(k)\mathbf{|y)}$, which nests the marginal posterior probability
attached to the $kth$ model: $p(k\mathbf{|y)}$. That is, posterior model
probabilities are an automatic outcome of the simulation scheme. Moreover,
the computation of \textit{any }expectation of interest incorporates all
uncertainty associated with both the parameters of each model and the model
itself; hence model averaging\textit{\ }is automatic. See \cite{green03}, 
\cite{fan2011reversible} and \cite{geyer2011introduction} for reviews of
RJMCMC, including: its theoretical and implementation properties, its links
to other `multi-model' samplers, and the scope of its application.

\subsection{Computation in Bayesian Prediction\label{pred}}

Conditional on the assumed model that underpins the likelihood function
being `correctly specified' --- i.e. $p(\mathbf{y}|\boldsymbol{\theta })$
coinciding with the true data generating process (DGP) --- the `gold
standard' for Bayesian prediction is the specification of (\ref{gen_expect})
with $g(\boldsymbol{\theta }\mathbf{)}=$ $p(y_{n+1}^{\ast }|\boldsymbol{%
\theta }\mathbf{,y})$, which yields:%
\begin{equation}
p(y_{n+1}^{\ast }|\mathbf{y})=\int_{{\Theta }}p(y_{n+1}^{\ast }|\boldsymbol{%
\theta }\mathbf{,y})p(\boldsymbol{\theta }\mathbf{|y})d\boldsymbol{\theta }.
\label{exact_predict}
\end{equation}%
The distribution $p(y_{n+1}^{\ast }|\mathbf{y})$ summarizes all uncertainty
about $y_{n+1}$, conditional on both the assumed model --- which underpins
the structure of both the conditional predictive, $p(y_{n+1}^{\ast }|%
\boldsymbol{\theta }\mathbf{,y})$, and the posterior itself --- and the
prior beliefs that inform $p(\boldsymbol{\theta }\mathbf{|y}).$ Point and
interval predictions of $y_{n+1}$, and indeed any other distributional
summary, can be extracted from (\ref{exact_predict}). In the case where the
model itself is uncertain, and a finite set of models, $\mathcal{M}_{1}$, $%
\mathcal{M}_{2}$,...,$\mathcal{M}_{K},$ is assumed to span the model space,
the principle of model averaging\textit{\ }can be used to produce a
`model-averaged' predictive, $p_{MA}(y_{n+1}^{\ast }|\mathbf{y})$ as 
\begin{equation}
p_{MA}(y_{n+1}^{\ast }|\mathbf{y})=\sum\limits_{k=1}^{K}p(y_{n+1}^{\ast }|%
\mathbf{y},\mathcal{M}_{k})p(\mathcal{M}_{k}|\mathbf{y})\text{,}  \label{ma}
\end{equation}%
where $p(y_{n+1}^{\ast }|\mathbf{y},\mathcal{M}_{k})$ denotes the density in
(\ref{exact_predict}), but conditioned explicitly on the $kth$ model in the
set. In the typical case where (\ref{exact_predict}) and (\ref{ma}) are
unavailable analytically, \textit{any }of the computational methods that
have been discussed thus far could be used to compute either $%
p(y_{n+1}^{\ast }|\mathbf{y})$ or $p_{MA}(y_{n+1}^{\ast }|\mathbf{y})$, and
any summaries thereof. In some ways then, this could be viewed as completing
the Bayesian prediction story.

However, Bayesian prediction is more than just a special case of the general
Bayesian computational problem defined by (\ref{gen_expect}). Predicting
outcomes that have not yet been observed is arguably the most stringent test
to which any statistical methodology can be put. Moreover, out-of-sample
accuracy is the ultimate arbiter in this setting, and any computational
method used to produce predictions needs to be assessed in this light.

Such is the motivation of recent works that pose the following question: In
cases (such as those highlighted in Section \ref{approx}) where the exact
posterior is inaccessible and the exact predictive in (\ref{exact_predict})
is thus also unavailable, what are the implications for predictive accuracy
of adopting an approximation to\textbf{\ }$p(\boldsymbol{\theta }\mathbf{|y})
$ and, hence, $p(y_{n+1}^{\ast }|\mathbf{y})$? \cite{frazier2019approximate}%
, for instance, document that an `approximate predictive' produced by
replacing $p(\boldsymbol{\theta }\mathbf{|y})$ with an ABC-based posterior,
is numerically indistinguishable (in the cases investigated) from the exact
predictive and, thus, yields equivalent predictive accuracy. Further, under
certain regularity conditions, the exact and approximate predictives are
shown to be asymptotically (in $n$) equivalent. The tenor of related work
exploring prediction in approximate inference (including VB) settings --- 
\cite{park2014variational}, \cite{canale2016}, \cite{koop2018variational}, 
\cite{quiroz2018gaussian} and \cite{konkamking2019} --- is somewhat similar
to that of \citeauthor{frazier2019approximate}; that is, computing $p(%
\boldsymbol{\theta }\mathbf{|y})$ via an approximate method does not \textit{%
necessarily} reduce predictive accuracy.

However, an arguably more challenging question is: How does one even think
about Bayesian prediction --- and the use of computation therein --- once
one acknowledges that, in reality: \textit{i)} any given model is
misspecified; and \textit{ii)} any finite set of models does not span the
truth? The `probably approximately correct (PAC)-Bayes' approach\ in machine
learning (reviewed in \citealp{guedj2019}) replaces the likelihood function
in the Bayesian update with the exponential of a general loss function, with
a view to producing predictions that are more targeted to the problem at
hand, rather than being tied to a particular model specification. This
approach yields so-called `Gibbs posteriors' (\citealp{Zhang2006a}; %
\citealp{Zhang2006b}; \citealp{jiang2008}), and mimics --- in a prediction
setting --- the generalized \textit{inferential} methods proposed in (%
\textit{inter alia}) \cite{bissiri:etal:2016}, \cite{giummole2017objective}, 
\cite{holmes2017assigning}, \cite{lyddon2019general} and \cite%
{syring2019calibrating}. In addition, PAC-Bayes uses\ updates based on
`tempered', or `power' likelihoods, in which robustness to model
misspecification is sought by raising the likelihood associated with an
assumed model to a particular power.

Other work tackles{\ this issue by estimating weighted combinations of
predictives via either forecast accuracy criteria or the criterion of
predictive calibration (\citealp{Dawid1982}; %
\citealp{gneiting2007probabilistic}), without assuming that the true model
lies within the set of constituent predictives ({\citealp{Billio2013}; %
\citealp{casarin2015}; \citealp{Pett2016}; \citealp{Bassetti2018}; %
\citealp{BASTURK2019})}. In a similar spirit, \cite{loaiza2019focused}}
propose the use of \textit{focused Bayesian prediction}, in which the
Bayesian update is driven by a criterion that captures a user-specified
measure of predictive accuracy. The authors do indeed find that focusing on
the loss that matters produces superior predictive performance relative to
using a misspecified likelihood update.

In summary, once attention is on the particular version of (\ref{gen_expect}%
) that yields a predictive distribution,\textbf{\ }it is the updating
criterion function itself that is increasingly being viewed as key, with
uncertainty about the unknown parameters --- and the computational method
used to quantify that uncertainty --- assuming a somewhat secondary role.

\section{The Future\label{future}}

Our journey with Bayesian computation began in \textit{1763}: with a
posterior probability defined in terms of a scalar $\theta $, {whose
solution challenged} Bayes. {We now end our journey in \textit{2020}: having
referenced posterior distributions defined over thousands, possibly millions
of unknowns, and computational problems with a degree of complexity --- and
scale --- to match. Along the way, we have seen the huge variety of
imaginative computational solutions that have been brought to bear on all
such problems, over the span of 250 years. A natural question to ask then
is: `what is there left to do?' }

Judging from the wealth of contributions to `\textit{Bayes Comp 2020}', in
January 2020 (http://users.stat.\newline
ufl.edu/\symbol{126}jhobert/BayesComp2020/Conf\_Website/), the answer to
that question is: `a great deal!'; and \textit{most certainly} as pertains
to matters of scale. Indeed, of the 129 abstracts recorded on the conference
website ({and acknowledging some double counting}), 16 make explicit use of
the term \textit{scalability }(or something comparable); 10 refer to the
ability of a proposed computational method to deal effectively with \textit{%
large data sets}, and 22 refer to \textit{high-dimensional problems} of one
form or another. There are attempts to scale\textit{\ }most categories of
computational methods reviewed in this paper, and the scope of the empirical
problems to which these advances are applied --- from probabilistic topic
models, health studies on hypertension and sepsis, problems in neuroimaging,
genomics, biology, epidemiology, ecology, psychology and econometrics,
through to probabilistic record linkage and geographic profiling --- is
extremely broad, highlighting again the critical importance of Bayesian
computation to diverse fields.

But if \textit{scale }may be viewed as a key focus of this selection of
research, there is another theme that can also be discerned. A second glance
at these conference proceedings, and at recent journal publications,
pinpoints a growing interest in the impact of \textit{model} \textit{%
misspecification }on computational methods; specifically: \textit{1)} what
are the implications for computation if an assumed parametric model is
misspecified?; and, \textit{2)} what are the implications for computation if
the conventional likelihood-based paradigm is eschewed altogether (as is
already happening in some prediction settings)? Remembering that our
interest here is on the implications for \textit{computation per se }of both
misspecified and non-likelihood settings, we note that specific attempts to
address \textit{1) }and \textit{2)} directly are still quite small in
number, if growing.\footnote{%
We refer to \cite{kleijn2012} and \cite{Muller2013} for general treatments
of Bayesian likelihood-based inference in misspecified models; to some of
the literature cited in Section \ref{pred}, plus \cite{CH03},\cite%
{gallant2016reflections}, \cite{chib2018bayesian}, and \cite%
{miller2019robust}, for various generalizations of the standard Bayesian
paradigm, including posterior up-dates driven by problem-specific loss (or
moment) functions. It can be argued, however, that in none of these papers
are the implications of model misspecification and/or non-likelihood
up-dates for computation \textit{per se }the primary focus.} We complete our
review by briefly summarizing \textit{five} recent (sets of) papers in this
vein:

\begin{enumerate}
\item[\textit{i)}] \textit{First}, \cite{lyddon2019general} and \cite%
{syring2019calibrating} use bootstrap principles to compute so-called 
\textit{general Bayesian posteriors}, in which the likelihood function
associated with an assumed (and potentially misspecified) model is replaced
by a more general loss function that is not tied to a particular model
specification. \cite{huggins2019using} also use the bootstrap to construct
`bagged' posteriors (\textit{BayesBag}), and thereby conduct Bayesian
inference that is robust to model misspecification.

\item[\textit{ii)}] \textit{Second}, \cite{frazier2020model} analyze the
theoretical properties of ABC under model misspecification; outlining when
ABC concentrates posterior mass on an appropriately defined pseudo-true
value, and when it does not. The nonstandard asymptotic behaviour of the ABC
posterior, including its failure to yield credible sets with valid
frequentist coverage, is highlighted. The authors also devise techniques for
diagnosing model misspecification in the context of ABC. \cite%
{frazier2019robust} devise a version of BSL that is robust to model
misspecification, and demonstrate that this version can be much more
computationally efficient than standard BSL when the model is misspecified.

\item[\textit{iii)}] \textit{Third,} \cite{wangblei2019b} investigate VB
under model misspecification. They demonstrate that the VB posterior both
concentrates on the value that minimizes the Kullback-Leibler (KL)
divergence from the true DGP, and is asymptotically normal; as is the VB
posterior mean. These results generalize the asymptotic results for VB of 
\cite{wangblei2019a}, derived under correct specification, to the
misspecification case.

\item[\textit{iv)}] \textit{Fourth}, \cite{knoblauch2019generalized} propose
what they term \textit{generalized variational inference, }by extending the
specification of the Bayesian paradigm to accommodate general loss functions
(thereby avoiding the reliance on potentially misspecified likelihoods) and
building an optimization-based computational tool within that setting.

\item[\textit{v)}] \textit{Fifth}, building on earlier work in the context
of HMC (and which is cited therein), \cite{bornn2019moment} derive an MCMC
scheme for sampling on the lower-dimensional manifold implied by the moment
conditions that they adopt within a Bayesian framework. Whilst this work
embeds the moments within a nonparametric Bayesian set-up --- and we have
not discussed computation in nonparametric settings in this review --- we
make note of this work as an example of a fundamental shift in computational
design that is required when moving to a particular \textit{non-likelihood}
Bayesian up-date. Whilst not motivated by this same problem, the work on
extending VB to manifolds by \cite{tran2019variational} is also relevant
here.
\end{enumerate}

Bayesian computation in parametric models is thus beginning to confront ---
and adapt to --- the reality of misspecified DGPs, and the generalizations
beyond the standard likelihood-based up-date that are evolving. Allied with
the growing ability of computational methods to also deal with the scale%
\textit{\ }of modern problems, the future of the paradigm {in the 21st
century }thus seems assured. {And with this, the 18th century Bayes (and his
loyal champion, Price) would no doubt be duly impressed!}

{\footnotesize \ \baselineskip8pt \setlength{\bibsep}{8pt} 
\bibliographystyle{apalikeit}
\bibliography{Bayes_comp}

\begin{thebibliography}{}

\bibitem[Albert and Chib, 1993]{albert:chib:1993b}
Albert, J. and Chib, S. (1993).
\newblock {B}ayesian analysis of binary and polychotomous response data.
\newblock {\em J. American Statist. Assoc.}, 88:669--679.

\bibitem[Andrieu \emph{et~al.}, 2011]{andrieu:doucet:holenstein:2010}
Andrieu, C., Doucet, A., and Holenstein, R. (2011).
\newblock Particle {M}arkov chain {M}onte {C}arlo.
\newblock {\em J. Royal Statist. Society Series B}, 72(2):269--342.
\newblock With discussion.

\bibitem[Andrieu \emph{et~al.}, 2004]{andrieu2004computational}
Andrieu, C., Doucet, A., and Robert, C. (2004).
\newblock Computational advances for and from {B}ayesian analysis.
\newblock {\em Statist. Science}, 19(1):118--127.

\bibitem[Andrieu and Roberts, 2009]{andrieu:roberts:2009}
Andrieu, C. and Roberts, G. (2009).
\newblock The pseudo-marginal approach for efficient {M}onte {C}arlo
  computations.
\newblock {\em Ann. Statist.}, 37(2):697--725.

\bibitem[Baker \emph{et~al.}, 2019]{baker2019}
Baker, J., Fearnhead, P., Fox, E., and Nemeth, C. (2019).
\newblock Control variates for stochastic gradient mcmc.
\newblock {\em Statist. Comp.}, 29:599–615.

\bibitem[Banterle \emph{et~al.}, 2019]{banterle2019accelerating}
Banterle, M., Grazian, C., Lee, A., and Robert, C.~P. (2019).
\newblock Accelerating {M}etropolis-{H}astings algorithms by delayed
  acceptance.
\newblock {\em Foundations of Data Science}, 1(2):103--128.

\bibitem[Bardenet \emph{et~al.}, 2017]{bardenet2017markov}
Bardenet, R., Doucet, A., and Holmes, C. (2017).
\newblock On {M}arkov chain {M}onte {C}arlo methods for tall data.
\newblock {\em J. Machine Learning Res.}, 18(1):1515--1557.

\bibitem[Barnard and Bayes, 1958]{Bayes:Biometrika1958}
Barnard, G. and Bayes, T. (1958).
\newblock Studies in the history of probability and statistics: {IX}. {T}homas
  {B}ayes's essay towards solving a problem in the doctrine of chances.
\newblock {\em Biometrika}, 45(3/4):293--315.

\bibitem[Barthelm{\'e} \emph{et~al.}, 2018]{barthelme2018divide}
Barthelm{\'e}, S., Chopin, N., and Cottet, V. (2018).
\newblock Divide and conquer in {ABC}: Expectation-propagation algorithms for
  likelihood-free inference.
\newblock {\em Handbook of Approximate {B}ayesian Computation}, pages 415--34.
\newblock Chapman \& Hall/CRC. Eds. Sisson, S., Fan, Y., Beaumont, M.

\bibitem[Barthelmé and Chopin, 2014]{barthelme:chopin:2014}
Barthelmé, S. and Chopin, N. (2014).
\newblock Expectation propagation for likelihood-free inference.
\newblock {\em J. American Statist. Assoc.}, 109(505):315--333.

\bibitem[Barthelmé and Chopin, 2015]{barthelme:chopin:2015}
Barthelmé, S. and Chopin, N. (2015).
\newblock The {P}oisson transform for unnormalised statistical models.
\newblock {\em Stat. Comput.}, 25:767--780.

\bibitem[Bassetti \emph{et~al.}, 2018]{Bassetti2018}
Bassetti, F., Casarin, R., and Ravazzolo, F. (2018).
\newblock {B}ayesian nonparametric calibration and combination of predictive
  distributions.
\newblock {\em J. American Statist. Assoc.}, 113(522):675--685.

\bibitem[Bauwens and Richard, 1985]{bauwens:1985}
Bauwens, L. and Richard, J. (1985).
\newblock A 1-1 {P}oly-$t$ random variable generator with application to
  {M}onte {C}arlo integration.
\newblock {\em J. Econometrics}, 29(1):19--46.

\bibitem[Bayes, 1764]{bayes:1764}
Bayes, T. (1764).
\newblock An essay towards solving a problem in the doctrine of chances.
\newblock {\em Philosophical Transactions of the {R}oyal Society of {L}ondon
  for 1973}, 53:370--418.

\bibitem[Baştürk \emph{et~al.}, 2019]{BASTURK2019}
Baştürk, N., Borowska, A., Grassi, S., Hoogerheide, L., and van Dijk, H.
  (2019).
\newblock Forecast density combinations of dynamic models and data driven
  portfolio strategies.
\newblock {\em Journal of Econometrics}, 210(1):170--186.

\bibitem[Beaumont, 2003]{beaumont:2003}
Beaumont, M. (2003).
\newblock Estimation of population growth or decline in genetically monitored
  populations.
\newblock {\em Genetics}, 164:1139--1160.

\bibitem[Beaumont, 2010]{beaumont:2010}
Beaumont, M. (2010).
\newblock Approximate {B}ayesian computation in evolution and ecology.
\newblock {\em Annual Review of Ecology, Evolution, and Systematics},
  41:379--406.

\bibitem[Beaumont \emph{et~al.}, 2009]{beaumont:cornuet:marin:robert:2009}
Beaumont, M., Cornuet, J.-M., Marin, J.-M., and Robert, C. (2009).
\newblock Adaptive approximate {B}ayesian computation.
\newblock {\em Biometrika}, 96(4):983--990.

\bibitem[Beaumont \emph{et~al.}, 2002]{be02}
Beaumont, M., Zhang, W., and Balding, D. (2002).
\newblock Approximate {B}ayesian computation in population genetics.
\newblock {\em Genetics}, 162(4):2025--2035.

\bibitem[B{\'e}dard \emph{et~al.}, 2012]{bedard2012scaling}
B{\'e}dard, M., Douc, R., and Moulines, E. (2012).
\newblock Scaling analysis of multiple-try {MCMC} methods.
\newblock {\em Stochastic Processes and their Applications}, 122(3):758--786.

\bibitem[Berger, 1985]{berger:1985}
Berger, J. (1985).
\newblock {\em Statistical Decision Theory and {B}ayesian Analysis}.
\newblock Springer-Verlag, New York, second edition.

\bibitem[Bernton \emph{et~al.}, 2019]{Bernton2019}
Bernton, E., Jacob, P.~E., Gerber, M., and Robert, C.~P. (2019).
\newblock Approximate {B}ayesian computation with the wasserstein distance.
\newblock {\em J. Royal Statist. Society Series B}, 81(2):235--269.

\bibitem[Besag, 1974]{besag:1974}
Besag, J. (1974).
\newblock Spatial interaction and the statistical analysis of lattice systems.
\newblock {\em J. Royal Statist. Society Series B}, 36(2):192--326.
\newblock With discussion.

\bibitem[Besag and Green, 1993]{besag:green:1993}
Besag, J. and Green, P. (1993).
\newblock Spatial statistics and {B}ayesian computation.
\newblock {\em J. Royal Statist. Society Series B}, 55(1):25--37.
\newblock With discussion.

\bibitem[{Betancourt}, 2018]{betancourt:2018}
{Betancourt}, M. (2018).
\newblock A conceptual introduction to {H}amiltonian {M}onte {C}arlo.
\newblock {\em https://arxiv.org/abs/1701.02434v2}.

\bibitem[Bierkens \emph{et~al.}, 2018]{bierkens2018piecewise}
Bierkens, J., Bouchard-C{\^o}t{\'e}, A., Doucet, A., Duncan, A.~B., Fearnhead,
  P., Lienart, T., Roberts, G., and Vollmer, S.~J. (2018).
\newblock Piecewise deterministic {M}arkov processes for scalable {M}onte
  {C}arlo on restricted domains.
\newblock {\em Statist. Prob. Letters}, 136:148--154.

\bibitem[Bierkens \emph{et~al.}, 2019]{bierkens2019}
Bierkens, J., Fearnhead, P., and Roberts, G. (2019).
\newblock The zig-zag process and super-efficient sampling for {B}ayesian
  analysis of big data.
\newblock {\em Ann. Statist.}, 47(3):1288--1320.

\bibitem[Billio \emph{et~al.}, 2013]{Billio2013}
Billio, M., Casarin, R., Ravazzolo, F., and van Dijk, H. (2013).
\newblock Time-varying combinations of predictive densities using nonlinear
  filtering.
\newblock {\em Journal of Econometrics}, 177(2):213--232.

\bibitem[Bishop, 2006]{bishop2006pattern}
Bishop, C.~M. (2006).
\newblock {\em Pattern Recognition and Machine Learning}.
\newblock Springer, New York.

\bibitem[Bissiri \emph{et~al.}, 2016]{bissiri:etal:2016}
Bissiri, P.~G., Holmes, C.~C., and Walker, S.~G. (2016).
\newblock A general framework for updating belief distributions.
\newblock {\em J. Royal Statist. Society Series B}, 78(5):1103--1130.

\bibitem[Blei \emph{et~al.}, 2017]{blei2017variational}
Blei, D.~M., Kucukelbir, A., and McAuliffe, J.~D. (2017).
\newblock Variational inference: A review for statisticians.
\newblock {\em J. American Statist. Assoc.}, 112(518):859--877.

\bibitem[Blum, 2010]{blum:2010}
Blum, M. (2010).
\newblock {A}pproximate {B}ayesian computation: a non-parametric perspective.
\newblock {\em J. American Statist. Assoc.}, 105(491):1178--1187.

\bibitem[Blum \emph{et~al.}, 2013]{blum:etal:2013}
Blum, M. G.~B., Nunes, M.~A., Prangle, D., and Sisson, S.~A. (2013).
\newblock A comparative review of dimension reduction methods in approximate
  {B}ayesian computation.
\newblock {\em Statist. Science}, 28(2):189--208.

\bibitem[Bornn \emph{et~al.}, 2017]{bornn2017use}
Bornn, L., Pillai, N.~S., Smith, A., and Woodard, D. (2017).
\newblock The use of a single pseudo-sample in approximate {B}ayesian
  computation.
\newblock {\em Statist. Comp.}, 27(3):583--590.

\bibitem[Bornn \emph{et~al.}, 2019]{bornn2019moment}
Bornn, L., Shephard, N., and Solgi, R. (2019).
\newblock Moment conditions and {B}ayesian non-parametrics.
\newblock {\em J. Royal Statist. Society Series B}, 81(1):5--43.

\bibitem[Braun and McAuliffe, 2010]{braun2010variational}
Braun, M. and McAuliffe, J. (2010).
\newblock Variational inference for large-scale models of discrete choice.
\newblock {\em J. American Statist. Assoc.}, 105(489):324--335.

\bibitem[Brooks \emph{et~al.}, 2011]{brooks:etal:2011}
Brooks, S., Gelman, A., Jones, G., and Meng, X. (2011).
\newblock {\em Handbook of {M}arkov {C}hain {M}onte {C}arlo}.
\newblock Taylor \& Francis.

\bibitem[Buchholz and Chopin, 2019]{buchholz2019improving}
Buchholz, A. and Chopin, N. (2019).
\newblock Improving approximate {B}ayesian computation via quasi-{M}onte
  {C}arlo.
\newblock {\em J. Comput. Graph. Statist.}, 28(1):205--219.

\bibitem[Calvet and Czellar, 2015]{calvet2015accurate}
Calvet, L.~E. and Czellar, V. (2015).
\newblock Accurate methods for approximate {B}ayesian computation filtering.
\newblock {\em J. Finan. Econometrics}, 13(4):798--838.

\bibitem[Canale and Ruggiero, 2016]{canale2016}
Canale, A. and Ruggiero, M. (2016).
\newblock {B}ayesian nonparametric forecasting of monotonic functional time
  series.
\newblock {\em Electronic Journal of Statistics}, 10(2):3265--3286.

\bibitem[Capp\'e \emph{et~al.}, 2004]{cappe:guillin:marin:robert:2004}
Capp\'e, O., Guillin, A., Marin, J., and Robert, C. (2004).
\newblock Population {M}onte {C}arlo.
\newblock {\em J. Comput. Graph. Statist.}, 13(4):907--929.

\bibitem[Carpenter \emph{et~al.}, 2017]{carpenter:etal:2017}
Carpenter, B., Gelman, A., Hoffman, M., Lee, D., Goodrich, B., Betancourt, M.,
  Brubaker, M., Guo, J., Li, P., and Riddell, A. (2017).
\newblock Stan: A probabilistic programming language.
\newblock {\em Journal of Statistical Software, Articles}, 76(1).

\bibitem[Carter and Kohn, 1994]{carter:kohn:1994}
Carter, C.~K. and Kohn, R. (1994).
\newblock On {G}ibbs sampling for state space models.
\newblock {\em Biometrika}, 81(3):541--553.

\bibitem[Casarin \emph{et~al.}, 2015]{casarin2015}
Casarin, R., Leisen, F., Molina, G., and ter Horst, E. (2015).
\newblock A {B}ayesian beta {M}arkov random field calibration of the term
  structure of implied risk neutral densities.
\newblock {\em {B}ayesian Analysis}, 10(4):791--819.

\bibitem[Casella and George, 1992]{casella:george:1992}
Casella, G. and George, E. (1992).
\newblock An introduction to {G}ibbs sampling.
\newblock {\em American Statist.}, 46:167--174.

\bibitem[Casella \emph{et~al.}, 2001]{casella:lavine:2001}
Casella, G., Lavine, M., and Robert, C.~P. (2001).
\newblock Explaining the perfect sampler.
\newblock {\em American Statist.}, 55(4):299--305.

\bibitem[Casella and Robert, 1996]{casella:robert:1996}
Casella, G. and Robert, C. (1996).
\newblock {R}ao-{B}lackwellisation of sampling schemes.
\newblock {\em Biometrika}, 83(1):81--94.

\bibitem[Ceruzzi, 2003]{ceruzzi:2003}
Ceruzzi, P. (2003).
\newblock {\em A History of Modern Computing}.
\newblock MIT Press, second edition.

\bibitem[Chen \emph{et~al.}, 2011]{chen2011}
Chen, S., Dick, J., and Owen, A.~B. (2011).
\newblock Consistency of {M}arkov chain quasi-{M}onte {C}arlo on continuous
  state spaces.
\newblock {\em Ann. Statist.}, 39(2):673--701.

\bibitem[Chernozhukov and Hong, 2003]{CH03}
Chernozhukov, V. and Hong, H. (2003).
\newblock An {MCMC} approach to classical estimation.
\newblock {\em J. Econometrics}, 115(2):293--346.

\bibitem[Chib, 1995]{chib:1995}
Chib, S. (1995).
\newblock Marginal likelihood from the {G}ibbs output.
\newblock {\em J. American Statist. Assoc.}, 90(432):1313--1321.

\bibitem[Chib, 2011]{chib2011introduction}
Chib, S. (2011).
\newblock Introduction to simulation and {MCMC} methods.
\newblock {\em The Oxford Handbook of {B}ayesian Econometrics}, pages 183--217.
\newblock OUP. Eds. Geweke, J., Koop, G. and van Dijk, H.

\bibitem[Chib and Greenberg, 1995]{chibandgreenberg:1995}
Chib, S. and Greenberg, E. (1995).
\newblock Understanding the {M}etropolis--{H}astings algorithm.
\newblock {\em American Statist.}, 49:327--335.

\bibitem[Chib and Greenberg, 1996]{chib_greenberg_1996}
Chib, S. and Greenberg, E. (1996).
\newblock Markov chain {M}onte {C}arlo simulation methods in econometrics.
\newblock {\em Econometric Theory}, 12(3):409–431.

\bibitem[Chib and Jeliazkov, 2001]{chib:jeliazkov:2001}
Chib, S. and Jeliazkov, I. (2001).
\newblock Marginal likelihood from the {M}etropolis--{H}astings output.
\newblock {\em J. American Statist. Assoc.}, 96(453):270--281.

\bibitem[Chib and Ramamurthy, 2010]{CHIB201019}
Chib, S. and Ramamurthy, S. (2010).
\newblock Tailored randomized block {MCMC} methods with application to {DSGE}
  models.
\newblock {\em Journal of Econometrics}, 155(1):19 -- 38.

\bibitem[Chib \emph{et~al.}, 2018]{chib2018bayesian}
Chib, S., Shin, M., and Simoni, A. (2018).
\newblock {B}ayesian estimation and comparison of moment condition models.
\newblock {\em J. American Statist. Assoc.}, 113(524):1656--1668.

\bibitem[Chib \emph{et~al.}, 2020]{chib_shin_tan_2020}
Chib, S., Sin, M., and Tan, F. (2020).
\newblock Estimating high-dimensional {DSGE} models with the {TaRB-MH}
  algorithm.
\newblock Technical report, Olin Business School, Washington University.

\bibitem[Chopin and Robert, 2010]{chopin:robert:2010}
Chopin, N. and Robert, C. (2010).
\newblock Properties of nested sampling.
\newblock {\em Biometrika}, 97(3):741--755.

\bibitem[Christen and Fox, 2005]{christen:fox:2005}
Christen, J. and Fox, C. (2005).
\newblock Markov chain {M}onte {C}arlo using an approximation.
\newblock {\em J. Comput. Graph. Statist.}, 14(4):795--810.

\bibitem[Clart{\'e} \emph{et~al.}, 2020]{clarte2019component}
Clart{\'e}, G., Robert, C.~P., Ryder, R., and Stoehr, J. (2020).
\newblock Component-wise approximate {B}ayesian computation via {G}ibbs-like
  steps.
\newblock {\em Biometriak}.
\newblock To appear.

\bibitem[Craiu and Meng, 2005]{craiu:meng:2005}
Craiu, R.~V. and Meng, X.-L. (2005).
\newblock Multiprocess parallel antithetic coupling for backward and forward
  {M}arkov chain {M}onte {C}arlo.
\newblock {\em Ann. Statist.}, 33(2):661--697.

\bibitem[Craiu and Meng, 2011]{craiu2011perfection}
Craiu, R.~V. and Meng, X.-L. (2011).
\newblock Perfection within reach: exact {MCMC} sampling.
\newblock {\em Handbook of Markov Chain Monte Carlo}, pages 199--226.
\newblock Chapman \& Hall/CRC. Eds. Brooks, S., Gelman, A., Jones, G., Meng,
  X-L.

\bibitem[Damien \emph{et~al.}, 1999]{damien:wakefield:walker:1999}
Damien, P., Wakefield, J., and Walker, S. (1999).
\newblock {G}ibbs sampling for {B}ayesian non-conjugate and hierarchical models
  by using auxiliary variables.
\newblock {\em J. Royal Statist. Society Series B}, 61(2):331--344.

\bibitem[Davis and Rabinowitz, 1975]{davis1975numerical}
Davis, P. and Rabinowitz, P. (1975).
\newblock {\em Numerical Methods of Integration}.
\newblock Academic Press, New York.

\bibitem[Dawid, 1982]{Dawid1982}
Dawid, A.~P. (1982).
\newblock The well-calibrated {B}ayesian.
\newblock {\em Journal of the American Statistical Association},
  77(379):605--610.

\bibitem[De~Bruijn, 1961]{bruijn1961asymptotic}
De~Bruijn, N. (1961).
\newblock {\em Asymptotic methods in analysis. 2nd Edn., Sect. 4.2}.
\newblock Amsterdam: North-Holland.

\bibitem[Dean \emph{et~al.}, 2014]{dean2014parameter}
Dean, T.~A., Singh, S.~S., Jasra, A., and Peters, G.~W. (2014).
\newblock Parameter estimation for hidden {M}arkov models with intractable
  likelihoods.
\newblock {\em Scandinavian Journal of Statistics}, 41(4):970--987.

\bibitem[Dehideniya \emph{et~al.}, 2019]{dehideniya2019synthetic}
Dehideniya, M., Overstall, A.~M., Drovandi, C.~C., and McGree, J.~M. (2019).
\newblock A synthetic likelihood-based {Laplace} approximation for efficient
  design of biological processes.
\newblock {\em https://arXiv:1903.04168}.

\bibitem[Deligiannidis \emph{et~al.}, 2018]{deligiannidis2018correlated}
Deligiannidis, G., Doucet, A., and Pitt, M.~K. (2018).
\newblock The correlated pseudomarginal method.
\newblock {\em J. Royal Statist. Society Series B}, 80(5):839--870.

\bibitem[Dellaportas and Kontoyiannis, 2012]{Dellaportas2012}
Dellaportas, P. and Kontoyiannis, I. (2012).
\newblock Control variates for estimation based on reversible {M}arkov chain
  {M}onte {C}arlo samplers.
\newblock {\em J. Royal Statist. Society Series B}, 74(1):133--161.

\bibitem[Devroye, 1986]{devroye:1986}
Devroye, L. (1986).
\newblock {\em Non-Uniform Random Variate Generation}.
\newblock Springer-Verlag, New York.

\bibitem[Diebolt and Robert, 1994]{dieb:robe:1994}
Diebolt, J. and Robert, C.~P. (1994).
\newblock Estimation of finite mixture distributions by {B}ayesian sampling.
\newblock {\em J. Royal Statist. Society Series B}, 56(2):363--375.

\bibitem[Dongarra and Sullivan, 2000]{dongarra2000guest}
Dongarra, J. and Sullivan, F. (2000).
\newblock Guest editors’ introduction: The top 10 algorithms.
\newblock {\em Computing in Science \& Engineering}, 2(1):22--23.

\bibitem[Douc and Robert, 2011]{douc:robert:2011}
Douc, R. and Robert, C.~P. (2011).
\newblock A vanilla {R}ao–{B}lackwellization of {M}etropolis–{H}astings
  algorithms.
\newblock {\em Ann. Statist.}, 39(1):261--277.

\bibitem[Doucet \emph{et~al.}, 2001]{doucet:defreitas:gordon:2001}
Doucet, A., {{de Freitas}}, N., and Gordon, N. (2001).
\newblock {\em Sequential {M}onte {C}arlo Methods in Practice}.
\newblock Springer-Verlag, New York.

\bibitem[Doucet \emph{et~al.}, 2000]{doucet:godsill:andrieu:2000}
Doucet, A., Godsill, S., and Andrieu, C. (2000).
\newblock On sequential {M}onte-{C}arlo sampling methods for {B}ayesian
  filtering.
\newblock {\em Statist. Comp.}, 10:197--208.

\bibitem[Doucet \emph{et~al.}, 2015]{doucet2015efficient}
Doucet, A., Pitt, M.~K., Deligiannidis, G., and Kohn, R. (2015).
\newblock Efficient implementation of {M}arkov chain {M}onte {C}arlo when using
  an unbiased likelihood estimator.
\newblock {\em Biometrika}, 102(2):295--313.

\bibitem[Drovandi \emph{et~al.}, 2011]{drovandi:pettitt:faddy:2011}
Drovandi, C., Pettitt, A., and Faddy, M. (2011).
\newblock Approximate {B}ayesian computation using indirect inference.
\newblock {\em J. Royal Statist. Society Series A}, 60(3):503--524.

\bibitem[Drovandi \emph{et~al.}, 2015]{drovandi2015bayesian}
Drovandi, C.~C., Pettitt, A.~N., and Lee, A. (2015).
\newblock {B}ayesian indirect inference using a parametric auxiliary model.
\newblock {\em Statist. Science}, 30(1):72--95.

\bibitem[Dunson and Johndrow, 2019]{dunson2019hastings}
Dunson, D. and Johndrow, J. (2019).
\newblock The {H}astings algorithm at fifty.
\newblock {\em Biometrika}, 107(1):1--23.

\bibitem[Everitt \emph{et~al.}, 2020]{everitt:etal:2020}
Everitt, R., Culliford, R., Medina-Aguayo, F., and Wilson, D. (2020).
\newblock Sequential {M}onte {C}arlo with transformations.
\newblock {\em Stat. Comput.}, 30:663--676.

\bibitem[Fan and Sisson, 2011]{fan2011reversible}
Fan, Y. and Sisson, S.~A. (2011).
\newblock Reversible jump {MCMC}.
\newblock {\em Handbook of Markov Chain Monte Carlo}, pages 67--92.
\newblock Chapman \& Hall/CRC. Eds. Brooks, S., Gelman, A., Jones, G., Meng,
  X-L.

\bibitem[Fearnhead, 2011]{fearnhead2011mcmc}
Fearnhead, P. (2011).
\newblock {MCMC} for state-space models.
\newblock {\em Handbook of Markov Chain Monte Carlo}, pages 513--529.
\newblock Chapman \& Hall/CRC. Eds. Brooks, S., Gelman, A., Jones, G., Meng,
  X-L.

\bibitem[Fearnhead, 2018]{fearnhead2018asymptotics}
Fearnhead, P. (2018).
\newblock Asymptotics of {ABC}.
\newblock {\em Handbook of Approximate {B}ayesian Computation}, pages 269--288.
\newblock Chapman \& Hall/CRC. Eds. Sisson, S., Fan, Y., Beaumont, M.

\bibitem[Fearnhead \emph{et~al.}, 2018]{fearnhead:etal:2018}
Fearnhead, P., Bierkens, J., Pollock, M., and Roberts, G.~O. (2018).
\newblock Piecewise deterministic {M}arkov processes for continuous-time
  {M}onte {C}arlo.
\newblock {\em Statist. Sci.}, 33(3):386--412.

\bibitem[Fearnhead and Prangle, 2012]{fearnhead:prangle:2012}
Fearnhead, P. and Prangle, D. (2012).
\newblock Constructing summary statistics for approximate {B}ayesian
  computation: {S}emi-automatic approximate {B}ayesian computation.
\newblock {\em J. Royal Statist. Society Series B}, 74(3):419--474.
\newblock With discussion.

\bibitem[Feroz \emph{et~al.}, 2019]{Feroz_2019}
Feroz, F., Hobson, M.~P., Cameron, E., and Pettitt, A.~N. (2019).
\newblock Importance nested sampling and the multinest algorithm.
\newblock {\em The Open Journal of Astrophysics}, 2(1).

\bibitem[Fiel and Wyse, 2012]{fiel:wyse:2012}
Fiel, N. and Wyse, J. (2012).
\newblock Estimating the evidence: a review.
\newblock {\em Statistica Neerlandica}, 66(3):288--308.

\bibitem[Fienberg, 2006]{fienberg:2006}
Fienberg, S. (2006).
\newblock When did {B}ayesian inference become ``{B}ayesian``?
\newblock {\em {B}ayesian Analysis}, 1(1):1--40.

\bibitem[Fourment \emph{et~al.}, 2018]{fourment201819}
Fourment, M., Magee, A.~F., Whidden, C., Bilge, A., Matsen, I., Frederick, A.,
  and Minin, V.~N. (2018).
\newblock 19 dubious ways to compute the marginal likelihood of a phylogenetic
  tree topology.
\newblock {\em https://arXiv:1811.11804}.

\bibitem[Frazier, 2020]{frazier2020robust}
Frazier, D.~T. (2020).
\newblock Robust and efficient {A}pproximate {B}ayesian {C}omputation: A
  minimum distance approach.
\newblock {\em arXiv preprint arXiv:2006.14126}.

\bibitem[Frazier and Drovandi, 2019]{frazier2019robust}
Frazier, D.~T. and Drovandi, C. (2019).
\newblock Robust approximate {B}ayesian inference with synthetic likelihood.
\newblock {\em https://arXiv:1904.04551}.

\bibitem[Frazier \emph{et~al.}, 2019a]{frazier2019approximate}
Frazier, D.~T., Maneesoonthorn, W., Martin, G.~M., and McCabe, B.~P. (2019a).
\newblock Approximate {B}ayesian forecasting.
\newblock {\em Intern. J. Forecasting}, 35(2):521--539.

\bibitem[Frazier \emph{et~al.}, 2018]{FMRR2016}
Frazier, D.~T., Martin, G.~M., Robert, C.~P., and Rousseau, J. (2018).
\newblock Asymptotic properties of approximate {B}ayesian computation.
\newblock {\em Biometrika}, 105(3):593--607.

\bibitem[Frazier \emph{et~al.}, 2019b]{frazier2019bayesian}
Frazier, D.~T., Nott, D.~J., Drovandi, C., and Kohn, R. (2019b).
\newblock {B}ayesian inference using synthetic likelihood: {A}symptotics and
  adjustments.
\newblock {\em https://arXiv:1902.04827}.

\bibitem[Frazier \emph{et~al.}, 2020]{frazier2020model}
Frazier, D.~T., Robert, C.~P., and Rousseau, J. (2020).
\newblock Model misspecification in approximate {B}ayesian computation:
  consequences and diagnostics.
\newblock {\em J. Royal Statist. Society Series B}.

\bibitem[Frazier, 2018]{frazier2018tutorial}
Frazier, P. (2018).
\newblock A tutorial on {B}ayesian optimization.
\newblock {\em https://arxiv.org/abs/1807.02811}.

\bibitem[Frigessi \emph{et~al.}, 2000]{Frigessi2000}
Frigessi, A., Gasemyr, J., and Rue, H. (2000).
\newblock Antithetic coupling of two {G}ibbs sampler chains.
\newblock {\em The Annals of Statistics}, 28(4):1128--1149.

\bibitem[Fr\"{u}hwirth-Schnatter, 1994]{fruhwirth-schnatter:1994}
Fr\"{u}hwirth-Schnatter, S. (1994).
\newblock Data augmentation and dynamic linear models.
\newblock {\em J. Time Ser. Anal.}, 15(2):183--202.

\bibitem[Fr{\"u}hwirth-Schnatter, 2004]{fruhwirth:2004}
Fr{\"u}hwirth-Schnatter, S. (2004).
\newblock Estimating marginal likelihoods for mixture and {M}arkov switching
  models using bridge sampling techniques.
\newblock {\em The Econometrics Journal}, 7(1):143--167.

\bibitem[Gallant, 2016]{gallant2016reflections}
Gallant, A.~R. (2016).
\newblock Reflections on the probability space induced by moment conditions
  with implications for {B}ayesian inference.
\newblock {\em J. Finan. Econometrics}, 14(2):229--247.

\bibitem[Gallant and Tauchen, 1996]{gallant1996moments}
Gallant, A.~R. and Tauchen, G. (1996).
\newblock Which moments to match?
\newblock {\em Econometric theory}, 12(4):657--681.

\bibitem[Gelfand and Dey, 1994]{gelfand:dey:1994}
Gelfand, A. and Dey, D. (1994).
\newblock {B}ayesian model choice: {A}symptotics and exact calculations.
\newblock {\em J. Royal Statist. Society Series B}, 56(3):501--514.

\bibitem[Gelfand and Smith, 1990]{gelfand:smith90}
Gelfand, A. and Smith, A. (1990).
\newblock {S}ampling based approaches to calculating marginal densities.
\newblock {\em J. Amer. Statist. Assoc.}, 85(410):398--409.

\bibitem[Gelman and Meng, 1998]{gelman:meng:1998}
Gelman, A. and Meng, X. (1998).
\newblock Simulating normalizing constants: From importance sampling to bridge
  sampling to path sampling.
\newblock {\em Statist. Science}, 13:163--185.

\bibitem[Gelman and Shirley, 2011]{gelman2011inference}
Gelman, A. and Shirley, K. (2011).
\newblock Inference from simulations and monitoring convergence.
\newblock {\em Handbook of Markov chain Monte Carlo}, pages 163--174.
\newblock Chapman \& Hall/CRC. Eds. Brooks, S., Gelman, A., Jones, G., Meng,
  X-L.

\bibitem[Geman and Geman, 1984]{geman:1984}
Geman, S. and Geman, D. (1984).
\newblock Stochastic relaxation, {G}ibbs distributions and the {B}ayesian
  restoration of images.
\newblock {\em IEEE Trans. Pattern Anal. Mach. Intell.}, 6:721--741.

\bibitem[George, 2000]{george:2000}
George, E. (2000).
\newblock {T}he variable selection problem.
\newblock {\em J. American Statist. Assoc.}, 95(452):1304--1308.

\bibitem[Gerber and Chopin, 2015]{gerber:chopin:2015}
Gerber, M. and Chopin, N. (2015).
\newblock Sequential quasi {M}onte {C}arlo.
\newblock {\em J. Royal Statist. Society Series B}, 77(3):509--579.

\bibitem[Geweke, 1989a]{geweke:1989}
Geweke, J. (1989a).
\newblock {B}ayesian inference in econometric models using {M}onte {C}arlo
  integration.
\newblock {\em Econometrica}, 57(6):1317--1340.

\bibitem[Geweke, 1989b]{geweke1989exact}
Geweke, J. (1989b).
\newblock Exact predictive densities for linear models with {ARCH}
  disturbances.
\newblock {\em J. Econometrics}, 40(1):63--86.

\bibitem[Geweke, 1999]{geweke99}
Geweke, J. (1999).
\newblock Using simulation methods for {B}ayesian econometric models:
  Inference, development, and communication.
\newblock {\em Econometric Reviews}, 18(1):1--73.
\newblock With discussion.

\bibitem[Geweke \emph{et~al.}, 2011]{geweke2011handbook}
Geweke, J., Koop, G., and van Dijk, H. (2011).
\newblock {\em The Oxford Handbook of {B}ayesian Econometrics}.
\newblock OUP.

\bibitem[Geyer, 1991]{geyer91pt}
Geyer, C. (1991).
\newblock {M}arkov chain {M}onte {C}arlo maximum likelihood.
\newblock In Keramigas, E., editor, {\em Computing Science and Statistics:
  Proceedings of the 32rd Symposium on the Interface}, pages 156--163, Fairfax.
  Interface Foundation.

\bibitem[Geyer, 1993]{geyer:1993}
Geyer, C. (1993).
\newblock Estimating normalizing constants and reweighting mixtures in {M}arkov
  chain {M}onte {C}arlo.
\newblock Technical Report 568, School of Statistics, Univ. of Minnesota.

\bibitem[Geyer, 2011a]{geyer2011st}
Geyer, C.~J. (2011a).
\newblock Importance sampling, simulated tempering, and umbrella sampling.
\newblock {\em Handbook of Markov chain Monte Carlo}, pages 295--311.
\newblock Chapman \& Hall/CRC. Eds. Brooks, S., Gelman, A., Jones, G., Meng,
  X-L.

\bibitem[Geyer, 2011b]{geyer2011introduction}
Geyer, C.~J. (2011b).
\newblock Introduction to {M}arkov chain {M}onte {C}arlo.
\newblock {\em Handbook of Markov chain Monte Carlo}, pages 3--48.
\newblock Chapman \& Hall/CRC. Eds. Brooks, S., Gelman, A., Jones, G., Meng,
  X-L.

\bibitem[Ghosal \emph{et~al.}, 1995]{ghosal1995convergence}
Ghosal, S., Ghosh, J.~K., and Samanta, T. (1995).
\newblock On convergence of posterior distributions.
\newblock {\em The Annals of Statistics}, 23(6):2145--2152.

\bibitem[Ghosal and Van~der Vaart, 2017]{ghosal2017fundamentals}
Ghosal, S. and Van~der Vaart, A. (2017).
\newblock {\em Fundamentals of Nonparametric {B}ayesian Inference}, volume~44.
\newblock Cambridge University Press.

\bibitem[Giordani \emph{et~al.}, 2011]{giordani2011bayesian}
Giordani, P., Pitt, M., and Kohn, R. (2011).
\newblock {B}ayesian inference for time series state space models.
\newblock {\em The Oxford Handbook of {B}ayesian Econometrics}, pages 61--124.
\newblock OUP. Eds. Geweke, J., Koop, G. and van Dijk, H.

\bibitem[Giummol{\`e} \emph{et~al.}, 2017]{giummole2017objective}
Giummol{\`e}, F., Mameli, V., Ruli, E., and Ventura, L. (2017).
\newblock Objective {B}ayesian inference with proper scoring rules.
\newblock {\em TEST}, 28(3):1--28.

\bibitem[Glynn, 2016]{Glynn:exact2016}
Glynn, P.~W. (2016).
\newblock Exact simulation vs. exact estimation.
\newblock {\em Proceedings of the 2016 Winter Simulation Conference},
  WSC16:193–205.

\bibitem[Glynn and Rhee, 2014]{glynn:rhee:2014}
Glynn, P.~W. and Rhee, C.-H. (2014).
\newblock Exact estimation for {M}arkov chain equilibrium expectations.
\newblock {\em J. Appl. Probab.}, 51(A):377--389.

\bibitem[Gneiting \emph{et~al.}, 2007]{gneiting2007probabilistic}
Gneiting, T., Balabdaoui, F., and Raftery, A.~E. (2007).
\newblock Probabilistic forecasts, calibration and sharpness.
\newblock {\em Journal of the Royal Statistical Society: Series B (Statistical
  Methodology)}, 69(2):243--268.

\bibitem[Golightly \emph{et~al.}, 2015]{Golightly2015}
Golightly, A., Henderson, D.~A., and Sherlock, C. (2015).
\newblock Delayed acceptance particle {MCMC} for exact inference in stochastic
  kinetic models.
\newblock {\em Statist. Comp.}, 25(5):1039--1055.

\bibitem[Goodfellow \emph{et~al.}, 2014]{goodfellow2014generative}
Goodfellow, I., Pouget-Abadie, J., Mirza, M., Xu, B., Warde-Farley, D., Ozair,
  S., Courville, A., and Bengio, Y. (2014).
\newblock Generative adversarial nets.
\newblock In {\em Advances in neural information processing systems}, pages
  2672--2680.

\bibitem[Gordon \emph{et~al.}, 1993]{gordon:salmon:smith:1993}
Gordon, N., Salmond, J., and Smith, A. (1993).
\newblock A novel approach to non-linear/non-{G}aussian {B}ayesian state
  estimation.
\newblock {\em IEEE Proceedings on Radar and Signal Processing},
  140(2):107--113.

\bibitem[Gouri{\'e}roux \emph{et~al.}, 1993]{gourieroux:monfort:renault:1993}
Gouri{\'e}roux, C., Monfort, A., and Renault, E. (1993).
\newblock Indirect inference.
\newblock {\em J. Applied Econometrics}, 8:85--118.

\bibitem[Gramacy \emph{et~al.}, 2010]{Gramacy2010}
Gramacy, R., Samworth, R., and King, R. (2010).
\newblock Importance tempering.
\newblock {\em Statist. Comp.}, 20(1):1--7.

\bibitem[Green, 1995]{green:1995}
Green, P. (1995).
\newblock Reversible jump {MCMC} computation and {B}ayesian model
  determination.
\newblock {\em Biometrika}, 82(4):711--732.

\bibitem[Green \emph{et~al.}, 2015]{greenetal2015}
Green, P., Latuszynski, K., Pereyra, M., and Robert, C. (2015).
\newblock {B}ayesian computation: a summary of the current state, and samples
  backwards and forwards.
\newblock {\em Statist. Comp.}, 25:835--862.

\bibitem[Green, 2003]{green03}
Green, P.~J. (2003).
\newblock Trans-dimensional {M}arkov chain {M}onte {C}arlo.
\newblock {\em Oxford Statistical Science Series}, 27:179--198.

\bibitem[Gronau \emph{et~al.}, 2017]{gronau2017tutorial}
Gronau, Q.~F., Sarafoglou, A., Matzke, D., Ly, A., Boehm, U., Marsman, M.,
  Leslie, D.~S., Forster, J.~J., Wagenmakers, E.-J., and Steingroever, H.
  (2017).
\newblock A tutorial on bridge sampling.
\newblock {\em Journal of Mathematical Psychology}, 81:80--97.

\bibitem[Gubernatis, 2005]{Gub2005}
Gubernatis, J.~E. (2005).
\newblock {M}arshall {R}osenbluth and the {M}etropolis algorithm.
\newblock {\em Physics of Plasmas}, 12(5):057303.

\bibitem[Guedj, 2019]{guedj2019}
Guedj, B. (2019).
\newblock A primer on {PAC-B}ayesian learning.
\newblock {\em arXiv preprint arXiv:1901.05353}.

\bibitem[Gutmann and Corander, 2016]{gutmann2016bayesian}
Gutmann, M.~U. and Corander, J. (2016).
\newblock {B}ayesian optimization for likelihood-free inference of
  simulator-based statistical models.
\newblock {\em The Journal of Machine Learning Research}, 17(1):4256--4302.

\bibitem[Gutmann and Hyv{\"a}rinen, 2012]{gutmann:hyvaarinen:2012}
Gutmann, M.~U. and Hyv{\"a}rinen, A. (2012).
\newblock Noise-contrastive estimation of unnormalized statistical models, with
  applications to natural image statistics.
\newblock {\em Journal of Machine Learning Research}, 13(Feb):307--361.

\bibitem[Hajargasht and Wo{\'z}niak, 2018]{hajargasht2018accurate}
Hajargasht, G. and Wo{\'z}niak, T. (2018).
\newblock Accurate computation of marginal data densities using variational
  {B}ayes.
\newblock {\em https://arXiv:1805.10036}.

\bibitem[Hammersley and Handscomb, 1964]{hammersley:handscomb:1964}
Hammersley, J. and Handscomb, D. (1964).
\newblock {\em {M}onte {C}arlo Methods}.
\newblock John Wiley, New York.

\bibitem[Hastings, 1970]{hastings:1970}
Hastings, W. (1970).
\newblock {M}onte {C}arlo sampling methods using {M}arkov chains and their
  application.
\newblock {\em Biometrika}, 57(1):97--109.

\bibitem[Higdon, 1998]{higdon1998}
Higdon, D.~M. (1998).
\newblock Auxiliary variable methods for {M}arkov chain {M}onte {C}arlo with
  applications.
\newblock {\em Journal of the American Statistical Association},
  93(442):585--595.

\bibitem[Hitchcock, 2003]{hitchcock:2003}
Hitchcock, D.~H. (2003).
\newblock A history of the {M}etropolis--{H}astings algorithm.
\newblock {\em American Statist.}, 57(4):254--257.

\bibitem[Hoffman and Gelman, 2014]{hoffman2014no}
Hoffman, M.~D. and Gelman, A. (2014).
\newblock The {N}o-{U}-{T}urn sampler: adaptively setting path lengths in
  {H}amiltonian {M}onte {C}arlo.
\newblock {\em Journal of Machine Learning Research}, 15(1):1593--1623.

\bibitem[Holmes and Walker, 2017]{holmes2017assigning}
Holmes, C. and Walker, S. (2017).
\newblock Assigning a value to a power likelihood in a general {B}ayesian
  model.
\newblock {\em Biometrika}, 104(2):497--503.

\bibitem[Hoogerheide \emph{et~al.}, 2009]{hoogerheide2009simulation}
Hoogerheide, L.~F., van Dijk, H.~K., and van Oest, R.~D. (2009).
\newblock Simulation based {B}ayesian econometric inference: principles and
  some recent computational advances.
\newblock {\em Handbook of Computational Econometrics}, pages 215--280.
\newblock John Wiley \& Sons. Eds. van Dijk, H. and van Oest, R.

\bibitem[Hooper, 2013]{Hooper:2013}
Hooper, M. (2013).
\newblock Richard {P}rice, {B}ayes’ theorem, and {G}od.
\newblock {\em Significance}, 10(1):36--39.

\bibitem[Huber, 2016]{huber2016perfect}
Huber, M.~L. (2016).
\newblock {\em Perfect simulation}.
\newblock Chapman \& Hall/CRC.

\bibitem[Huggins \emph{et~al.}, 2019]{huggins2019validated}
Huggins, J.~H., Kasprzak, M., Campbell, T., and Broderick, T. (2019).
\newblock Validated variational inference via practical posterior error bounds.
\newblock {\em https://arXiv:1910.04102}.

\bibitem[Huggins and Miller, 2019]{huggins2019using}
Huggins, J.~H. and Miller, J.~W. (2019).
\newblock Using bagged posteriors for robust inference and model criticism.
\newblock {\em https://arXiv:1912.07104}.

\bibitem[Iba, 2000]{iba0}
Iba, Y. (2000).
\newblock Population-based {M}onte {C}arlo algorithms.
\newblock {\em Trans.~Japanese Soc.~Artificial Intell.}, 16(2):279--286.

\bibitem[Jacob \emph{et~al.}, 2011]{jacob:robert:smith:2010}
Jacob, P., Robert, C., and Smith, M. (2011).
\newblock Using parallel computation to improve independent
  {M}etropolis--{H}astings based estimation.
\newblock {\em J. Comput. Graph. Statist.}, 20(3):616--635.

\bibitem[Jacob \emph{et~al.}, 2020]{jacob2019unbiased}
Jacob, P.~E., O’Leary, J., and Atchad{\'e}, Y.~F. (2020).
\newblock Unbiased {M}arkov chain {M}onte {C}arlo methods with couplings.
\newblock {\em J. Royal Statist. Society Series B}, 82:1--32.
\newblock With discussion.

\bibitem[Jacquier \emph{et~al.}, 1994]{jacquier94}
Jacquier, R., Polson, N.~G., and Rossi, P.~E. (1994).
\newblock {B}ayesian analysis of stochastic volatility models.
\newblock {\em J. Business and Economic Statistics}, 12(4):371--389.
\newblock With discussion.

\bibitem[Jahan \emph{et~al.}, 2020]{Jahan2020}
Jahan, F., Ullah, I., and Mengersen, K. (2020).
\newblock A review of {B}ayesian statistical approaches for {B}ig {D}ata.
\newblock In Mengersen, K., Pudlo, P., and Robert, C., editors, {\em Case
  Studies in Applied {B}ayesian Science}, pages 17--44. Springer.

\bibitem[Jasra, 2015]{jasra2015approximate}
Jasra, A. (2015).
\newblock Approximate {B}ayesian computation for a class of time series models.
\newblock {\em International Statistical Review}, 83(3):405--435.

\bibitem[Jasra \emph{et~al.}, 2012]{jasra:etal:2012}
Jasra, A., Singh, S., Martin, J., and McCoy, E. (2012).
\newblock Filtering via approximate {B}ayesian computation.
\newblock {\em Statist. Comp.}, 22:1223--1237.

\bibitem[Jennings and Madigan, 2017]{jennings2017astroabc}
Jennings, E. and Madigan, M. (2017).
\newblock Astro{ABC}: an approximate {B}ayesian computation sequential {M}onte
  {C}arlo sampler for cosmological parameter estimation.
\newblock {\em Astronomy and Computing}, 19:16--22.

\bibitem[Jiang and Tanner, 2008]{jiang2008}
Jiang, W. and Tanner, M.~A. (2008).
\newblock Gibbs posterior for variable selection in high-dimensional
  classification and data-mining.
\newblock {\em Annals of Statistics}, 36(5):2207--2231.

\bibitem[Johndrow \emph{et~al.}, 2019]{johndrow2019mcmc}
Johndrow, J.~E., Smith, A., Pillai, N., and Dunson, D.~B. (2019).
\newblock {MCMC} for imbalanced categorical data.
\newblock {\em J. American Statist. Assoc.}, 114(527):1394--1403.

\bibitem[Joyce and Marjoram, 2008]{joyce:marjoram:2008}
Joyce, P. and Marjoram, P. (2008).
\newblock Approximately sufficient statistics and {B}ayesian computation.
\newblock {\em Statistical Applications in Genetics and Molecular Biology},
  7(1):article 26.

\bibitem[Kabisa \emph{et~al.}, 2016]{kabisa2016online}
Kabisa, S., Dunson, D.~B., and Morris, J.~S. (2016).
\newblock Online variational {B}ayes inference for high-dimensional correlated
  data.
\newblock {\em J. Comput. Graph. Statist.}, 25(2):426--444.

\bibitem[Kass and Raftery, 1995]{kass:raftery:1995}
Kass, R. and Raftery, A. (1995).
\newblock {B}ayes factors.
\newblock {\em J. American Statist. Assoc.}, 90:773--795.

\bibitem[Kim \emph{et~al.}, 1998]{kim1998svl}
Kim, S., Shephard, N., and Chib, S. (1998).
\newblock Stochastic volatility: likelihood inference and comparison with
  {ARCH} models.
\newblock {\em The Review of Economic Studies}, 65(3):361--393.

\bibitem[Kleijn and van~der Vaart, 2012]{kleijn2012}
Kleijn, B. and van~der Vaart, A. (2012).
\newblock The {B}ernstein-von-{M}ises theorem under misspecification.
\newblock {\em Electron. J. Statist.}, 6:354--381.

\bibitem[Kloek and van Dijk, 1978]{kloek1978bayesian}
Kloek, T. and van Dijk, H.~K. (1978).
\newblock {B}ayesian estimates of equation system parameters: an application of
  integration by {M}onte {C}arlo.
\newblock {\em Econometrica}, 46(1):1--19.

\bibitem[Knoblauch \emph{et~al.}, 2019]{knoblauch2019generalized}
Knoblauch, J., Jewson, J., and Damoulas, T. (2019).
\newblock Generalized variational inference.
\newblock {\em https://arXiv:1904.02063}.

\bibitem[Kon Kam~King \emph{et~al.}, 2019]{konkamking2019}
Kon Kam~King, G., Canale, A., and Ruggiero, M. (2019).
\newblock {B}ayesian functional forecasting with locally-autoregressive
  dependent processes.
\newblock {\em {B}ayesian Anal.}, 14(4):1121--1141.

\bibitem[Koop and Korobilis, 2018]{koop2018variational}
Koop, G. and Korobilis, D. (2018).
\newblock Variational {B}ayes inference in high-dimensional time-varying
  parameter models.
\newblock {\em SSRN 3246472}.

\bibitem[Koop, 2003]{koop2003bayesian}
Koop, G.~M. (2003).
\newblock {\em {B}ayesian Econometrics}.
\newblock John Wiley \& Sons Inc.

\bibitem[Lazar, 2003]{lazar:2003}
Lazar, N.~A. (2003).
\newblock {B}ayesian empirical likelihood.
\newblock {\em Biometrika}, 90:319--326.

\bibitem[Lemieux, 2009]{lemieux2009monte}
Lemieux, C. (2009).
\newblock {\em Monte {C}arlo and quasi-{M}onte {C}arlo sampling}.
\newblock Springer Science \& Business Media.

\bibitem[Li and Fearnhead, 2018a]{LF2016b}
Li, W. and Fearnhead, P. (2018a).
\newblock {C}onvergence of regression-adjusted approximate {B}ayesian
  computation.
\newblock {\em Biometrika}, 105(2):301--318.

\bibitem[Li and Fearnhead, 2018b]{LF2016a}
Li, W. and Fearnhead, P. (2018b).
\newblock {O}n the asymptotic efficiency of approximate {B}ayesian computation
  estimators.
\newblock {\em Biometrika}, 105(2):285--299.

\bibitem[Lindley, 1980]{lindley1980bayesian}
Lindley, D. (1980).
\newblock Approximate {B}ayesian methods.
\newblock In {\em {B}ayesian statistics: Proceedings of the first international
  meeting held in Valencia (Spain), May 28 to June 2, 1979}.

\bibitem[Lintusaari \emph{et~al.}, 2017]{lintusaari2017fundamentals}
Lintusaari, J., Gutmann, M.~U., Dutta, R., Kaski, S., and Corander, J. (2017).
\newblock Fundamentals and recent developments in approximate {B}ayesian
  computation.
\newblock {\em Systematic biology}, 66(1):e66--e82.

\bibitem[Liu, 2001]{liu01}
Liu, J.~S. (2001).
\newblock {\em {M}onte {C}arlo Strategies in Scientific Computing}.
\newblock Springer Verlag, New-York.

\bibitem[Liu \emph{et~al.}, 2000]{liuliang2000}
Liu, J.~S., Liang, F., and Wong, W.~H. (2000).
\newblock The multiple-try method and local optimization in {M}etropolis
  sampling.
\newblock {\em J. American Statist. Assoc.}, 95(449):121--134.

\bibitem[Loaiza-Maya \emph{et~al.}, 2020]{loaiza2019focused}
Loaiza-Maya, R., Martin, G.~M., and Frazier, D.~T. (2020).
\newblock Focused {B}ayesian prediction.
\newblock {\em Forthcoming. Journal of Applied Econometrics}.

\bibitem[Luo and Tjelmeland, 2019]{luo2019multiple}
Luo, X. and Tjelmeland, H. (2019).
\newblock A multiple-try {M}etropolis-{H}astings algorithm with tailored
  proposals.
\newblock {\em Computational Statistics}, 34(3):1109--1133.

\bibitem[Lyddon \emph{et~al.}, 2019]{lyddon2019general}
Lyddon, S., Holmes, C., and Walker, S. (2019).
\newblock General {B}ayesian updating and the loss-likelihood bootstrap.
\newblock {\em Biometrika}, 106(2):465--478.

\bibitem[Lyne \emph{et~al.}, 2015]{lyne2015}
Lyne, A.-M., Girolami, M., Atchadé, Y., Strathmann, H., and Simpson, D.
  (2015).
\newblock On {R}ussian roulette estimates for {B}ayesian inference with
  doubly-intractable likelihoods.
\newblock {\em Statist. Sci.}, 30(4):443--467.

\bibitem[Marin \emph{et~al.}, 2011]{marin:pudlo:robert:ryder:2011}
Marin, J., Pudlo, P., Robert, C., and Ryder, R. (2011).
\newblock Approximate {B}ayesian computational methods.
\newblock {\em Statist. Comp.}, 21(2):279--291.

\bibitem[Marin and Robert, 2011]{marin:robert:2010}
Marin, J. and Robert, C. (2011).
\newblock Importance sampling methods for {B}ayesian discrimination between
  embedded models.
\newblock In Chen, M.-H., Dey, D., M{\"u}ller, P., Sun, D., and Ye, K.,
  editors, {\em Frontiers of Statistical Decision Making and {B}ayesian
  Analysis}. Springer-Verlag, New York.

\bibitem[Marin \emph{et~al.}, 2005]{marin:mengersen:robert:2005}
Marin, J.-M., Mengersen, K., and Robert, C. (2005).
\newblock {B}ayesian modelling and inference on mixtures of distributions.
\newblock {\em Handbook of Statistics}, pages 459--507.
\newblock Elsevier. Eds. Rao, C. and Dey, D.

\bibitem[Marinari and Parisi, 1992]{marinarietparisi92}
Marinari, E. and Parisi, G. (1992).
\newblock Simulated tempering: A new {M}onte {C}arlo scheme.
\newblock {\em Europhysics Letters}, 19(6):451--458.

\bibitem[Marjoram \emph{et~al.}, 2003]{marjoram:etal:2003}
Marjoram, P., Molitor, J., Plagnol, V., and Tavar{\'e}, S. (2003).
\newblock Markov chain {M}onte {C}arlo without likelihoods.
\newblock {\em Proc.~Natl.~Acad.~Sci.~USA}, 100(26):15324--15328.

\bibitem[Martin \emph{et~al.}, 2019]{martin2019auxiliary}
Martin, G.~M., McCabe, B.~P., Frazier, D.~T., Maneesoonthorn, W., and Robert,
  C.~P. (2019).
\newblock Auxiliary likelihood-based approximate {B}ayesian computation in
  state space models.
\newblock {\em J. Comput. Graph. Statist.}, 28(3):508--522.

\bibitem[Martino, 2018]{martino2018review}
Martino, L. (2018).
\newblock A review of multiple try {MCMC} algorithms for signal processing.
\newblock {\em Digital Signal Processing}, 75:134--152.

\bibitem[Martino and Riebler, 2019]{martino2019integrated}
Martino, S. and Riebler, A. (2019).
\newblock Integrated nested {L}aplace approximations ({INLA}).
\newblock {\em https://arXiv:1907.01248}.

\bibitem[Meng and Schilling, 2002]{meng:schilling:2002}
Meng, X. and Schilling, S. (2002).
\newblock Warp bridge sampling.
\newblock {\em J. Comput. Graph. Statist.}, 11(3):552--586.

\bibitem[Meng and Wong, 1996]{meng:wong:1996}
Meng, X. and Wong, W. (1996).
\newblock Simulating ratios of normalizing constants via a simple identity: a
  theoretical exploration.
\newblock {\em {S}tatist. Sinica}, 6:831--860.

\bibitem[Metropolis \emph{et~al.}, 1953]{metropolis:1953}
Metropolis, N., Rosenbluth, A.~W., Rosenbluth, M.~N., Teller, A.~H., and
  Teller, E. (1953).
\newblock Equations of state calculations by fast computing machines.
\newblock {\em J. Chem. Phys.}, 21:1087--1092.

\bibitem[Middleton \emph{et~al.}, 2018]{middleton2018unbiased}
Middleton, L., Deligiannidis, G., Doucet, A., and Jacob, P.~E. (2018).
\newblock Unbiased {M}arkov chain {M}onte {C}arlo for intractable target
  distributions.
\newblock {\em https://arXiv:1807.08691}.

\bibitem[Miller and Dunson, 2019]{miller2019robust}
Miller, J.~W. and Dunson, D.~B. (2019).
\newblock Robust {B}ayesian inference via coarsening.
\newblock {\em J. American Statist. Assoc.}, 114(527):1113--1125.

\bibitem[Mukherjee \emph{et~al.}, 2006]{mukherjee2006nsa}
Mukherjee, P., Parkinson, D., and Liddle, A. (2006).
\newblock {A nested sampling algorithm for cosmological model selection}.
\newblock {\em The Astrophysical Journal}, 638(2):L51--L54.

\bibitem[Muller, 2013]{Muller2013}
Muller, U.~K. (2013).
\newblock Risk of {B}ayesian inference in misspecified models, and the sandwich
  covariance matrix.
\newblock {\em Econometrica}, 81(5):1805--1849.

\bibitem[Naesseth \emph{et~al.}, 2019]{naesseth2019elements}
Naesseth, C.~A., Lindsten, F., Sch{\"o}n, T.~B.,\emph{et~al.} (2019).
\newblock Elements of sequential {M}onte {C}arlo.
\newblock {\em Foundations and Trends in Machine Learning}, 12(3):307--392.

\bibitem[Naylor and Smith, 1982]{naylor:smith:1982}
Naylor, J. and Smith, A. (1982).
\newblock Application of a method for the efficient computation of posterior
  distributions.
\newblock {\em Applied {S}tatistics}, 31(3):214--225.

\bibitem[Neal, 1994]{neal:1994}
Neal, R. (1994).
\newblock Contribution to the discussion of ``{A}pproximate {B}ayesian
  inference with the weighted likelihood bootstrap" by {M}ichael {A}. {N}ewton
  and {A}drian {E.} {R}aftery.
\newblock {\em J. Royal Statist. Society Series B}, 56(1):41--42.

\bibitem[Neal, 1996]{Neal1996}
Neal, R. (1996).
\newblock Sampling from multimodal distributions using tempered transitions.
\newblock {\em Statist. Comp.}, 6:353–366.

\bibitem[Neal, 1998]{Neal1998}
Neal, R. (1998).
\newblock Suppressing random walks in {M}arkov chain {M}onte {C}arlo using
  ordered overrelaxation.
\newblock {\em Learning in Graphical Models}, pages 205--228.
\newblock Chapman \& Hall/CRC. Ed. Jordan, M. I.

\bibitem[Neal, 1999]{neal:1999}
Neal, R. (1999).
\newblock Erroneous results in `{M}arginal likelihood from the {G}ibbs output'.
\newblock Technical report, University of Toronto.

\bibitem[Neal, 2001]{neal2001ais}
Neal, R. (2001).
\newblock {Annealed importance sampling}.
\newblock {\em Stat. Comput.}, 11(2):125--139.

\bibitem[Neal, 2003]{neal:2003}
Neal, R. (2003).
\newblock Slice sampling.
\newblock {\em Ann. Statist.}, 31(3):705--767.
\newblock With discussion.

\bibitem[Neal, 2011a]{neal2011mcmc}
Neal, R. (2011a).
\newblock {MCMC} using ensembles of states for problems with fast and slow
  variables such as {G}aussian process regression.
\newblock {\em https://arXiv:1101.0387}.

\bibitem[Neal, 2011b]{neal:2011}
Neal, R. (2011b).
\newblock {MCMC} using {H}amiltonian dynamics.
\newblock {\em Handbook of Markov Chain Monte Carlo}, pages 113--162.
\newblock Chapman \& Hall/CRC. Eds. Brooks, S., Gelman, A., Jones, G., Meng,
  X-L.

\bibitem[Neiswanger \emph{et~al.}, 2013]{neiswanger:wang:xing:2013}
Neiswanger, W., Wang, C., and Xing, E. (2013).
\newblock Asymptotically exact, embarrassingly parallel {MCMC}.
\newblock {\em https://arXiv:1311.4780}.

\bibitem[Nemeth and Fearnhead, 2019]{nemeth2019stochastic}
Nemeth, C. and Fearnhead, P. (2019).
\newblock Stochastic gradient {M}arkov chain {M}onte {C}arlo.
\newblock {\em https://arXiv:1907.06986}.

\bibitem[Newton and Raftery, 1994]{newton:raftery:1994}
Newton, M. and Raftery, A. (1994).
\newblock Approximate {B}ayesian inference by the weighted likelihood bootstrap
  (with discussion).
\newblock {\em J. Royal Statist. Society Series B}, 56(1):3--48.

\bibitem[Nguyen \emph{et~al.}, 2020]{nguyen2020approximate}
Nguyen, H.~D., Arbel, J., L{\"u}, H., and Forbes, F. (2020).
\newblock Approximate {B}ayesian computation via the energy statistic.
\newblock {\em IEEE Access}, 8:131683--131698.

\bibitem[Nott \emph{et~al.}, 2018]{nott2018high}
Nott, D., Ong, V. M.-H., Fan, Y., and Sisson, S. (2018).
\newblock High-dimensional {ABC}.
\newblock {\em Handbook of Approximate {B}ayesian Computation}, pages 211--242.
\newblock Chapman \& Hall/CRC. Eds. Sisson, S., Fan, Y., Beaumont, M.

\bibitem[Nott and Kohn, 2005]{nott2005adaptive}
Nott, D.~J. and Kohn, R. (2005).
\newblock Adaptive sampling for {B}ayesian variable selection.
\newblock {\em Biometrika}, 92(4):747--763.

\bibitem[Ong \emph{et~al.}, 2018a]{ong2018variational}
Ong, V.~M., Nott, D.~J., Tran, M.-N., Sisson, S.~A., and Drovandi, C.~C.
  (2018a).
\newblock Variational {B}ayes with synthetic likelihood.
\newblock {\em Statist. Comp.}, 28(4):971--988.

\bibitem[Ong \emph{et~al.}, 2018b]{ong2018likelihood}
Ong, V. M.-H., Nott, D.~J., Tran, M.-N., Sisson, S.~A., and Drovandi, C.~C.
  (2018b).
\newblock Likelihood-free inference in high dimensions with synthetic
  likelihood.
\newblock {\em Computational Statistics \& Data Analysis}, 128:271--291.

\bibitem[Ormerod and Wand, 2010]{ormerod2010explaining}
Ormerod, J.~T. and Wand, M.~P. (2010).
\newblock Explaining variational approximations.
\newblock {\em American Statist.}, 64(2):140--153.

\bibitem[Owen, 2017]{owen2017statistically}
Owen, A.~B. (2017).
\newblock Statistically efficient thinning of a {M}arkov chain sampler.
\newblock {\em J. Comput. Graph. Statist.}, 26(3):738--744.

\bibitem[Park and Nassar, 2014]{park2014variational}
Park, M. and Nassar, M. (2014).
\newblock Variational {B}ayesian inference for forecasting hierarchical time
  series.
\newblock In {\em Divergence Methods in Probabilistic Inference (DMPI) workshop
  at International Conference on Machine Learning (ICML)}. Citeseer.

\bibitem[Peters \emph{et~al.}, 2012]{peters2012likelihood}
Peters, G.~W., Sisson, S.~A., and Fan, Y. (2012).
\newblock Likelihood-free {B}ayesian inference for $\alpha$-stable models.
\newblock {\em Comput. Statist. Data Anal.}, 56(11):3743--3756.

\bibitem[Pettenuzzo and Ravazzolo, 2016]{Pett2016}
Pettenuzzo, D. and Ravazzolo, F. (2016).
\newblock Optimal portfolio choice under decision-based model combinations.
\newblock {\em Journal of Applied Econometrics}, 31(7):1312--1332.

\bibitem[Pitt \emph{et~al.}, 2012]{pitt2012some}
Pitt, M.~K., dos Santos~Silva, R., Giordani, P., and Kohn, R. (2012).
\newblock On some properties of {M}arkov chain {M}onte {C}arlo simulation
  methods based on the particle filter.
\newblock {\em J. Econometrics}, 171(2):134--151.

\bibitem[Pollock \emph{et~al.}, 2020]{pollock:etal:2020}
Pollock, M., Fearnhead, P., Johansen, A.~M., and Roberts, G.~O. (2020).
\newblock Quasi-stationary {M}onte {C}arlo and the {ScaLE} algorithm.
\newblock {\em J. Royal Statist. Society Series B}, 82(5):1167--1221.
\newblock (With discussion.).

\bibitem[Polson \emph{et~al.}, 1992]{carlin:polson:stoffer:1992}
Polson, N.~G., Carlin, B.~P., and Stoffer, D.~S. (1992).
\newblock A {M}onte {C}arlo approach to nonnormal and nonlinear state-space
  modeling.
\newblock {\em J. American Statist. Assoc.}, 87(418):493--500.

\bibitem[Price \emph{et~al.}, 2018]{price2018bayesian}
Price, L.~F., Drovandi, C.~C., Lee, A., and Nott, D.~J. (2018).
\newblock {B}ayesian synthetic likelihood.
\newblock {\em J. Comput. Graph. Statist.}, 27(1):1--11.

\bibitem[Price, 1764]{price:1764}
Price, R. (1764).
\newblock A demonstration of the second rule in the essay towards the solution
  of a problem in the doctine of chances.
\newblock {\em Philosophical Transactions of the {R}oyal Society of {L}ondon},
  54:296--325.

\bibitem[Priddle \emph{et~al.}, 2019]{priddle2019efficient}
Priddle, J.~W., Sisson, S.~A., Frazier, D.~T., and Drovandi, C. (2019).
\newblock Efficient {B}ayesian synthetic likelihood with whitening
  transformations.
\newblock {\em arXiv preprint arXiv:1909.04857}.

\bibitem[Pritchard \emph{et~al.}, 1999]{pritchard:seielstad:perez:feldman:1999}
Pritchard, J., Seielstad, M., Perez-Lezaun, A., and Feldman, M. (1999).
\newblock Population growth of human {Y} chromosomes: a study of {Y} chromosome
  microsatellites.
\newblock {\em Mol. Biol. Evol.}, 16:1791--1798.

\bibitem[Propp and Wilson, 1996]{proppetwilson96}
Propp, J.~G. and Wilson, D.~B. (1996).
\newblock Exact sampling with coupled {M}arkov chains and applications to
  statistical mechanics.
\newblock In {\em Proceedings of the Seventh International Conference on Random
  Structures and Algorithms (Atlanta, GA, 1995)}, volume~9, pages 223--252.

\bibitem[Quiroz \emph{et~al.}, 2019]{quiroz2019speeding}
Quiroz, M., Kohn, R., Villani, M., and Tran, M.-N. (2019).
\newblock Speeding up {MCMC} by efficient data subsampling.
\newblock {\em J. American Statist. Assoc.}, 114(526):831--843.

\bibitem[Quiroz \emph{et~al.}, 2018a]{quiroz2018gaussian}
Quiroz, M., Nott, D.~J., and Kohn, R. (2018a).
\newblock Gaussian variational approximation for high-dimensional state space
  models.
\newblock {\em https://arXiv:1801.07873}.

\bibitem[Quiroz \emph{et~al.}, 2018b]{quiroz2018speeding}
Quiroz, M., Tran, M.-N., Villani, M., and Kohn, R. (2018b).
\newblock Speeding up {MCMC} by delayed acceptance and data subsampling.
\newblock {\em J. Comput. Graph. Statist.}, 27(1):12--22.

\bibitem[Raftery, 1996]{raftery:1996}
Raftery, A. (1996).
\newblock Hypothesis testing and model selection.
\newblock In Gilks, W., Spiegelhalter, D., and Richardson, S., editors, {\em
  {M}arkov Chain {M}onte {C}arlo in Practice}, pages 163--188. Chapman and
  Hall, New York, London.

\bibitem[Rischard \emph{et~al.}, 2018]{rischard2018unbiased}
Rischard, M., Jacob, P.~E., and Pillai, N. (2018).
\newblock Unbiased estimation of log normalizing constants with applications to
  {B}ayesian cross-validation.

\bibitem[Ritter and Tanner, 1992]{ritter:tanner:1992}
Ritter, C. and Tanner, M. (1992).
\newblock Facilitating the {G}ibbs sampler: The {G}ibbs stopper and the
  {G}riddy-{G}ibbs sampler.
\newblock {\em J. American Statist. Assoc.}, 87(419):861--868.

\bibitem[Robert, 2001]{robert:2001}
Robert, C. (2001).
\newblock {\em The {B}ayesian Choice}.
\newblock Springer-Verlag, New York, second edition.

\bibitem[Robert, 2007]{robert:2007}
Robert, C. (2007).
\newblock {\em The {B}ayesian Choice}.
\newblock Springer-Verlag, New York.

\bibitem[Robert and Casella, 2004]{robert:casella:2004}
Robert, C. and Casella, G. (2004).
\newblock {\em {M}onte {C}arlo Statistical Methods}.
\newblock Springer-Verlag, New York, second edition.

\bibitem[Robert and Casella, 2011]{robert:casella:2011}
Robert, C. and Casella, G. (2011).
\newblock A history of {M}arkov chain {M}onte {C}arlo---subjective
  recollections from incomplete data.
\newblock {\em Statist. Science}, 26(1):102--115.

\bibitem[Robert, 2015]{robert2015metropolishastings}
Robert, C.~P. (2015).
\newblock {\em The Metropolis–Hastings Algorithm}, pages 1--15.
\newblock Wiley StatsRef: Statistics Reference Online. American Cancer Society.

\bibitem[Robert \emph{et~al.}, 2018]{robert2018accelerating}
Robert, C.~P., Elvira, V., Tawn, N., and Wu, C. (2018).
\newblock Accelerating {MCMC} algorithms.
\newblock {\em Wiley Interdisciplinary Reviews: Computational Statistics},
  10(5):e1435.

\bibitem[Roberts \emph{et~al.}, 1997]{roberts:gelman:gilks:1997}
Roberts, G., Gelman, A., and Gilks, W. (1997).
\newblock Weak convergence and optimal scaling of random walk {M}etropolis
  algorithms.
\newblock {\em Ann. Applied Prob.}, 7(1):110--120.

\bibitem[Roberts and Rosenthal, 1999]{roberts:rosenthal:1999}
Roberts, G. and Rosenthal, J. (1999).
\newblock Convergence of slice sampler {M}arkov chains.
\newblock {\em J. Royal Statist. Society Series B}, 61(3):643--660.

\bibitem[Roberts and Sahu, 1997]{roberts:sahu:1997}
Roberts, G. and Sahu, S. (1997).
\newblock Updating schemes, covariance structure, blocking and parametrisation
  for the {G}ibbs sampler.
\newblock {\em J. Royal Statist. Society Series B}, 59(2):291--317.

\bibitem[Roberts and Rosenthal, 1998]{roberts1998optimal}
Roberts, G.~O. and Rosenthal, J.~S. (1998).
\newblock Optimal scaling of discrete approximations to {L}angevin diffusions.
\newblock {\em J. Royal Statist. Society Series B}, 60(1):255--268.

\bibitem[Roberts and Rosenthal, 2009]{roberts2009examples}
Roberts, G.~O. and Rosenthal, J.~S. (2009).
\newblock Examples of adaptive {MCMC}.
\newblock {\em J. Comput. Graph. Statist.}, 18(2):349--367.

\bibitem[Roberts \emph{et~al.}, 1996]{roberts1996exponential}
Roberts, G.~O., Tweedie, R.~L.,\emph{et~al.} (1996).
\newblock Exponential convergence of {L}angevin distributions and their
  discrete approximations.
\newblock {\em Bernoulli}, 2(4):341--363.

\bibitem[Rodrigues \emph{et~al.}, 2019]{rodrigues2019likelihood}
Rodrigues, G., Nott, D.~J., and Sisson, S. (2019).
\newblock Likelihood-free approximate {G}ibbs sampling.
\newblock {\em https://arXiv:1906.04347}.

\bibitem[Rosenthal, 2011]{rosenthal2011optimal}
Rosenthal, J.~S. (2011).
\newblock Optimal proposal distributions and adaptive {MCMC}.
\newblock {\em Handbook of Markov Chain Monte Carlo}, pages 93--111.
\newblock Chapman \& Hall/CRC. Eds. Brooks, S., Gelman, A., Jones, G., Meng,
  X-L.

\bibitem[Rubin, 1981]{rubin1981}
Rubin, D.~B. (1981).
\newblock The {B}ayesian bootstrap.
\newblock {\em Ann. Statist.}, 9(1):130--134.

\bibitem[Rue and Held, 2005]{rue:held:2005}
Rue, H. and Held, L. (2005).
\newblock {\em Gaussian {M}arkov Random Fields: {T}heory and Applications},
  volume 104 of {\em Monographs on Statistics and Applied Probability}.
\newblock Chapman \& Hall, London.

\bibitem[Rue \emph{et~al.}, 2009]{rue:martino:chopin:2009}
Rue, H., Martino, S., and Chopin, N. (2009).
\newblock Approximate {B}ayesian inference for latent {G}aussian models using
  integrated nested {L}aplace approximations.
\newblock {\em J. Royal Statist. Society Series B}, 71(2):319--392.

\bibitem[Scott \emph{et~al.}, 2016]{scott2016bayes}
Scott, S.~L., Blocker, A.~W., Bonassi, F.~V., Chipman, H.~A., George, E.~I.,
  and McCulloch, R.~E. (2016).
\newblock Bayes and big data: The consensus {M}onte {C}arlo algorithm.
\newblock {\em International Journal of Management Science and Engineering
  Management}, 11(2):78--88.

\bibitem[Sisson and Fan, 2011]{sisson2011likelihood}
Sisson, S. and Fan, Y. (2011).
\newblock Likelihood-free {M}arkov chain {M}onte {C}arlo.
\newblock {\em Handbook of Markov Chain Monte Carlo}, pages 313--333.
\newblock Chapman \& Hall/CRC. Eds. Brooks, S., Gelman, A., Jones, G., Meng,
  X-L.

\bibitem[Sisson \emph{et~al.}, 2019]{sisson2018handbook}
Sisson, S.~A., Fan, Y., and Beaumont, M. (2019).
\newblock {\em Handbook of Approximate {B}ayesian Computation}.
\newblock Chapman \& Hall/CRC.

\bibitem[Sisson \emph{et~al.}, 2007]{sisson:fan:tanaka:2007}
Sisson, S.~A., Fan, Y., and Tanaka, M. (2007).
\newblock Sequential {M}onte {C}arlo without likelihoods.
\newblock {\em Proc. Natl. Acad. Sci. USA}, 104(6):1760--1765.

\bibitem[Skilling, 2007]{skilling:2007}
Skilling, J. (2007).
\newblock Nested sampling for {B}ayesian computations.
\newblock {\em {B}ayesian Analysis}, 1(4):833--859.

\bibitem[Smith and Roberts, 1993]{smith:roberts:1993}
Smith, A. and Roberts, G. (1993).
\newblock {B}ayesian computation via the {G}ibbs sampler and related {M}arkov
  chain {M}onte {C}arlo methods.
\newblock {\em J. Royal Statist. Society Series B}, 55(1):3--24.
\newblock With discussion.

\bibitem[Stigler, 1986a]{stigler:1986}
Stigler, S. (1986a).
\newblock {\em The History of {S}tatistics}.
\newblock Belknap, Cambridge.

\bibitem[Stigler, 1986b]{stigler:Laplace1774}
Stigler, S. (1986b).
\newblock Memoir on inverse probability.
\newblock {\em Statistical Science}, 1(3):359--363.

\bibitem[Stigler, 1975]{stigler:1975}
Stigler, S.~M. (1975).
\newblock Studies in the history of probability and statistics. {XXXIV}
  {N}apoleonic statistics: The work of {L}aplace.
\newblock {\em Biometrika}, 62(2):503--517.

\bibitem[Stigler, 2018]{stigler2018}
Stigler, S.~M. (2018).
\newblock Richard {P}rice, the first {B}ayesian.
\newblock {\em Statist. Sci.}, 33(1):117--125.

\bibitem[Stoehr, 2017]{stoehr2017review}
Stoehr, J. (2017).
\newblock A review on statistical inference methods for discrete {M}arkov
  random fields.
\newblock {\em https://arXiv:1704.03331}.

\bibitem[Strachan and van Dijk, 2014]{strahan:2014}
Strachan, R. and van Dijk, H. (2014).
\newblock Divergent priors and well behaved {B}ayes factors.
\newblock {\em Central European Journal of Economic Modelling and Econometrics,
  CEJEME}, 6(1):1--31.

\bibitem[Syring and Martin, 2019]{syring2019calibrating}
Syring, N. and Martin, R. (2019).
\newblock Calibrating general posterior credible regions.
\newblock {\em Biometrika}, 106(2):479--486.

\bibitem[Tanner and Wong, 1987]{tanner87}
Tanner, M.~A. and Wong, W. (1987).
\newblock The calculation of posterior distributions by data augmentation.
\newblock {\em J. American Statist. Assoc.}, 82(398):528--550.
\newblock With discussion.

\bibitem[Tavar{\'e} \emph{et~al.}, 1997]{tavare:balding:griffith:donnelly:1997}
Tavar{\'e}, S., Balding, D., Griffith, R., and Donnelly, P. (1997).
\newblock Inferring coalescence times from {DNA} sequence data.
\newblock {\em Genetics}, 145:505--518.

\bibitem[Tawn \emph{et~al.}, 2020]{tawn2019weight}
Tawn, N.~G., Roberts, G.~O., and Rosenthal, J.~S. (2020).
\newblock Weight-preserving simulated tempering.
\newblock {\em Statist. Comp.}, 30:27--41.

\bibitem[Tierney, 1994]{tierney94}
Tierney, L. (1994).
\newblock {M}arkov chains for exploring posterior distributions.
\newblock {\em Ann. Statist.}, 22(4):1701--1762.
\newblock With discussion and a rejoinder by the author.

\bibitem[Tierney and Kadane, 1986]{tierney:kadane:1986}
Tierney, L. and Kadane, J. (1986).
\newblock Accurate approximations for posterior moments and marginal densities.
\newblock {\em J. American Statist. Assoc.}, 81(393):82--86.

\bibitem[Tierney \emph{et~al.}, 1989]{tierney:kass:kadane:1989}
Tierney, L., Kass, R., and Kadane, J. (1989).
\newblock Fully exponential {L}aplace approximations to expectations and
  variances of non-positive functions.
\newblock {\em J. American Statist. Assoc.}, 84(407):710--716.

\bibitem[Tierney and Mira, 1998]{tierney:mira:1998}
Tierney, L. and Mira, A. (1998).
\newblock Some adaptive {M}onte {C}arlo methods for {B}ayesian inference.
\newblock {\em {S}tatistics in Medicine}, 18:2507--2515.

\bibitem[Tokdar and Kass, 2010]{Tokdar2010}
Tokdar, S. and Kass, R. (2010).
\newblock Importance sampling: A review.
\newblock {\em Wiley Interdisciplinary Reviews: Computational Statistics}, 2:54
  -- 60.

\bibitem[Tran, 2018]{tran2018common}
Tran, K.~T. (2018).
\newblock A common derivation for {M}arkov chain {M}onte {C}arlo algorithms
  with tractable and intractable targets.
\newblock {\em arXiv:}, 1607.01985.

\bibitem[Tran \emph{et~al.}, 2019]{tran2019variational}
Tran, M.-N., Nguyen, D.~H., and Nguyen, D. (2019).
\newblock Variational {B}ayes on manifolds.
\newblock {\em https://arXiv:1908.03097}.

\bibitem[Tran \emph{et~al.}, 2017]{tran2017variational}
Tran, M.-N., Nott, D.~J., and Kohn, R. (2017).
\newblock Variational {B}ayes with intractable likelihood.
\newblock {\em J. Comput. Graph. Statist.}, 26(4):873--882.

\bibitem[van~der Vaart, 1998]{vandervaart:1998}
van~der Vaart, A. (1998).
\newblock {\em Asymptotic Statistics}.
\newblock Cambridge University Press.

\bibitem[Vanslette \emph{et~al.}, 2019]{vanslette2019simple}
Vanslette, K., Alsheikh, A.~A., and Youcef-Toumi, K. (2019).
\newblock Why simple quadrature is just as good as {M}onte {C}arlo.
\newblock {\em https://arXiv:1908.00947}.

\bibitem[Wand, 2017]{wand2017fast}
Wand, M.~P. (2017).
\newblock Fast approximate inference for arbitrarily large semiparametric
  regression models via message passing.
\newblock {\em J. American Statist. Assoc.}, 112(517):137--168.

\bibitem[Wang and Dunson, 2013]{wang:dunson:2013}
Wang, X. and Dunson, D. (2013).
\newblock Parallel {MCMC} via {W}eierstrass sampler.
\newblock {\em https://arXiv:1312.4605}.

\bibitem[Wang and Blei, 2019a]{wangblei2019b}
Wang, Y. and Blei, D. (2019a).
\newblock Variational {B}ayes under model misspecification.
\newblock In {\em Advances in Neural Information Processing Systems}, pages
  13357--13367.

\bibitem[Wang and Blei, 2019b]{wangblei2019a}
Wang, Y. and Blei, D.~M. (2019b).
\newblock Frequentist consistency of variational {B}ayes.
\newblock {\em J. American Statist. Assoc.}, 114(527):1147--1161.

\bibitem[Wilkinson, 2013]{wilkinson:2013}
Wilkinson, R. (2013).
\newblock Approximate {B}ayesian computation {(ABC)} gives exact results under
  the assumption of model error.
\newblock {\em Statistical Applications in Genetics and Molecular Biology},
  12(2):129--141.

\bibitem[Wiqvist \emph{et~al.}, 2018]{wiqvist2018accelerating}
Wiqvist, S., Picchini, U., Forman, J.~L., Lindorff-Larsen, K., and Boomsma, W.
  (2018).
\newblock Accelerating delayed-acceptance {M}arkov chain {M}onte {C}arlo
  algorithms.
\newblock {\em https://arXiv:1806.05982}.

\bibitem[Wood, 2010]{wood:2010}
Wood, S. (2010).
\newblock Statistical inference for noisy nonlinear ecological dynamic systems.
\newblock {\em Nature}, 466(7310):1102--–1104.

\bibitem[Wood, 2019]{wood2019simplified}
Wood, S. (2019).
\newblock Simplified integrated nested {L}aplace approximation.
\newblock {\em Biometrika}, 107(1):223--230.

\bibitem[Yu \emph{et~al.}, 2019]{yu2019assessment}
Yu, X., Nott, D.~J., Tran, M.-N., and Klein, N. (2019).
\newblock Assessment and adjustment of approximate inference algorithms using
  the law of total variance.
\newblock {\em https://arXiv:1911.08725}.

\bibitem[Zanella and Roberts, 2019]{zanellaroberts2019}
Zanella, G. and Roberts, G. (2019).
\newblock Scalable importance tempering and {B}ayesian variable selection.
\newblock {\em J. Royal Statist. Society Series B}, 81(3):489--517.

\bibitem[Zellner, 1971]{zellner:1971}
Zellner, A. (1971).
\newblock {\em An Introduction to {B}ayesian Econometrics}.
\newblock John Wiley, New York.

\bibitem[Zhang \emph{et~al.}, 2018]{zhang2018advances}
Zhang, C., B{\"u}tepage, J., Kjellstr{\"o}m, H., and Mandt, S. (2018).
\newblock Advances in variational inference.
\newblock {\em IEEE transactions on pattern analysis and machine intelligence},
  41(8):2008--2026.

\bibitem[Zhang and Gao, 2017]{zhang2017convergence}
Zhang, F. and Gao, C. (2017).
\newblock Convergence rates of variational posterior distributions.
\newblock {\em https://arXiv:1712.02519}.

\bibitem[Zhang, 2006a]{Zhang2006a}
Zhang, T. (2006a).
\newblock From eps-entropy to {KL} entropy: analysis of minimum information
  complexity density estimation.
\newblock {\em Annals of Statistics}, 34:2180–2210.

\bibitem[Zhang, 2006b]{Zhang2006b}
Zhang, T. (2006b).
\newblock Information-theoretic upper and lower bounds for statistical
  estimation.
\newblock {\em IEEE Trans. Info. Theory}, 52(4):1307–1321.

\end{thebibliography}
}

\end{document}